\documentclass[12pt,twoside,a4paper]{article}
\usepackage[margin=2cm]{geometry}
\usepackage[T1]{fontenc}
\usepackage[utf8]{inputenc}
\usepackage{mathrsfs}
\usepackage{amssymb,amsmath,,amsthm,color}
\usepackage{thmtools,thm-restate}
\usepackage{epsfig,subfigure}
\usepackage{enumerate}
\usepackage{caption}
\usepackage{setspace,graphicx,setspace,multirow,lscape,longtable,threeparttable,dcolumn}
\usepackage{proba,float}
\usepackage{booktabs}
\usepackage[authoryear]{natbib}
\usepackage{threeparttable}
\usepackage[normalem]{ulem}
\usepackage{arydshln}
\usepackage{xcolor}
\usepackage{adjustbox}
\usepackage{threeparttable}

\numberwithin{equation}{section}
\useunder{\uline}{\ul}{}

\def\1{\mathds{1}}

\DeclareSymbolFont{sfoperators}{OT1}{cmss}{m}{n}
\DeclareSymbolFontAlphabet{\mathsf}{sfoperators}

\makeatletter
\def\operator@font{\mathgroup\symsfoperators}
\makeatother

\begin{document}

\onehalfspacing

\title{Short-Term Covid-19 Forecast for Latecomers}

\author{
Marcelo C. Medeiros$^1$\footnote{Corresponding author. Department of Economics, Pontifical Catholic University of Rio de Janeiro. Address: Rua Marquês de São Vicente 225, Rio de Janeiro, RJ, Brazil, 22451-900. email: mcm@econ.puc-rio.br} \and
Alexandre Street$^2$ \and
Davi Valladão$^3$ \and
Gabriel Vasconcelos$^4$ \and
Eduardo Zilberman$^{5,1}$
}
\date{
\begin{small}
    $^1$Department of Economics, Pontifical Catholic University of Rio de Janeiro\\
    $^2$Department of Electrical Engineering, Pontifical Catholic University of Rio de Janeiro\\
    $^3$Department of Industrial Engineering, Pontifical Catholic University of Rio de Janeiro\\
    $^4$Bank of Communications - BBM/BOCOM\\
    $^5$Gávea Investimentos\\[2ex]%
\end{small}
    \today
}

\maketitle

\begin{abstract}
The number of new Covid-19 cases is still high in several countries, despite the vaccination of the population. A number of countries are experiencing new and worse waves. Therefore, the availability of reliable forecasts for the number of cases and deaths in the coming days is of fundamental importance. We propose a simple statistical method for short-term real-time forecasting of the number of Covid-19 cases and fatalities in countries that are latecomers -- i.e., countries where cases of the disease started to appear some time after others. In particular, we propose a penalized (LASSO) regression with an error correction mechanism to construct a model of a latecomer in terms of the other countries that were at a similar stage of the pandemic some days before. By tracking the number of cases in those countries, we forecast through an adaptive rolling-window scheme the number of cases and deaths in the latecomer. We apply this methodology to 45 different countries and we show detailed results for four of them: Brazil, Chile, Mexico, and Portugal. We show that the methodology performs very well when compared to alternative methods. These forecasts aim to foster a better short-run management of the health system capacity and can be applied not only to countries but to different regions within a country, as well. Finally, the modeling framework derived in the paper can be applied to other infectious diseases.


\noindent
\textbf{Acknowledgements:} The authors wish to thank the Associate Editor and two anonymous referees for very helpful comments, and CAPES, CNPq and FAPERJ for partial financial support. Finally, the authors are extremely grateful to all the \texttt{Covid19Analystics (www.covid19analytics.com.br)} team for the great work and the many enlightening discussions.
\end{abstract}

\section{Introduction}
Being able to forecast accurately the number of cases and deaths of infectious diseases, like Covid-19, in the very short-run, say the next few days or weeks, is crucial to manage properly the health system. Depending on the next days pressure on the health system capacity, one can make more informed decisions on how to allocate hospital beds and ventilators, on whether to set more field hospitals, on whether to train more health workers, and so on and so forth. Furthermore, with accurate forecasting models can help tracing the evolution of new variants of the disease and the effects of the vaccination of the populations.

In this paper, we propose a statistical method to forecast in real-time the very short-run evolution of the number of Covid-19 cases and deaths in countries that are latecomers. Given that these latecomers were hit by the Covid-19 pandemic only after other countries, we can use information from these other countries when they were at a similar stage of the pandemic, a few days or weeks before. A latecomer country is one where Covid-19 cases/deaths were identified after others. Note that the labeling of a given country as a latecomer depends on the number of countries we would like to be ahead. We use a penalized Least Absolute Selection and Shrinkage Operator (LASSO) regression proposed by \citet{tibshirani1996regression}, to construct an error correction model (\texttt{ECM}) of a latecomer in terms of the other countries. The idea behind the \texttt{ECM} is to adjust the short-run dynamics of the latecomer to departures from the equilibrium (long-run relation) between the latecomer country and its peers. By tracking the number of cases in those countries, we can forecast the evolution of the disease in the latecomer few days ahead. The forecasts for the number of deaths are constructed as a linear regression on the number of cases. As the pandemic evolves, one can run the model on a daily basis, and in a adaptive rolling window scheme, to obtain updated forecasts for the next days. The model is easily estimated and confidence intervals can be computed in a straightforward manner by simulation techniques.

The rolling window (adaptive) scheme is important to acknowledge the dynamic nature of the pandemic and to attenuate the effects of outliers and potential structural breaks (due to, for example, more or less testing after a given period, policy changes, start of vaccination campaigns countries, new variants of the virus, or changes in the relations between countries used as explanatory variables and the latecomer). Nonetheless, it is important to emphasize that despite this attenuation, one might expect a worsen of the forecasts a few days after a structural break as the model adapts. Hence, and needless to say, the use of the proposed forecasting method should be complemented with evaluations on how the pandemic is evolving.\footnote{In the case of Brazil and other less developed countries, for example, the proposed method might not anticipate the acceleration in cases and fatalities after the Covid-19 reaches areas with high urban density that lack proper sanitation. But as the model adapts, we expect to get more reliable forecasts under this new stage of the pandemic's evolution.}

We apply the methodology to 45 different countries and present detailed results for four of them: Brazil, Chile, Mexico, and Portugal. The 45 countries are selected based on the following criteria: (1) must have at least six countries ahead 14 days or more; (2) has a population of at least 10 million people and (3) had at least 30,000 reported cases of Covid-19 by the end of 2020. We show that it has been performing very well in forecasting the out-of-sample number of cases and deaths up to the next 14 days. The number of cases used correspond to those that are detected by the health system, which is the proper measure to track if the concern is to evaluate the impact on its capacity.

Tracking the evolution of the Covid-19 has been posing several challenges. The proposed method overcomes some of them. First, standard epidemiological models used to track the evolution of an epidemic are hard to discipline quantitatively to a new disease. Despite the enormous effort worldwide to understand transmission, recovery and death rates, many parameters one needs to calibrate remain uncertain \citep{atkeson2020}, and behavioral responses of individuals as well as containment policies should affect these parameters \citep{eichenbaum2020macroeconomics}.\footnote{Epidemiologists and researchers from other fields rushed to improve those models and provide simulations on the spread of the disease, some of them taking into account counteracting policy and/or behavioral responses. A very incomplete list includes \cite{berger2020SEIR}, \cite{KUCHARSKI2020lancenet}, \cite{walker2020imperialcollege}, \cite{WU2020}, and \cite{bastos2020brasil}.} The proposed forecasting method, instead, has the advantage of being model-free, and makes projections based solely on available data.

Second, even if possible to discipline those epidemiological models reliably, they speak to the evolution of the infected population. From the perspective of managing health resources, the relevant figure is the number infected individuals that end up pressuring the health system. Note that a lot of individuals who end up being infected are asymptomatic or do not need to access the health system. Hence, the evolution of the virus among the sub-population that actually needs health care, which possesses certain characteristics that differ from the rest of the population, might be different from the evolution in the whole population. The proposed method avoid this problem by forecasting directly the number of infected individuals who are detected by the health system or specific parts of it that are of interest to be monitored, e.g., specific regions within a country.

Finally, alternative methods to track the evolution of the Covid-19, and forecast the pressure on health resources, such as massive testing, are expensive and unavailable to many countries. The  methodology and our codes are immediately and cheaply reproducible to any latecomer that tracks the number of Covid-19 cases (and deaths). Note also that the proposed methodology can be as well applied to different regions within a country. This is particularly useful in large countries as Brazil or India, where the disease have hit distinct regions with delays.\footnote{We post on a daily basis updated forecasts for Brazil, methodology updates, and codes. The domain is https://covid19analytics.com.br/, and one can check there for updates.}

The aforementioned challenges are even harder to overcome in poor and developing countries, mostly latecomers, due to the lack of high-quality research, reliable data and limited budget. By tracking the very short-run evolution of the number of Covid-19 cases (and deaths) in real-time, we hope that this methodology can be useful to inform policymakers and the general public. In the authors' point of view, an adaptive and accurate data-driven forecast is critical to foster better management of the health system, especially in those countries that lack proper resources.

\subsection{Main Takeaways}

The \texttt{ECM} method proposed in this paper provides forecasts for cases and deaths with, in general, lower mean absolute percentage errors (MAPE) than two benchmark models. Namely, a simple quadratic trend regression (\texttt{trend}) that has been shown to be quite precise for short-term forecasts of Covid-19 fatalities \citep{bench2020} and an integrated autoregressive model of order one (\texttt{AR}).\footnote{We tried many ARIMA specifications and the ARIMA(1,1,0) was the one with superior performance.}

Over the 14 forecasting horizons, on average the \texttt{ECM} approach produces smaller MAPEs than the \texttt{AR} benchmark in 51.10\% of the countries studies when the dependent variable is the number Covid-19 cases and in 65.50\% in case of deaths. When comparing with the \texttt{trend} model the percentages are 85.24\% and 82.06\%, respectively for cases and deaths. When comparing with the \texttt{AR} model, the gains of the \texttt{ECM} alternative are more noticeable for forecasting horizons larger than two days. On the other hand, the \texttt{ECM} overperforms the \texttt{trend} benchmark for all horizons. Finally, the results improve remarkably when we consider simple average combination of the \texttt{ECM} and \texttt{AR} forecasts.

When we turn attention to the four countries that we provide an in-depth analysis, the following results emerge. For Brazil, the \texttt{ECM} delivers MAPEs for Covid-19 cases between 0.44\% (1-day-ahead) and 3.94\% (14-days-ahead) while the \texttt{trend} model MAPEs range between 0.76\% and 4.86\%. The \texttt{AR} benchmark MAPEs lie between 0.38\% and 6.40\%. Although, the performance of the \texttt{ECM} is inferior to the one from the \texttt{AR} model for 1-day-ahead, it gets way better than the later for horizons for two or more days. For Chile, the \texttt{ECM} yield MAPEs between 0.32\% and 3.99\%, whereas the \texttt{trend} approach has MAPEs between 0.48\% and 4.09\%. The MAPEs from the \texttt{AR} models range between 0.32\% and 5.22\%. It is clear the superiority of the \texttt{ECM} in the case of Chile. For Mexico, our approach delivers MAPEs ranging from 0.23\% and 2.41\% while the \texttt{trend} alternative yields MAPEs between 0.41\% and 2.78\%. The MAPEs from the \texttt{AR} model lie between 0.23\% and 4.16\%. Finally, we report the results for Portugal. The \texttt{ECM} delivers MAPEs in the range between 0.25\% and 5.18\%, while the pure trend model has MAPEs between 0.74\% and 6.83\%. However, in this case the \texttt{AR} has the best performance with MAPEs between 0.22\% and 4.17\%. Nevertheless, when we consider a simple average of the forecasts from the \texttt{ECM} and \texttt{AR} models, the MAPEs reduce to 0.21\% and 3.94\%.

Now, we analyze the results concerning deaths. For Brazil, the MAPEs of the \texttt{ECM} range between 0.49\% and 4.13\%. The \texttt{trend} (\texttt{AR}) model yield MAPEs between 0.83\% (0.35\%) and 4.96\% (5.39\%). For Chile the \texttt{ECM} delivers the minimum MAPE of 0.73\% for 1-day-ahead forecasts and the maximum of 5.77\% for the two-weeks-ahead horizon. The \texttt{trend} benchmark has MAPEs between 1.44\% and 10.60\%. The minimum MAPE of the \texttt{AR} model is 0.58\%. The maximum MAPE is of the order of $10^6\%$ due to a major outlier. When this forecast is removed, the maximum MAPE becomes 8.25\%. For Mexico, our model produces MAPEs between 0.61\% and 3.17\%. On the other hand, the \texttt{trend} method has MAPEs in the range between 0.55\% and 8.23\%. The MAPEs of the \texttt{AR} benchmark lie between 0.39\% and 6.70\%. For Portugal, the minimum (1-day-ahead) and maximum (14-days-ahead) MAPEs of the \texttt{ECM} are 0.30\% and 4.14\% . The \texttt{trend} model yields MAPEs between 0.77\% and 7.09\% while the \texttt{AR} alternative has MAPEs ranging from 0.14\% and 3.59\%. It is clear that for Portugal, the \texttt{AR} benchmark is the model with the lowest MAPEs. However, as before, the simple average combination of the \texttt{ECM} and \texttt{AR} models brings the MAPEs to 0.17\% and 3.53\%. Although the forecast combination is not able to beat the \texttt{AR} model for 1-day-ahead, it outperforms the \texttt{AR} benchmark for horizons larger than 4 days.

Since the outbreak of the Covid-19 pandemic, a large number of papers dealing with short-term forecasts of cases and deaths counts has been published in a wide collection of academic journals. The models range from different versions of epidemiological compartmental models to pure statistical and machine learning approaches. The models can be as simple as a pure trend regression or as complicated as deep learning neural networks. Nevertheless, a number of studies provide strong evidence that is quite difficult to beat the simplest alternatives. Our approach keeps the simplicity of several statistical models, explores equilibrium relations between a latecomer country and its peers, and shows robustness against many breaks in the dynamics of the series over the years of 2020 and 2021.

\citet{bench2020} compare the predictive accuracy of forecasts for the number of fatalities produced by several forecasting teams and collected by the United States Centers for Disease Control and Prevention (CDC) and a simple benchmark alternative. The set of models include both statistical (dynamic growth model) and compartmental approaches (different versions of SEIRD models). The authors find that a simple quadratic trend regression outperforms all the alternatives for horizons up to one week ahead. For longer horizons, some of the models sometimes outperform the simple benchmark. However, the authors show that the ensemble of models outperform all the other alternatives. Similar quadratic trend models are also considered with some adjustments by \citet{fJzZxS2021} and \citet{sLoL2021}. Due to the previous satisfactory performance of the quadratic trend model, we also adopt it as a benchmark specification in this paper.

\citet{doornik21} consider a flexible trend model and apply it to forecast confirmed cases and deaths in a large number of countries. As ours, their model has no epidemiological component. For confirmed cases, the authors report MAPEs between 0.4\% and 2.1\% for one-day-ahead and from 1.7\% to 7.6\% for four-days-ahead. In the case of death counts, the MAPEs are higher. Although not directly compared to our MAPEs, as we are not analysing the same countries, the MAPEs reported here are lower than the ones in the above mentioned paper. See also \citet{petro21} for a similar approach to \citet{doornik21}.

Nonlinear machine learning methods such as, Long Short-Term Memory and Deep Neural Networks, Random Forests and Support Vector Machines, have also been considered to forecast cases and death counts in the short-run. For example, \citet{ribeiro20} estimate a vast amount of statistical and machine learning methods to forecast future cases in Brazil. Not only their models are much more complex than the ones considered here, but their MAPEs range between 0.87\%–3.51\%, 1.02\%–5.63\%, and 0.95\%–6.90\% for one, three, and six-days-ahead forecasts, respectively. These numbers are systematically larger than the MAPEs from our \texttt{ECM} specification. Other examples of application of nonlinear machine learning models are \citet{zeroual20}, \citet{chimmula20}, and \citet{silva20}, among many others.\footnote{None of these papers provide convincing evidence of the superiority of complicated machine learning models to simpler alternatives.}

\subsection{Organization of the Paper}

In addition to this Introduction, this paper is organized as follows. Section 2 presents the methodology. Section 3 gives some guidance to practitioners. Section 4 describes the results. Finally, Section 5 concludes.

\section{Methodology}

Let $\tau=1,2,\ldots,\mathcal{T}$, represents the number of days after the 100th confirmed case of Covid-19 in a given country/region.\footnote{We choose the threshold of 100 cases to avoid modeling the evolution in the very early days of the pandemic when reliable data were not available in several countries.} Define $y_{\tau}$ as the natural logarithm of the number of confirmed cases $\tau$ days after the 100th case of the disease in this specific country/region. In addition, let $\boldsymbol{x}_{\tau}$ be a vector containing the natural logarithm of the number of reported cases for $p$ other countries also $\tau$ days after the 100th case has been reported and a quadratic trend, i.e, $\boldsymbol{x}_{\tau}$ also includes $\tau$ and $\tau^2$. The idea is that, in the regular time scale, $\boldsymbol{x}_{\tau}$ may be ahead of time of $y_{\tau}$. For example, in France and Spain, the 100th was reported on February 29, 2020 and March 2, 2020, respectively. On the other hand, in Brazil, a latecomer, the 100th case was confirmed only on March 14, 2020. Therefore, the idea is to use, for instance, data from France on February 29 and Spain on March 2 to explain the number of cases in Brazil on March 14. Note that we do not claim any causal link between the $p$ countries and the latecomers. Our proposal explores the fact the evolution of the disease in different countries share similar patterns.\footnote{These similarities are also explored in compartmental models. See also, \citet{covid2020}.}

\begin{figure}%[H]
\centering
\includegraphics[width=\textwidth]{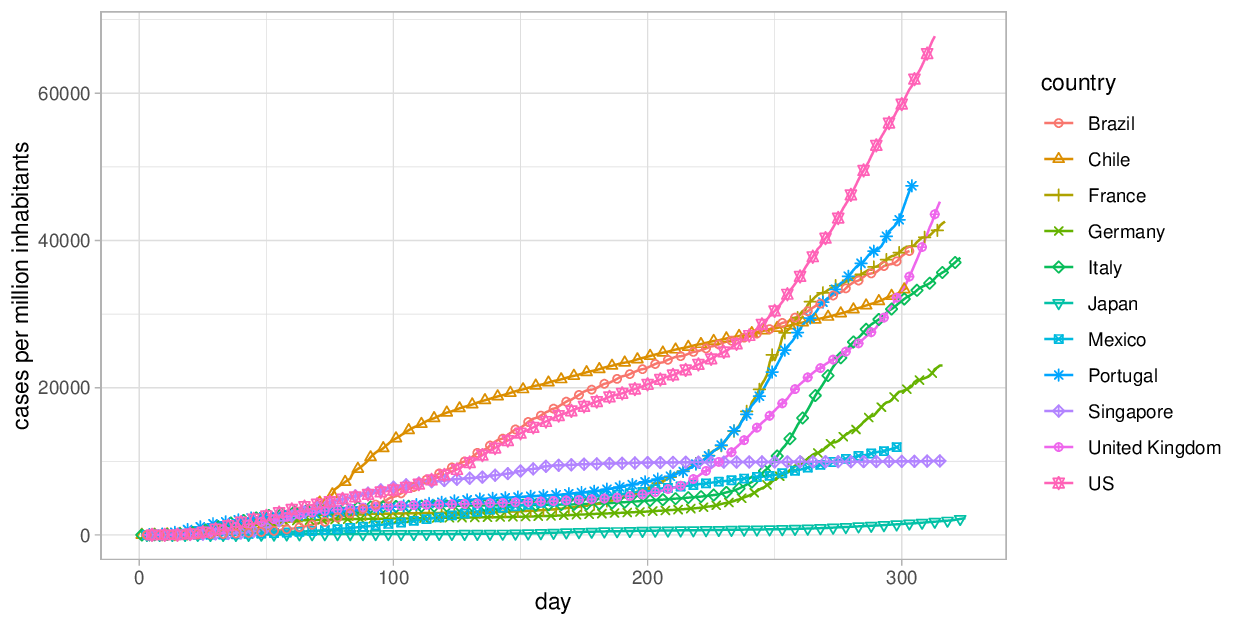}
\caption{Evolution of cases in different countries in epidemic time.}
\label{F:fig1}
\begin{minipage}{\linewidth}
\begin{footnotesize}
The figure illustrates the evolution of the cases of Covid-19 in different countries according to the epidemic calendar, i.e., the $x$-axis represents days from the first confirmed case of Covid-19. It is clear that come countries are in front of others in epidemic time.
\end{footnotesize}
\end{minipage}
\end{figure}

The statistical approach considered in this paper is a simple error correction model (\texttt{ECM}) which maps $\boldsymbol{x}_{\tau}$ into $y_{\tau}$ as:
\begin{equation}\label{E:ecm}
\Delta{y_{\tau}} =\Delta\boldsymbol{x'}_{\tau}\boldsymbol{\pi} + \gamma\left(y_{\tau -1} - \boldsymbol{x}'_{\tau-1}\boldsymbol{\beta}\right)+ u_{\tau},
\end{equation}
where $u_{\tau}$ is zero-mean random noise with variance $\sigma^2$, $\Delta y_{\tau}=y_{\tau}-y_{\tau-1}$, $\Delta \boldsymbol{x}_{\tau}=\boldsymbol{x}_{\tau}-\boldsymbol{x}_{\tau-1}$, and $\boldsymbol\pi$, $\gamma$, and $\boldsymbol\beta$ are unknown parameters to be estimated. As can be seen in Figure \ref{F:fig1}, the number of cases in $\tau$-scale in different countries display a strong common exponential trend. The logarithm transformation is important to turn the exponential trend into a linear one.

The model is estimated in two steps. In the first step, we estimate $\boldsymbol{\beta}$ in a long-run equilibrium model:
\begin{equation}\label{E:model}
y_{\tau}=\boldsymbol{x}'_{\tau}\boldsymbol{\beta}+\varepsilon_{\tau},
\end{equation}
where $\varepsilon_{\tau}$ is a zero-mean second-order stationary error term.

Due to the limited amount of data and the large dimension of $\boldsymbol{x}_t$ as compared to the sample size, we use the least absolute and shrinkage operator (LASSO) to recover the parameter vector. The goal of the LASSO is to balance the trade-off between bias and variance and is a useful tool to select the relevant peers in an environment with very few data points. Therefore, the estimator of the unknown parameter $\boldsymbol{\beta}_\mathcal{T}$ in equation \eqref{E:model} is defined as:
\begin{equation}
\widehat{\boldsymbol{\beta}}_{\mathcal{T}}=\arg\underset{\boldsymbol{\beta}}{\min}\left[\frac{1}{K}\sum_{\tau=\mathcal{T}-K+1}^{\mathcal{T}}\left(y_{\tau}-\boldsymbol{x}_{\tau}'\boldsymbol\beta\right)^2+\lambda\sum_{j=1}^p|\beta_j|\right],
\end{equation}
where $K$ is the number of days in the estimation window, and $\lambda>0$ is the penalty parameter. Theoretical justification for the use of LASSO to estimate the parameters in this setup with trends can be found in \citet{rMmcM2019}.

Once we estimated equation (\ref{E:model}), we proceed to a second step estimating the \texttt{ECM} by Ordinary Least Squares (OLS) with the variables selected in the first step with the LASSO. The final prediction $h$--step ahead from $\mathcal{T}$ reads as:
\begin{equation} \label{E:yecm}
     \widehat{y}_{\mathcal{T}+h} = \Delta\boldsymbol{x}'_{\mathcal{T}+h}\widehat{\boldsymbol{\pi}}_{\mathcal{T}} - \widehat{\gamma}_{\mathcal{T}}  \boldsymbol{x}'_{\mathcal{T}+h-1}\widehat{\boldsymbol{\beta}}_{\mathcal{T}} + \left(1+\widehat{\gamma}_{\mathcal{T}}\right)\widehat{y}_{\mathcal{T}+h-1},
     %\widehat{y}_{\tau+h} = \Delta\boldsymbol{x}'_{\tau+h}\widehat{\boldsymbol{\pi}}_{\mathcal{T}} - \widehat{\gamma}_{\mathcal{T}}  \boldsymbol{x}'_{\tau+h-1}\widehat{\boldsymbol{\beta}}_{\mathcal{T}} + \left(1+\widehat{\gamma}_{\mathcal{T}}\right)\widehat{y}_{\tau+h-1},
\end{equation}
where $\widehat{y}_{\mathcal{T}+h-1}$ is the forecast for the previous day.
Confidence intervals were obtained through simulation by assuming that the error term $u_{\tau}$  in \eqref{E:ecm} is normally distributed.

The intuition behind the proposed \texttt{ECM} is to model the dynamics and the reactions to departs from the equilibrium between $y_{\tau}$ and $\boldsymbol{x}_{\tau}$: the disease behaves somehow in a similar fashion in different countries. Note that we do not claim a causal link, for instance, from the cases in Germany to the cases in Brazil due to mobility among these two countries. What the model explores is that the evolution of diseases like Covid-19 seems to share similar patterns in different locations. Furthermore, as the first-stage LASSO regression is a model selection tool, if our hypothesis of common dynamics among countries is not valid, the LASSO will not select any country to explain the latecomer behavior and/or the residuals of the first stage LASSO regression present statistical evidence of non-stationarity.

Our interest relies on the forecasts for number of cases in levels not in logs: $Y_{\tau}:=\exp(y_{\tau})$. Therefore, for the horizon $\mathcal{T}+h$, the forecasts are constructed as:
\begin{equation}
\widehat{Y}_{\mathcal{T}+h}=\widehat{\alpha}_{\mathcal{T}} \, e^{\widehat{y}_{\mathcal{T}+h}},
\end{equation}
where $\widehat{\alpha}_\mathcal{T}=\frac{1}{K}\sum_{\tau=\mathcal{T}-K+1}^{\mathcal{T}}\exp(\widehat{u}_{\tau}):=\frac{1}{K}\sum_{\tau=\mathcal{T}-K+1}^{\mathcal{T}}\exp\left(y_{\tau}-\widehat{y}_{\tau}\right)$ is a correction which is essential to attenuate the induced bias when we take the exponential of the forecasted value of $y_{\mathcal{T}+h}$; see \citet{jW2019}. Note that, in the $\tau$-scale, the peers are ``in the future'' and we can plug-in actual values of $\boldsymbol{x}_{\mathcal{T}+h}$ to construct our forecasts. Note also that a rolling estimation window of $K=28$ days induce an adaptive forecasting framework suitable to capture the dynamic nature of the pandemic and to attenuate the effects of outliers and potential structural breaks. Finally, in order to give more weight to the newest observations, we inflated the data by repeating the last four observations where the last observation is repeated four times with a linear decay for the observations before.

It is worth emphasizing that this model is only a local approximation of a more complex and dynamic process. Therefore, its best use relies on fresh and updated data, and the rolling window scheme takes care of that. Although the model has been providing to have excellent adherence, the proposed forecasting method may be complemented with indexes, such as proxies for social distancing, to guide evaluation of the future dynamics affecting the number of cases and deaths. However, the inclusion of other regressors such as Google mobility has not showed to improve the quality of the forecasts.

\section{Guide to Practice}\label{S:Guide}
The implementation of the proposed forecasting method requires the choice of three tuning parameters: the penalty term in the first-stage LASSO regression ($\lambda$), the length of the estimation window ($K$) and the data inflation mechanism.

The penalty term is selected by the Bayesian Information Criterion (BIC) as discussed in \citet{mcMeM2016}. The degrees of freedom of the LASSO are determined by the nonzero estimated coefficients. Cross-validation can be also used to determine the penalty parameters. However, we prefer the BIC in order to avoid any extra computational burden.

The estimation window length ($K$) and the inflation scheme for the most recent observations can be estimated in a rolling window process. The motivation for the rolling window estimation and the data inflation is to give weight to more recent observations and attenuate potential structural breaks. Before computing the actual forecasts, one could select these tuning parameters from a rolling window using previous data and choosing the values that minimize the out-of-sample error measure (MAPE, for example). However, this procedure gives us the best model for past data, especially because we need a significant number of windows that go back several weeks to obtain stable estimates. Although a procedure like this could lead to some local improvements, it could also lead to situations where the forecast explodes, especially when a very small $K$ is selected with no data inflation. To avoid unnecessary data mining that could lead to unreliable results, we choose to use a fixed value of $K = 28$ and the inflation scheme for the four most recent lags. We understand that this sample size is enough to get a satisfactory model given the number of variables and it is not too big to include many structural breaks. The inflation scheme is just to give more weight to the most recent data, which is likely to be more similar to future data in the short-run. Small changes in the inflation strategy do not affect the overall results. However, no inflation yields higher errors.

Another important point is to check if the first-stage errors are stationary. This can be conducted by common unit-root tests. If the null hypothesis of unit-root is not rejected, the first stage is clearly misspecified and the forecasts will not be reliable. In this paper, we run unit-root tests after every LASSO estimation and there is no evidence of misspecification.

Finally, the \texttt{ECM} methodology proposed here is flexible enough to include other regressors, such as, for example, mobility data. However, in our experience the inclusion of such data did not bring any evident improvement in the performance of the model.

\section{Data}

We used the John Hopkins compiled data\footnote{John Hopkins data available at https://github.com/CSSEGISandData/COVID-19} for all countries with Covid-19 cases and the Brazilian Ministry of Health official data.\footnote{Brazilian Ministry of Health data available at https://covid.saude.gov.br/}. The data are organized in epidemiological time, i.e., the time dimension represents the number of days after the 100th case.

The models were computed on a rolling window scheme with 28 in-sample observations per window. For each country, the model estimation started when the number of confirmed cases of Covid-19 reached 20,000. The last in-sample day for every country was June 30, 2021. As described in Section \ref{S:Guide}, we setting the window length to 28 days turned out to be a good trade-off between the quality of the in-sample adjustment and robustness to potential structural breaks.

Figure \ref{F:fig1} illustrates the evolution of Covid-19 cases in several countries. The data is displayed in epidemiological time, i.e., the $x$-axis represents the number of days since the first registered case. It is clear from the figure that some countries are ahead of epidemiological time than others.

\section{Results}

\subsection{Benchmark Models}

In other to compare the performance of the \texttt{ECM} model proposed in this paper we consider the benchmark model as described in \citet{bench2020}. The model is a simple quadratic trend regression (\texttt{trend}) defined as:
\begin{equation}\label{eq:bench}
y_{\tau}=\alpha_0 + \alpha_1\tau + \alpha_2\tau^2+\epsilon_{\tau},
\end{equation}
where $\epsilon_{\tau}$ is a zero-mean error term. As mentioned before, although this bechmark is amazingly simple, it proved to be quite precise for short-term forecasts.

We also consider an integrated first-order autoregressive \texttt{AR} model:
\begin{equation}\label{eq:ar}
\Delta y_{\tau}=\phi_0 + \phi_1\Delta y_{\tau-1} +\eta_{\tau},
\end{equation}
where $\eta_{\tau}$ is a zero-mean error term.

\subsection{Mean Absolute Percentage Errors}

We start by reporting summary results for all 45 countries considered. Tables \ref{tab_cases_stat} and \ref{tab_deaths_stat} report summary statistics for the forecasts of cases and deaths, respectively. More specifically, the tables show descriptive statistics for the ratio of the forecasting mean absolute percentage error (MAPE) of either \texttt{ECM} or \texttt{trend} alternatives and the \texttt{AR} benchmark. Ratios below one indicate that the alternative models overperform the benchmark. For cases, the medians of the ratio of the MAPEs from the \texttt{ECM} and the \texttt{AR} models are below one for horizons between three and 12 days ahead. When deaths are considered the medians are below one for all horizons larger than two. Furthermore, the \texttt{ECM} is way better than the \texttt{trend} benchmark, which is frequently beaten by the \texttt{AR} model. The medians of the ratios over the 14 forecasting horizons and all countries are 0.99 for cases and 0.78 for deaths, indicating the superiority of the \texttt{ECM} over the \texttt{AR} model.

Now we turn attention to four specific countries. Tables \ref{tab_cases} and \ref{tab_deaths} present the forecasting results for the full out-of-sample period. The tables show the MAPE for forecasting horizons of 1 to 14 days ahead of the Covid-19 accumulated number of cases (Table \ref{tab_cases}) and deaths (Table \ref{tab_deaths}) for Brazil, Chile, Mexico and Portugal. Values in parenthesis are $p$-values for the Giacomini \& White test for superior predictive ability \citep{rGhW2006}. The null hypothesis of the test is that both forecasts have the same MAPE.

We start by comparing the models with respect to the forecasts for case counts. For Brazil, the \texttt{ECM} has lower MAPEs than the \texttt{AR} benchmark for all horizons larger than one-day-ahead and the differences are all significant. It is also clear from the table that the \texttt{ECM} outperforms the \texttt{trend} model. For Chile, the \texttt{ECM} is better than the benchmark for all horizons. The differences are statistically significant for 12 out 14 horizons. For Mexico, the \texttt{ECM} is superior to the benchmark for all horizons larger than one and the differences are significant. Finally, for Portugal, the \texttt{ECM} has a much worse performance and is superior to the \texttt{AR} alternative only for horizons between two- and six-days-ahead, but the differences are only significant for three-days-ahead. However, the \texttt{ECM} strongly  outperforms the \texttt{trend} model.

Turning the attention now to fatalities, the \texttt{ECM} model is clearly superior to the benchmark and the differences are much more pronounced.
For Brazil, the \texttt{ECM} is better than the \texttt{AR} for all horizons larger than one day and the differences are strongly significant. Our proposal also shows much better results than the \texttt{trend} alternative. For Chile and Mexico the results are similar. However, we should point out the for Chile there is a strong outlier in the forecasts of the \texttt{AR} models that distort the results with ratios close to zero for larger horizons. When we remove this specific outlier the ratio for one-day-ahead is 1.323 and range between 0.302 and 0.868, still indicating the superiority of the \texttt{ECM} for most horizons. For Portugal, the \texttt{AR} model delivers the best results for all horizons, followed by the \texttt{ECM}.

\begin{figure}%[H]
\centering
\includegraphics[width=\textwidth]{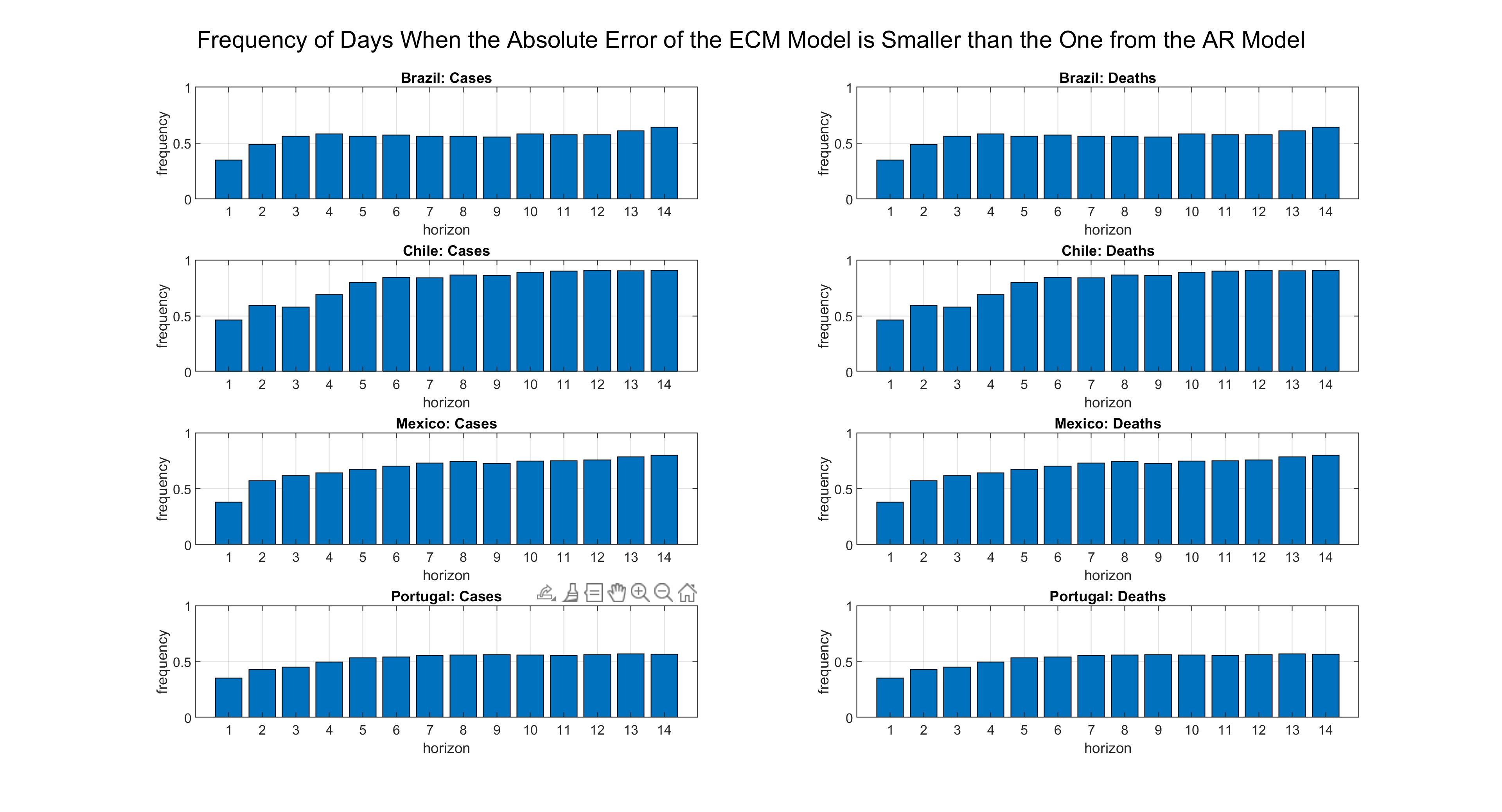}
\caption{Frequency of days when \texttt{ECM} is better than the \texttt{AR} model}
\begin{minipage}{\linewidth}
\begin{footnotesize}
The figure illustrates for different countries and horizons, the frequency of days when the absolute error of the \texttt{ECM} is smaller than the one from the \texttt{AR} specification.
\end{footnotesize}
\end{minipage}
\label{F:freq1}
\end{figure}

\begin{figure}%[H]
\centering
\includegraphics[width=\textwidth]{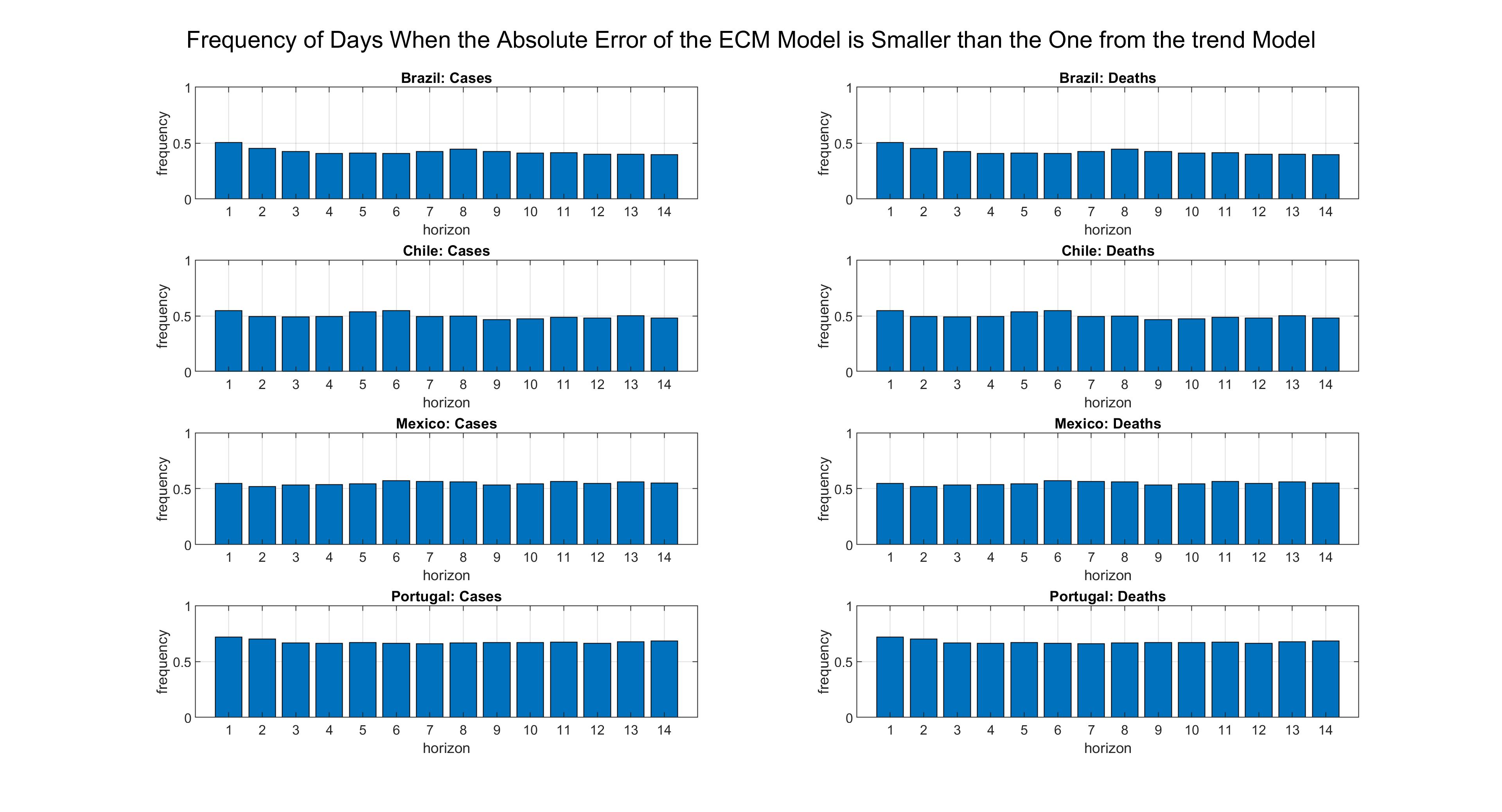}
\caption{Frequency of days when \texttt{ECM} is better than the \texttt{trend} model}
\begin{minipage}{\linewidth}
\begin{footnotesize}
The figure illustrates for different countries and horizons, the frequency of days when the absolute error of the \texttt{ECM} is smaller than the one from the \texttt{trend} specification.
\end{footnotesize}
\end{minipage}
\label{F:freq2}
\end{figure}

\begin{figure}%[H]
\centering
\includegraphics[width=\textwidth]{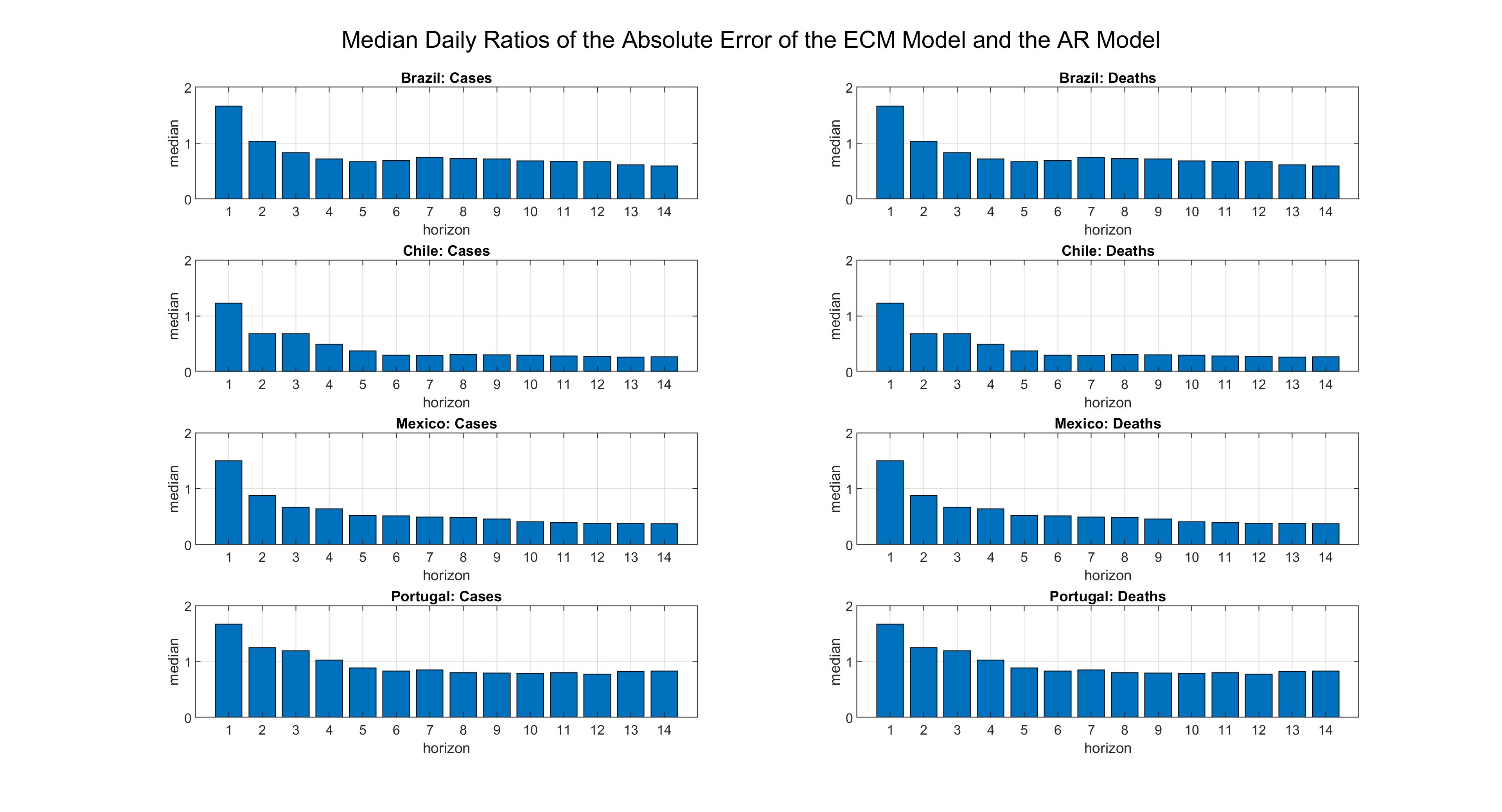}
\caption{Median of daily absolute error ratios}
\begin{minipage}{\linewidth}
\begin{footnotesize}
The figure illustrates for different countries and horizons, the median of the daily ratio of absolute errors of the \texttt{ECM} the one from the \texttt{AR} specification.
\end{footnotesize}
\end{minipage}
\label{F:ratio1}
\end{figure}

\begin{figure}%[H]
\centering
\includegraphics[width=\textwidth]{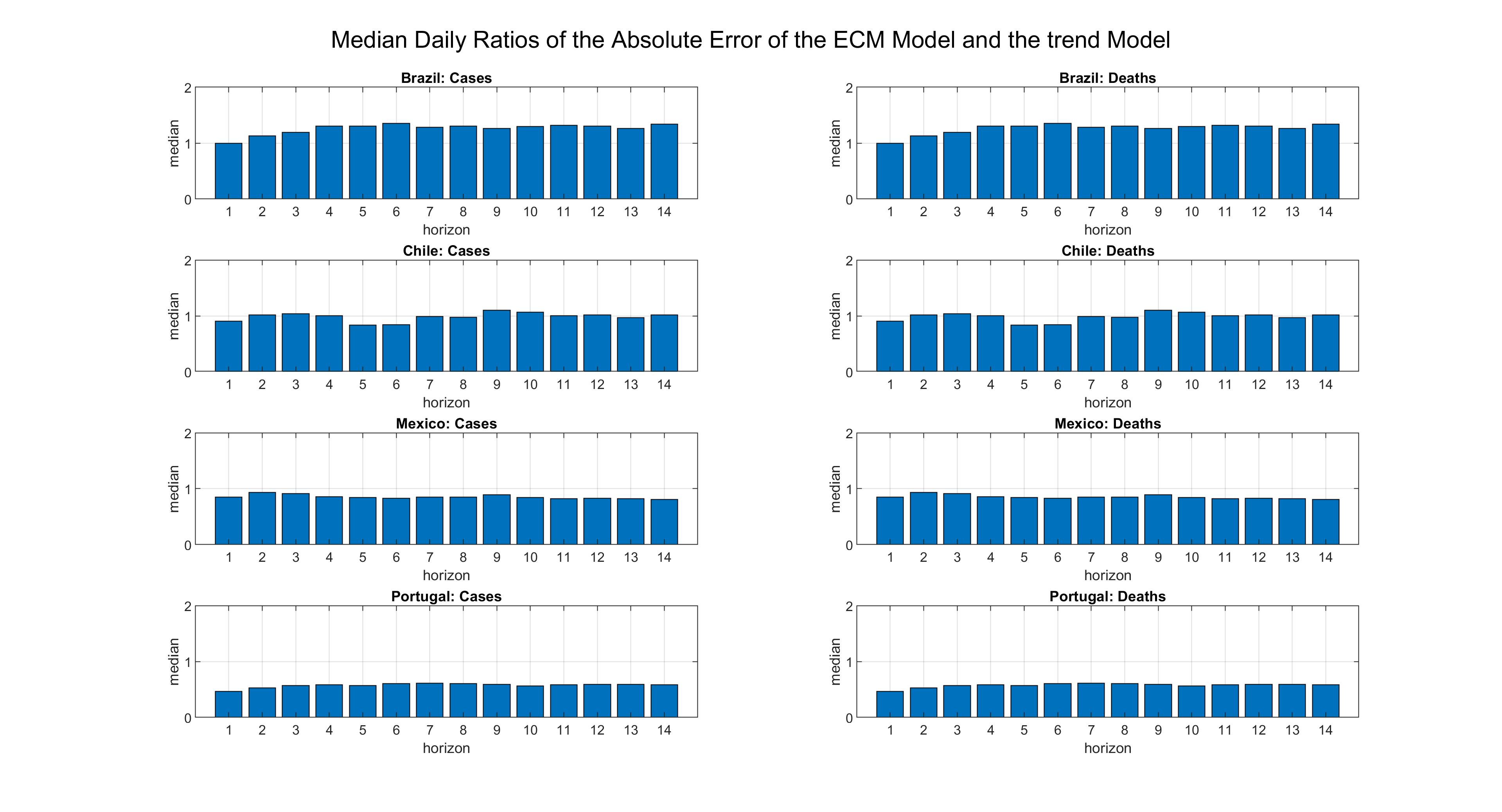}
\caption{Median of daily absolute error ratios}
\begin{minipage}{\linewidth}
\begin{footnotesize}
The figure illustrates for different countries and horizons, the median of the daily ratio of absolute errors of the \texttt{ECM} the one from the \texttt{AR} specification.
\end{footnotesize}
\end{minipage}
\label{F:ratio2}
\end{figure}

In order to complement the analysis, we compute the frequency of days when the \texttt{ECM} has a lower absolute percentage error than the benchmark and the median of the ratio of the daily absolute errors of the \texttt{ECM} and benchmark specification over the forecasting sample. Figures \ref{F:freq1} and \ref{F:freq2} report, for each forecasting horizon, the frequency of days over the sample when the daily absolute percentage error of the \texttt{ECM} is smaller than the one from the \texttt{AR} and \texttt{trend} alternatives, respectively. The first column in the figures present the results for cases whereas the second column shows the numbers of deaths. To quantify these gains, Figures  \ref{F:ratio1} and \ref{F:ratio2} present, for each horizon, the median of the ratios of the daily absolute errors. A number less than one favors the \texttt{ECM} model. As before, the first (second) column concerns cases (deaths). We prefer the use the median instead of the mean to avoid potential effects of outliers.

For Brazil, the results are favourable to the \texttt{ECM} as compared to the \texttt{AR} alternative for horizons larger than one-day-ahead. Both Figures \ref{F:freq2} and \ref{F:ratio2} point to the superiority of the benchmark. However, when looking at the results in Tables \ref{tab_cases} and \ref{tab_deaths} above, we may reach a different conclusion. Therefore, it is important to uncover the drivers to the best MAPE of the \texttt{ECM} over the entire out-of-sample period. The reason for this finding is that the \texttt{ECM} is way superior to the benchmark during the first 100 days of the sample. This was the period when the number of cases and deaths in Brazil was accelerating the most.

For Chile, the one-day-ahead forecasts of \texttt{ECM} are the best ones in 72.17\% (61.75\%) of the days when cases (deaths) are considered. These numbers drop to 46.52\% (53.91\%) one the forecasting horizon is set to 14 days. For almost all the horizons the proportion of days when the \texttt{ECM} is better than the benchmark is larger than 50\%. From the analysis of the results in Figure \ref{F:ratio1}, it is clear that the \texttt{ECM} is better than the benchmark for almost all horizons when cases are considered. When fatalities are analyzed, the \texttt{ECM} is always superior.

For Mexico, the results are very supportive to the \texttt{ECM} specification. The \texttt{ECM} is superior to the benchmark in more than 50\% of the days in almost every case considered in the analysis. The median ratios of the absolute percentage errors are always bellow one, when Covid-19 cases are considered. For deaths, the ratios are bellow one for horizons larger than four-days-ahead.

Finally, for Portugal, the superiority of the \texttt{ECM} draws attention. For all horizons considered, the \texttt{ECM} is better than the benchmark in terms of number of days with lower errors as well as in terms of the relative magnitude of the absolute percentage errors.

\subsection{Diagnostic Tests}

\begin{figure}%[H]
\centering
\includegraphics[width=\linewidth]{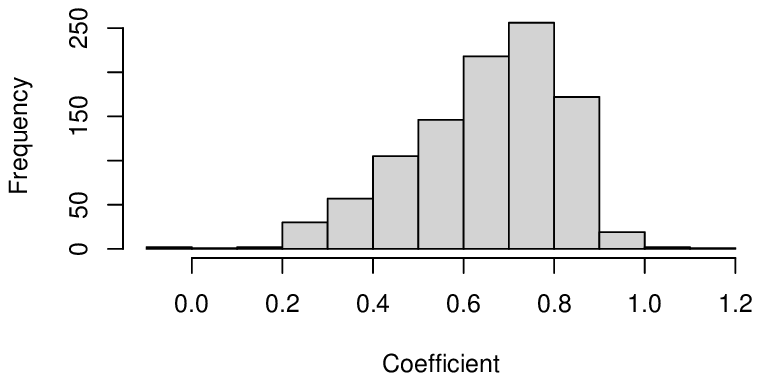}
\caption{First-stage Residual AR Coefficients}
\begin{minipage}{\linewidth}
\begin{footnotesize}
The figure illustrates the empirical distribution of the estimated AR coefficients of an AR(1) model estimated with the residuals of the first-stage LASSO regression.
\end{footnotesize}
\end{minipage}
\label{F:AR}
\end{figure}

We report two diagnostic tests. First, Figure \ref{F:AR} illustrates the empirical distribution of the estimated coefficient of a first-order autoregressive, \texttt{AR}(1), model estimated with the residuals from the first-stage LASSO regression. The distribution is over all the rolling windows and each one of the four countries analyzed in this paper. It is clear from the figure that apart from a single case, all the estimates are bellow one in absolute value. Unit-root tests also strongly reject the null of unit roots in all but one case. This specific negative case is related to a huge outlier in the data which distort the estimation of the \texttt{AR} coefficient and, consequently, the unit-root test. Based on this analyze we are quite confident that our methodology is adequate for the present data.

The second diagnostic is related to the data inflation heuristic. Table \ref{T:inflation} presents the MAPEs of the \texttt{ECM} with data inflation divided by the MAPEs of the \texttt{ECM} without data inflation. Numbers lower than one favors the inflation heuristic. For Brazil, Chile, and Portugal, it is clear that data inflation is superior to no inflation at all. For Mexico, we see improvements when the forecasts for cases are considered but not deaths. Changing the number of observations to inflate seems to have no significant effect and these extra results can be obtained upon request.

Finally, it is important to understand which variables are being selected by the LASSO during the first stage of the methodology. Table \ref{T:select} shows the frequency of selection of each variable over the rolling windows. Mexico is the latecomer country where each variable in the pool is selected at least once. Portugal seems to be the country with the most parsimonious model. Note also the frequency of selection of each variable differs from country to country.

\subsection{Robustness}

In this section we report a number of robustness checks. We start by reporting results concerning a number of alternative loss functions. In addition, we provide results showing the potential effects of vaccination on the performance of the \texttt{ECM} model.

\subsubsection{Alternative Loss Functions}

In addition to the MAPE, we present forecasting results for the following loss functions: (i) the median absolute deviation from the median (MAD), (ii) the mean absolute errors (MAE), and (iii) the mean squared errors. The results for the MAD, which is a loss robust to outliers, is shown in Tables \ref{tab_cases_mad}, \ref{tab_deaths_mad}, and \ref{tab_comb_mad}. The tables provide descriptive statistics of the MAD ratios with respect to the \texttt{AR} benchmark. With respect to cases, the \texttt{ECM} outperforms the \texttt{AR} alternative in more than 50\% of the countries for horizons between three and 12 days. Furthermore, the \texttt{ECM} is way better than the \texttt{trend} model. With respect to deaths, the performance of the \texttt{ECM} improves. In table \ref{tab_comb_mad} we report the results for the combination of \texttt{ECM} and \texttt{AR} models. It is clear from the table that the combination of the models strongly improve the results for both cases and deaths.

The results concerning the MAE are reported in Tables \ref{tab_cases_mae}, \ref{tab_deaths_mae}, and \ref{tab_comb_mae}. From Table \ref{tab_cases_mae}, the results for the \texttt{ECM} are not very encouraging. However, the combination of the \texttt{ECM} and the \texttt{AR} model is clearly superior to the \texttt{AR} benchmark. For deaths, the results for the \texttt{ECM} are much better.

Finally, the results for the MSE are shown in Tables \ref{tab_cases_mse}, \ref{tab_deaths_mse}, and \ref{tab_comb_mse}. The results for the \texttt{ECM} are not particularly good, specially due to some outliers. However, the combination of the \texttt{ECM} and the \texttt{AR} model seems to be very robust no matter which loss function is chosen.

\subsubsection{Forecast Errors and Vaccination}

One question of interest is whether the evolution of vaccination affects the performance of the \texttt{ECM}. To answer this question we run, for each country, the following regression:
\begin{equation}\label{E:vac}
\log\left(\frac{\left|\widehat{\epsilon}_{i,t+h}^{\texttt{ECM}}\right|}{\left|\widehat{\epsilon}_{i,t+h}^{\texttt{AR}}\right|}\right)=
\varrho_i + \rho_i V_{i,t+h} + e_{i,t+h},\quad i=1,\ldots,45,
\end{equation}
where $\widehat{\epsilon}_{i,t+h}^{\texttt{ECM}}$ and $\widehat{\epsilon}_{i,t+h}^{\texttt{AR}}$ are the out-of-sample forecasting errors of the \texttt{ECM} and \texttt{AR} models at the $h$-step-ahead horizon, respectively. $V_{i,t+h}$ is the proportion of the population fully vaccinated.

\begin{figure}%[H]
\centering
\includegraphics[width=\linewidth]{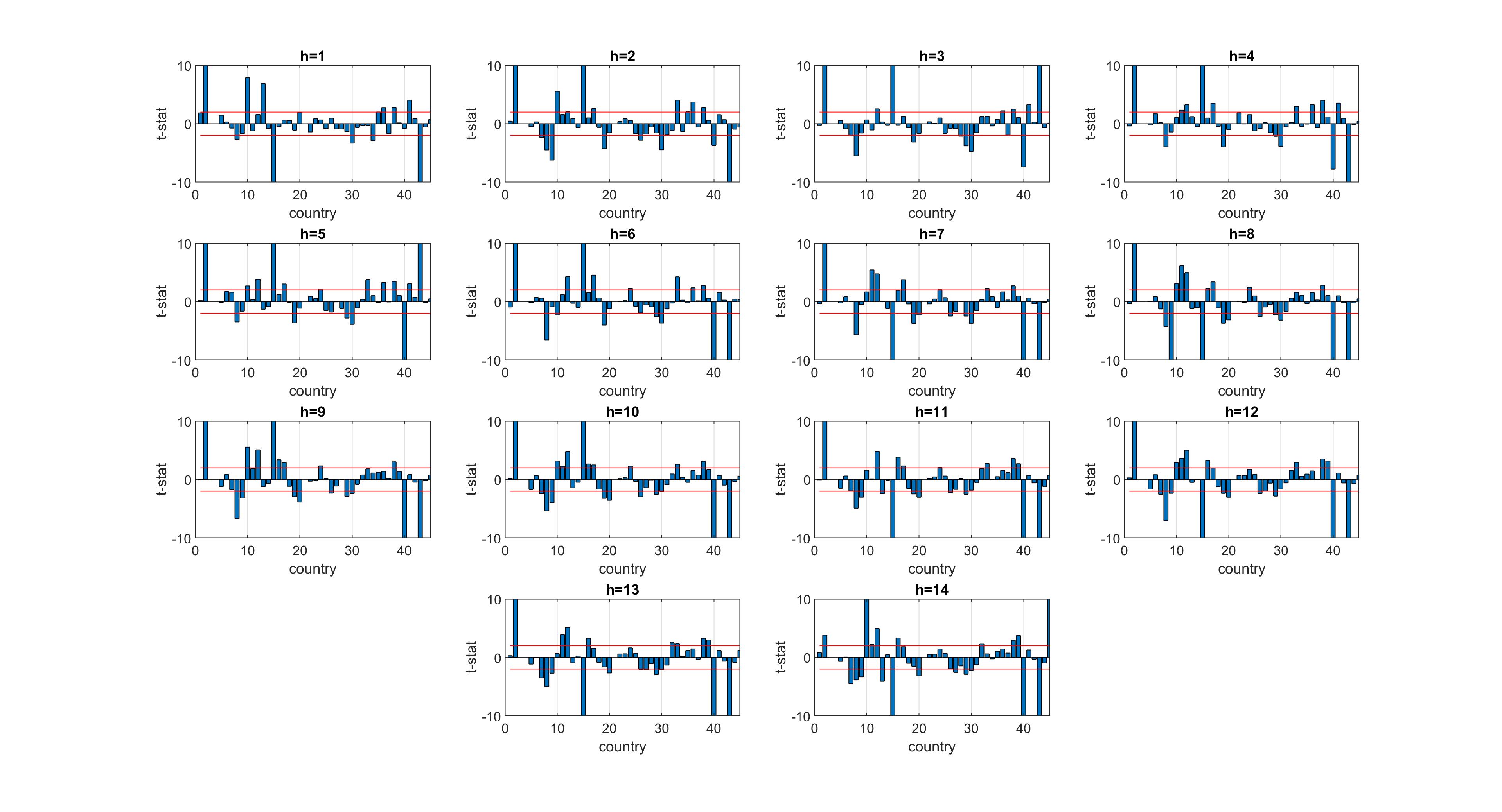}
\caption{t-statistic}
\begin{minipage}{\linewidth}
\begin{footnotesize}
The figure illustrates the t-statistic of the regression of $\log\left(\frac{\left|\widehat{\epsilon}_{t+h}^{\texttt{ECM}}\right|}{\left|\widehat{\epsilon}_{t+h}^{\texttt{AR}}\right|}\right)$ on the proportion of the population vaccinated in each country.
\end{footnotesize}
\end{minipage}
\label{F:vac}
\end{figure}

In Figure \ref{F:vac} we report the t-statistic for the null hypothesis $\mathcal{H}_0: \rho_i=0$. The red lines indicates the $+/- 1.96$ threshold. Standard errors for the coefficients are heteroskedastic and autocorrelation robust. As we can see from the plots, for most countries, the null hypothesis is not rejected, such that there is no evidence that vaccination is correlated with the performance of the models. For the case where the null is rejected, the effects can be either positive or negative. The adaptiveness of the locally regularized linear models capture the slow-varying dynamics in the relationship between dependent and independent variables. As such, we conclude that the cross-country differences in vaccination do not affect much the performance of the ECMs.

\section{Conclusion}

In this paper, we propose a statistical model to forecast in the very short-run the reported number of cases and deaths by the Covid-19 in countries/regions that are latecomers. We believe this is a useful tool to inform health management. Nonetheless, structural breaks might worsen forecasts a few days after such breaks. So the use of this tool should complemented with other external information, such as proxies for social distancing, to guide subjective or objective assessments on potential dynamic changes on the pandemic's evolution. We hope to keep improving the model by improving the methodology and incorporating more information. And we aim to keep forecasts, methodology and codes updated on a daily basis at https://covid19analytics.com.br/.

\appendix

\section{Additional Results}

In this appendix we report a number of additional results.

\subsection{Rolling MAPEs}

\begin{figure}%[H]
\centering
\subfigure[Cases]
{\includegraphics[width=0.49\textwidth]{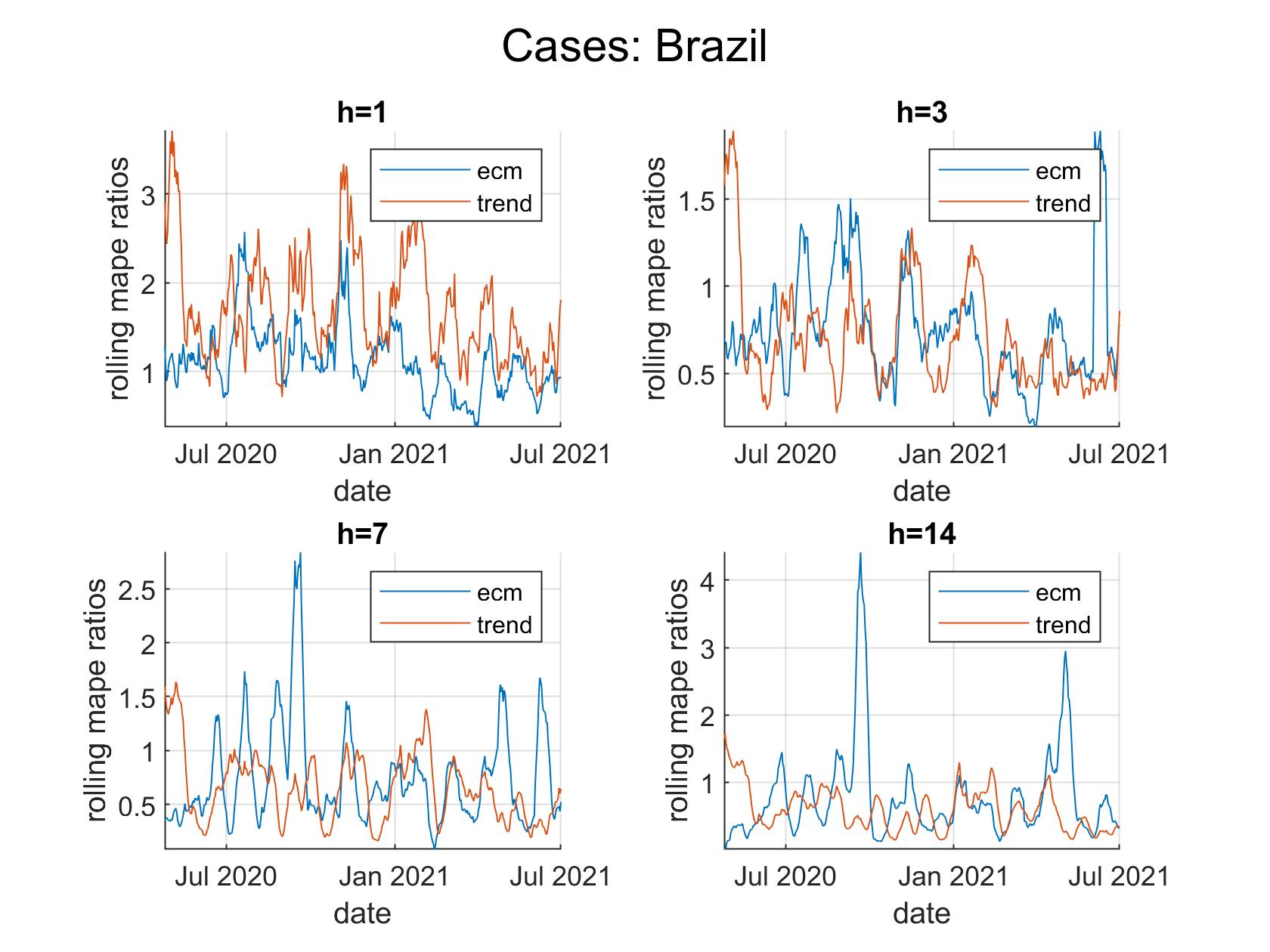}}
\subfigure[Deaths]
{\includegraphics[width=0.49\textwidth]{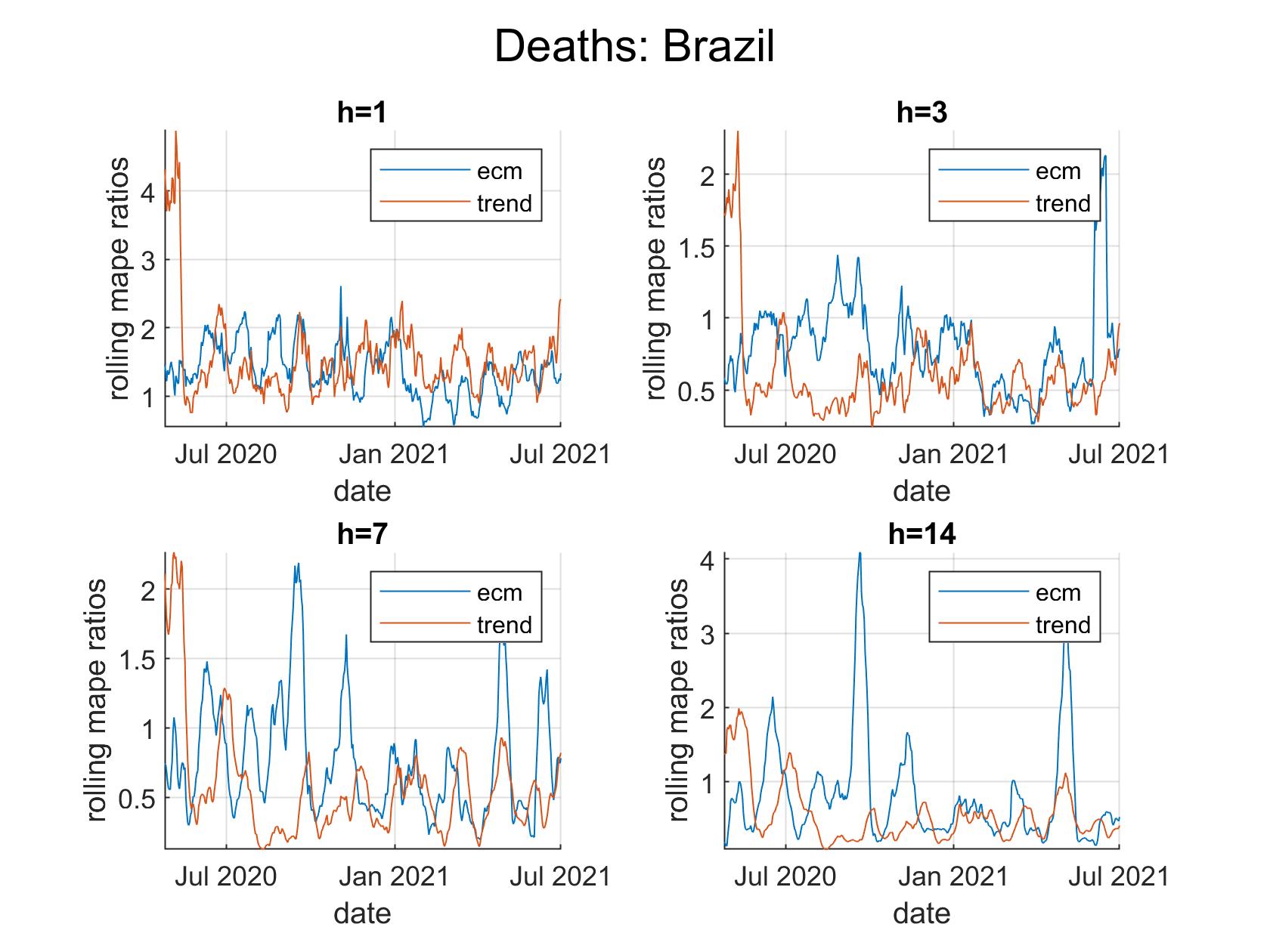}}
\caption{Rolling Mean absolute percentage error - Brazil.}
\begin{minipage}{\linewidth}
\begin{footnotesize}
The figure illustrates, for different horizons, the Mean Absolute Percentage Error (MAPE) computed over rolling windows with 14 observations.
\end{footnotesize}
\end{minipage}
\label{F:rMAPE1}
\end{figure}

\begin{figure}%[H]
\centering
\subfigure[Cases]
{\includegraphics[width=0.49\textwidth]{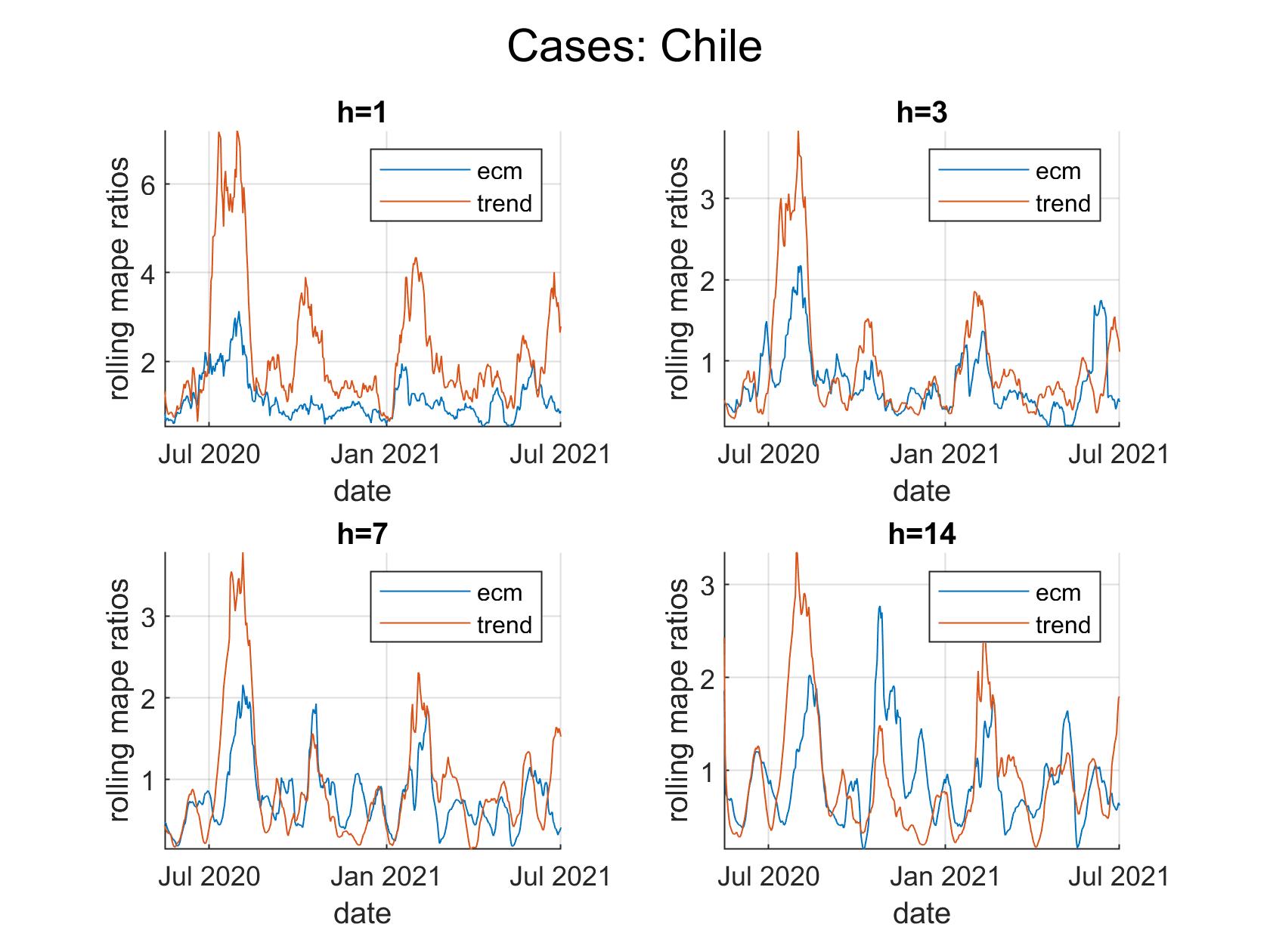}}
\subfigure[Deaths]
{\includegraphics[width=0.49\textwidth]{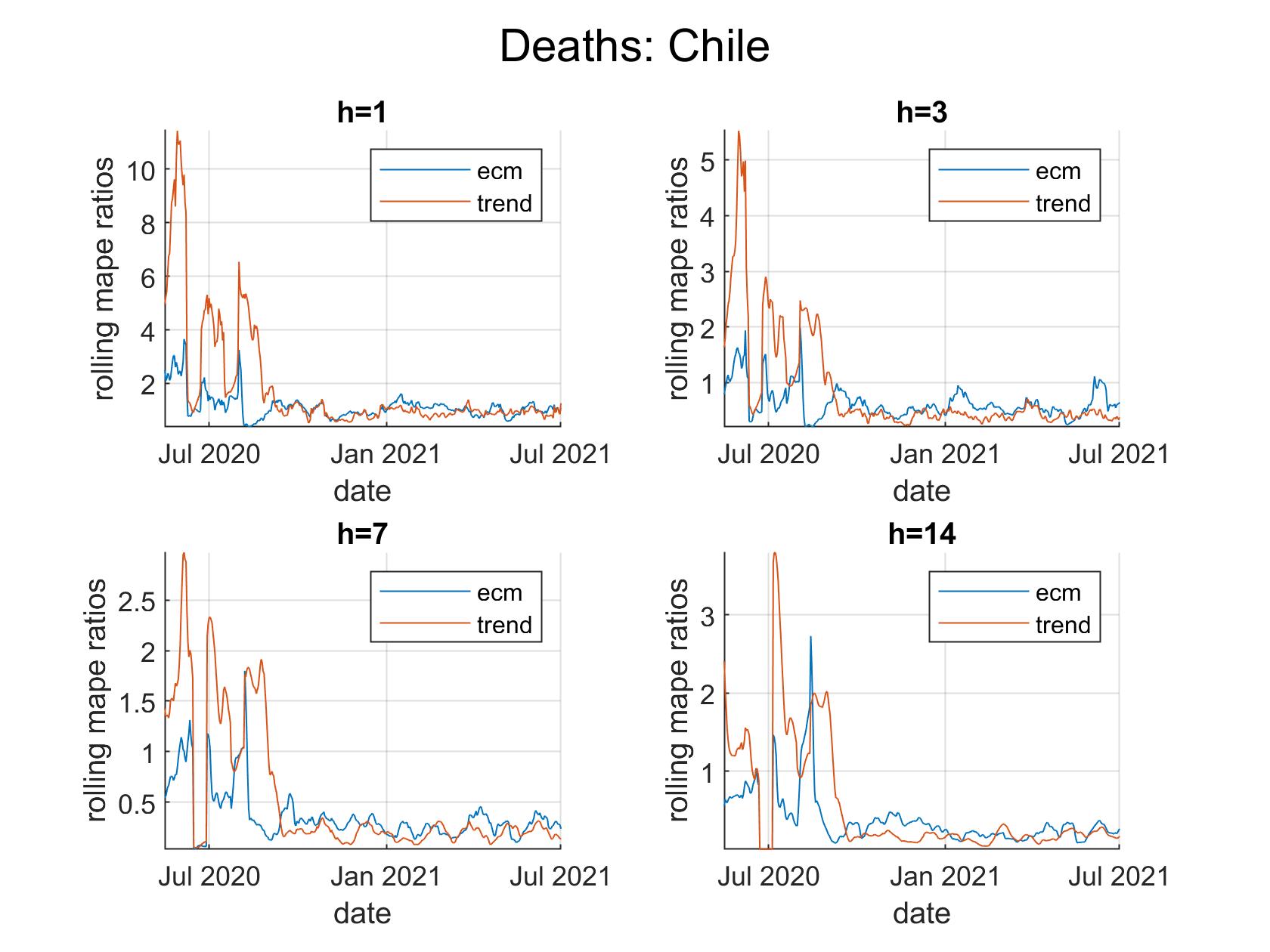}}
\caption{Rolling Mean absolute percentage error - Chile.}
\begin{minipage}{\linewidth}
\begin{footnotesize}
The figure illustrates, for different horizons, the Mean Absolute Percentage Error (MAPE) computed over rolling windows with 14 observations.
\end{footnotesize}
\end{minipage}
\label{F:rMAPE2}
\end{figure}

\begin{figure}%[H]
\centering
\subfigure[Cases]
{\includegraphics[width=0.49\textwidth]{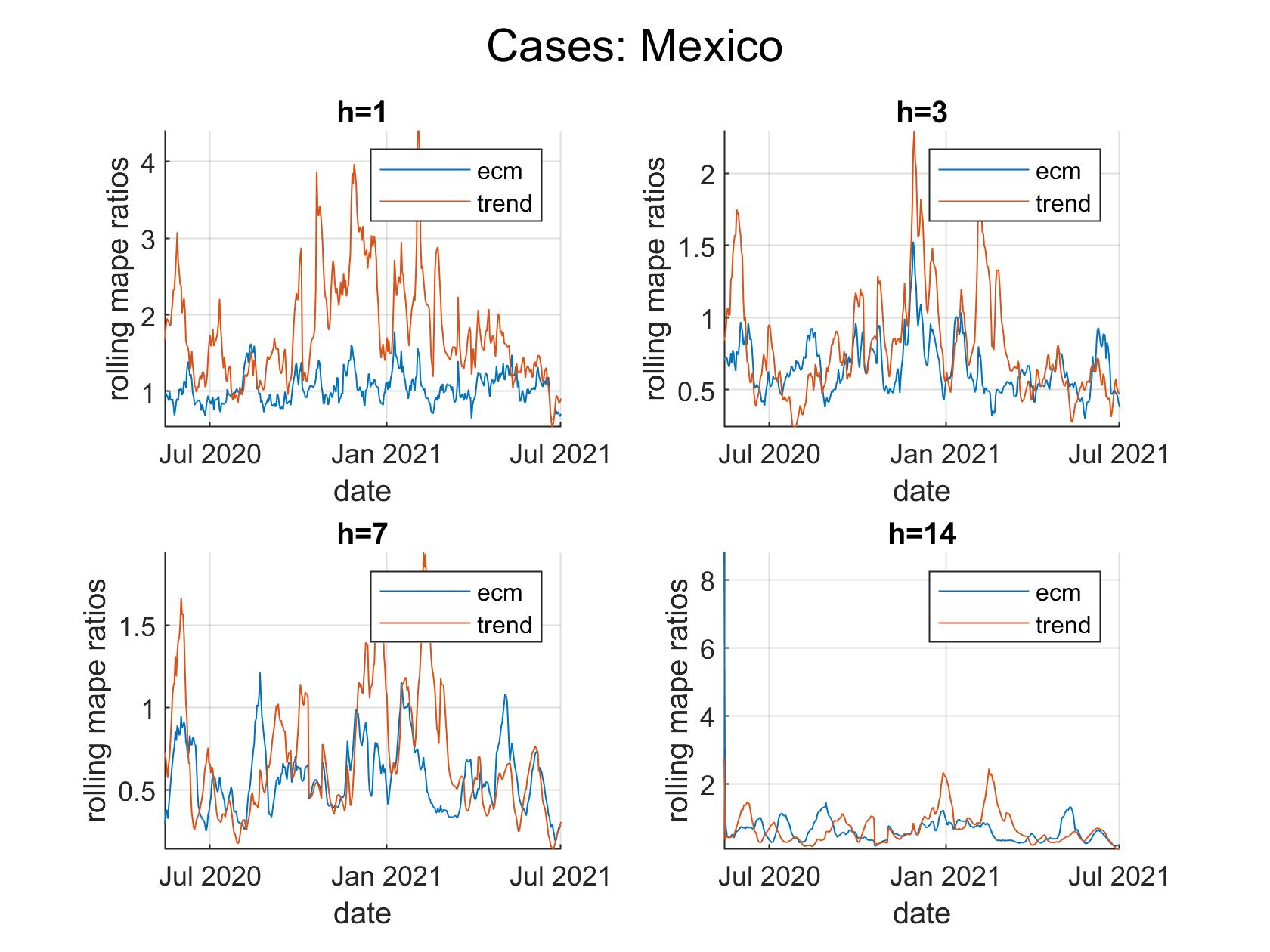}}
\subfigure[Deaths]
{\includegraphics[width=0.49\textwidth]{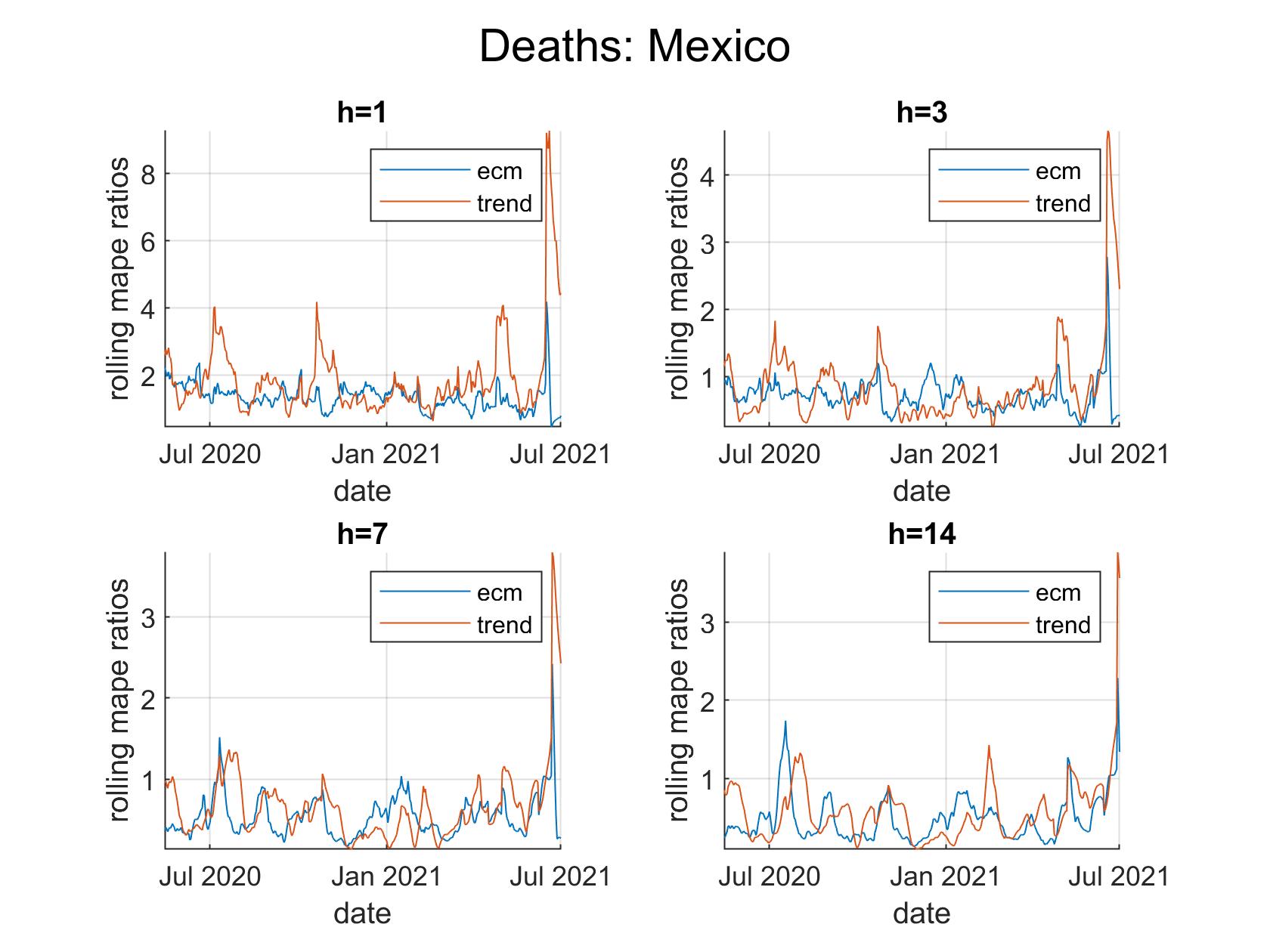}}
\caption{Rolling Mean absolute percentage error - Mexico.}
\begin{minipage}{\linewidth}
\begin{footnotesize}
The figure illustrates, for different horizons, the Mean Absolute Percentage Error (MAPE) computed over rolling windows with 14 observations.
\end{footnotesize}
\end{minipage}
\label{F:rMAPE3}
\end{figure}

\begin{figure}%[H]
\centering
\subfigure[Cases]
{\includegraphics[width=0.49\textwidth]{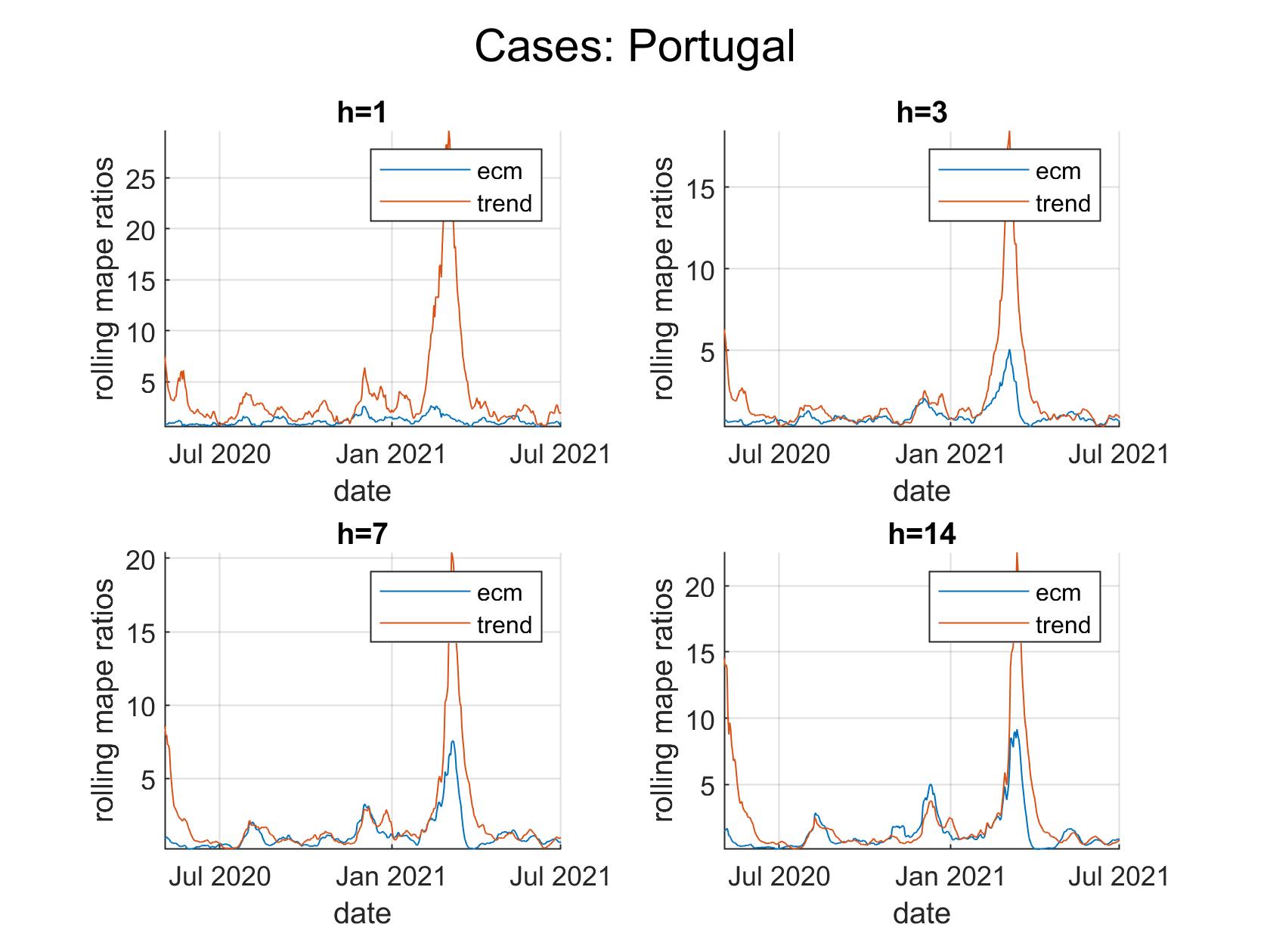}}
\subfigure[Deaths]
{\includegraphics[width=0.49\textwidth]{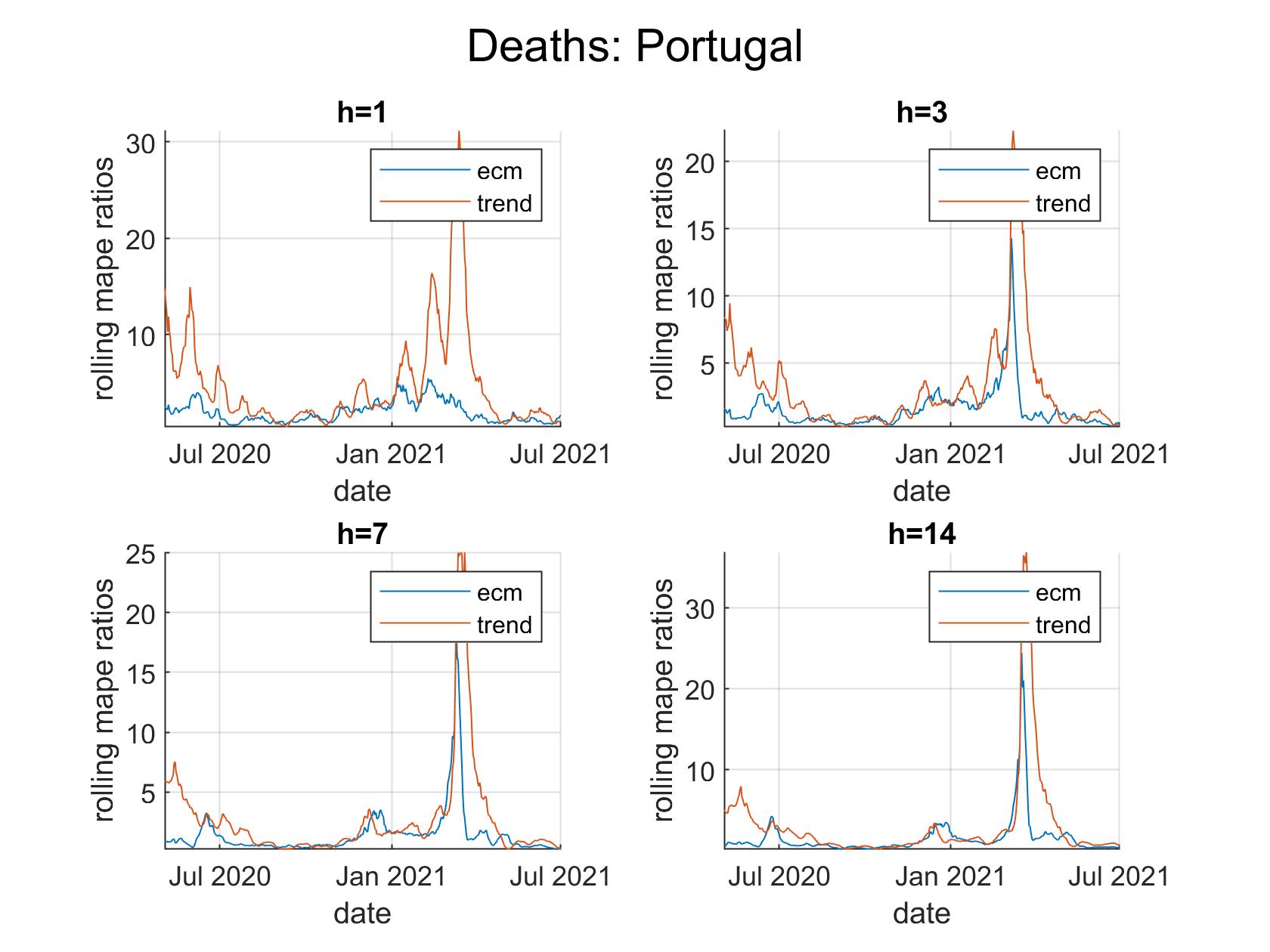}}
\caption{Rolling Mean absolute percentage error - Portugal.}
\begin{minipage}{\linewidth}
\begin{footnotesize}
The figure illustrates, for different horizons, the Mean Absolute Percentage Error (MAPE) computed over rolling windows with 14 observations.
\end{footnotesize}
\end{minipage}
\label{F:rMAPE4}
\end{figure}

In order to analyze how the errors unfold over the evolution of the pandemic, we plot rolling MAPEs over 14 days in Figures \ref{F:rMAPE1}--\ref{F:rMAPE4}. We report only results for selected horizons. It is clear from the figures, that both models improve over time. In some cases the reductions be larger than 50\%. For Chile, the gains of the \texttt{ECM} over the benchmark are more evident during the months of July and August when we look to cases. On the other hand, for deaths, the \texttt{ECM} is better than the benchmark for all windows. In case of Brazil, the superiority of the \texttt{ECM} forecasts for cases are more concentrated in the beginning of the sample, when the benchmark performs very poorly. A similar pattern is visible in the case of deaths. In the case of Mexico and for forecasts for case counts, the \texttt{ECM} is systematically superior to the benchmark over the sample and when we consider the one-day-ahead forecasts. For the other horizons, the gains are more concentrated in the beginning and in the end of the sample. Equivalent conclusions emerge when we look at the forecasts for deaths. In the case of Portugal, the benefits of using the \texttt{ECM} instead of the benchmark are clearer during the first half of the sample.

\subsection{Forecast Combination}

We consider a simple average combination of the forecasts from the \texttt{ECM} and \texttt{AR} models. The results are shown in Table \ref{tab_comb}. The table reports descriptive statistics for the ratio of the MAPEs of the combined models and the autoregressive benchmark. The upper panel presents the results for cases while the lower panel consider the forecasts for the number of deaths. Comparing with the results in Tables \ref{tab_cases_stat} and \ref{tab_deaths_stat}, the combination of models reduce the MAPE ratios in more than 20\% for cases and more than 10\% for deaths. The results for other losses are also reported in Tables \ref{tab_comb_mad}--\ref{tab_comb_mse}.

\subsection{Prediction Intervals}

We report results concerning prediction intervals from the \texttt{ECM}. Tables \ref{tab_ic90}--\ref{tab_ic99}. The tables report descriptive statistics for the frequency that the out-of-samples forecasting errors of the \texttt{ECM} violate the 90\%, 95\%, and 99\% prediction interval, respectively. The prediction intervals are computed under the assumption that the errors are normally distributed. As we can see the \texttt{ECM} produce very precise intervals for 95\%. The intervals are slightly conservative for 90\%. Finally, for the 99\% case, the \texttt{ECM} underestimated the prediction bands. Overall, we believe the results are satisfactory but they can be improved if the normality assumption is dropped and bootstrap is used to compute the prediction intervals.

\begin{landscape}

\begin{table}
\caption{Forecasting Results for Cases: Distribution of Mean Absolute Percentage Error Ratios}
\label{tab_cases_stat}
\centering
\begin{minipage}{0.9\linewidth}
\begin{footnotesize}
The table presents results with respect forecasting models for the number of cases of Covid-19.
The table shows descriptive statistics for the ratio of the forecasting mean absolute percentage error (MAPE)
of either ECM or Trend models and the AR benchmark.
The results are shown for forecasting horizons of 1 to 14 days ahead.
The models were computed on a rolling window scheme with 28 in-sample observations per window.
\end{footnotesize}
\end{minipage}
\resizebox{0.9\linewidth}{!}{
\begin{threeparttable}
\begin{tabular}{lcccccccccccccc}
\hline
\multicolumn{15}{c}{{\ul \textbf{Descriptive Statistics: Ratio of Forecasting Mean Absolute Percentage Errors}}}\\
{\ul \textbf{}}
& \multicolumn{14}{c}{{\ul \textbf{Cases: ECM x AR}}}\\
\textbf{Days ahead} & \textbf{1} & \textbf{2} & \textbf{3} & \textbf{4} & \textbf{5} & \textbf{6} & \textbf{7} & \textbf{8} & \textbf{9} & \textbf{10} & \textbf{11} & \textbf{12} & \textbf{13} & \textbf{14} \\ \hline
Min       & 0.885 & 0.536 & 0.103 & 0.006 & 0.000 & 0.000 & 0.000 & 0.000 & 0.000 & 0.000 & 0.000 & 0.000 & 0.000 & 0.000 \\
5\% prct  & 0.984 & 0.748 & 0.514 & 0.414 & 0.365 & 0.340 & 0.329 & 0.335 & 0.344 & 0.345 & 0.355 & 0.358 & 0.370 & 0.377 \\
10\% prct & 1.012 & 0.787 & 0.668 & 0.582 & 0.556 & 0.547 & 0.568 & 0.579 & 0.590 & 0.586 & 0.592 & 0.599 & 0.607 & 0.615 \\
25\% prct & 1.090 & 0.899 & 0.787 & 0.745 & 0.724 & 0.725 & 0.709 & 0.709 & 0.719 & 0.736 & 0.766 & 0.777 & 0.779 & 0.798 \\
Median    & 1.172 & 1.034 & 0.955 & 0.921 & 0.908 & 0.962 & 0.951 & 0.975 & 0.975 & 0.989 & 0.993 & 0.996 & 1.042 & 1.103 \\
75\% prct & 1.383 & 1.149 & 1.073 & 1.097 & 1.090 & 1.090 & 1.098 & 1.138 & 1.147 & 1.184 & 1.192 & 1.206 & 1.216 & 1.267 \\
90\% prct & 1.812 & 1.572 & 1.593 & 1.560 & 1.483 & 1.648 & 1.836 & 1.927 & 1.915 & 1.886 & 1.862 & 1.838 & 1.830 & 1.856 \\
95\% prct & 2.116 & 2.003 & 1.842 & 1.804 & 1.846 & 1.873 & 1.973 & 1.979 & 2.009 & 2.033 & 2.101 & 2.177 & 2.248 & 2.325 \\
Max       & 2.710 & 2.570 & 2.439 & 2.304 & 2.238 & 2.140 & 2.035 & 2.134 & 2.667 & 3.408 & 4.653 & 6.403 & 9.458 &12.029 \\
Mean      & 1.301 & 1.123 & 1.019 & 0.973 & 0.963 & 0.970 & 0.991 & 1.010 & 1.027 & 1.053 & 1.094 & 1.155 & 1.244 & 1.330 \\
Std       & 0.368 & 0.389 & 0.416 & 0.425 & 0.431 & 0.436 & 0.451 & 0.461 & 0.500 & 0.569 & 0.708 & 0.929 & 1.343 & 1.708 \\
\\
& \multicolumn{14}{c}{{\ul \textbf{Cases: Trend x AR}}}\\
\textbf{Days ahead} & \textbf{1} & \textbf{2} & \textbf{3} & \textbf{4} & \textbf{5} & \textbf{6} & \textbf{7} & \textbf{8} & \textbf{9} & \textbf{10} & \textbf{11} & \textbf{12} & \textbf{13} & \textbf{14} \\ \hline
Min       &  1.151 & 0.859 & 0.247 & 0.011 & 0.000 & 0.000 & 0.000 & 0.000 & 0.000 & 0.000 & 0.000 & 0.000 & 0.000 & 0.000 \\
5\% prct  &  1.552 & 0.917 & 0.585 & 0.455 & 0.394 & 0.358 & 0.339 & 0.352 & 0.361 & 0.380 & 0.391 & 0.398 & 0.405 & 0.413 \\
10\% prct &  1.694 & 1.072 & 0.780 & 0.704 & 0.668 & 0.635 & 0.619 & 0.616 & 0.624 & 0.622 & 0.622 & 0.626 & 0.632 & 0.639 \\
25\% prct &  1.975 & 1.299 & 1.011 & 0.902 & 0.879 & 0.917 & 0.934 & 0.921 & 0.908 & 0.906 & 0.891 & 0.853 & 0.811 & 0.783 \\
Median    &  2.685 & 1.726 & 1.468 & 1.322 & 1.292 & 1.256 & 1.253 & 1.230 & 1.195 & 1.167 & 1.151 & 1.159 & 1.151 & 1.157 \\
75\% prct &  3.452 & 2.373 & 1.910 & 1.691 & 1.598 & 1.600 & 1.607 & 1.562 & 1.531 & 1.508 & 1.524 & 1.538 & 1.531 & 1.526 \\
90\% prct &  5.728 & 3.157 & 2.623 & 2.349 & 2.229 & 2.128 & 2.062 & 2.024 & 1.956 & 1.904 & 1.856 & 1.815 & 1.786 & 1.777 \\
95\% prct &  6.920 & 4.615 & 3.644 & 3.083 & 2.722 & 2.510 & 2.338 & 2.183 & 2.059 & 1.975 & 1.950 & 1.963 & 1.990 & 2.004 \\
Max       & 11.652 & 7.210 & 5.460 & 4.737 & 4.170 & 3.815 & 3.572 & 3.386 & 3.240 & 3.130 & 3.070 & 3.055 & 3.074 & 3.104 \\
Mean      &  3.195 & 2.029 & 1.618 & 1.431 & 1.340 & 1.308 & 1.297 & 1.270 & 1.237 & 1.211 & 1.197 & 1.193 & 1.196 & 1.203 \\
Std       &  1.912 & 1.195 & 0.956 & 0.833 & 0.740 & 0.681 & 0.642 & 0.608 & 0.582 & 0.560 & 0.547 & 0.542 & 0.542 & 0.545 \\
\\
\hline
\end{tabular}
\end{threeparttable}}
\end{table}

\begin{table}
\caption{Forecasting Results for Deaths: Distribution Mean Absolute Percentage Error Ratios}
\label{tab_deaths_stat}
\centering
\begin{minipage}{0.9\linewidth}
\begin{footnotesize}
The table presents results with respect forecasting models for the number of deaths by Covid-19.
The table shows descriptive statistics for the ratio of the forecasting mean absolute percentage error (MAPE)
of either ECM or Trend models and the AR benchmark.
The results are shown for forecasting horizons of 1 to 14 days ahead.
The models were computed on a rolling window scheme with 28 in-sample observations per window.
\end{footnotesize}
\end{minipage}
\resizebox{0.9\linewidth}{!}{
\begin{threeparttable}
\begin{tabular}{lcccccccccccccc}
\hline
\multicolumn{15}{c}{{\ul \textbf{Descriptive Statistics: Ratio of Forecasting Mean Absolute Percentage Errors}}}\\
{\ul \textbf{}}
& \multicolumn{14}{c}{{\ul \textbf{Deaths: ECM x AR}}}\\
\textbf{Days ahead} & \textbf{1} & \textbf{2} & \textbf{3} & \textbf{4} & \textbf{5} & \textbf{6} & \textbf{7} & \textbf{8} & \textbf{9} & \textbf{10} & \textbf{11} & \textbf{12} & \textbf{13} & \textbf{14} \\ \hline
Min       & 0.049 & 0.000 & 0.000 & 0.105 & 0.119 & 0.136 & 0.153 & 0.076 & 0.027 & 0.007 & 0.002 & 0.000 & 0.000 & 0.000 \\
5\% prct  & 0.138 & 0.156 & 0.211 & 0.413 & 0.421 & 0.312 & 0.202 & 0.157 & 0.153 & 0.158 & 0.163 & 0.061 & 0.010 & 0.001 \\
10\% prct & 0.854 & 0.646 & 0.517 & 0.494 & 0.427 & 0.428 & 0.446 & 0.434 & 0.377 & 0.377 & 0.262 & 0.192 & 0.079 & 0.020 \\
25\% prct & 1.158 & 0.896 & 0.753 & 0.669 & 0.606 & 0.588 & 0.583 & 0.576 & 0.579 & 0.578 & 0.565 & 0.554 & 0.556 & 0.573 \\
Median    & 1.310 & 1.008 & 0.854 & 0.790 & 0.743 & 0.723 & 0.717 & 0.709 & 0.701 & 0.732 & 0.700 & 0.743 & 0.751 & 0.766 \\
75\% prct & 1.591 & 1.469 & 1.261 & 1.133 & 1.067 & 1.041 & 1.036 & 1.071 & 1.088 & 1.097 & 1.102 & 1.123 & 1.166 & 1.171 \\
90\% prct & 2.213 & 1.978 & 1.750 & 1.595 & 1.618 & 1.434 & 1.477 & 1.411 & 1.366 & 1.349 & 1.353 & 1.357 & 1.422 & 1.511 \\
95\% prct & 2.568 & 2.090 & 1.982 & 1.847 & 1.815 & 1.971 & 2.236 & 2.338 & 2.416 & 2.501 & 2.612 & 2.790 & 2.950 & 3.128 \\
Max       & 3.625 & 2.998 & 2.250 & 2.192 & 2.371 & 2.487 & 2.574 & 3.283 & 4.239 & 3.488 & 4.384 & 5.008 & 5.900 & 7.125 \\
Mean      & 1.409 & 1.118 & 0.978 & 0.928 & 0.886 & 0.863 & 0.867 & 0.882 & 0.901 & 0.883 & 0.898 & 0.916 & 0.945 & 1.013 \\
Std       & 0.661 & 0.551 & 0.480 & 0.448 & 0.455 & 0.473 & 0.527 & 0.601 & 0.709 & 0.649 & 0.758 & 0.860 & 0.996 & 1.194 \\
\\
& \multicolumn{14}{c}{{\ul \textbf{Deaths: Trend x AR}}}\\
\textbf{Days ahead} & \textbf{1} & \textbf{2} & \textbf{3} & \textbf{4} & \textbf{5} & \textbf{6} & \textbf{7} & \textbf{8} & \textbf{9} & \textbf{10} & \textbf{11} & \textbf{12} & \textbf{13} & \textbf{14} \\ \hline
Min       & 0.092 & 0.000 & 0.000 & 0.162 & 0.185 & 0.206 & 0.227 & 0.131 & 0.047 & 0.012 & 0.002 & 0.000 & 0.000 & 0.000 \\
5\% prct  & 0.309 & 0.326 & 0.393 & 0.559 & 0.518 & 0.468 & 0.287 & 0.218 & 0.212 & 0.218 & 0.229 & 0.102 & 0.016 & 0.001 \\
10\% prct & 1.048 & 0.768 & 0.688 & 0.621 & 0.640 & 0.609 & 0.614 & 0.605 & 0.590 & 0.516 & 0.402 & 0.247 & 0.103 & 0.026 \\
25\% prct & 1.680 & 1.120 & 0.933 & 0.849 & 0.769 & 0.745 & 0.746 & 0.723 & 0.688 & 0.659 & 0.644 & 0.627 & 0.631 & 0.636 \\
Median    & 2.169 & 1.466 & 1.200 & 1.074 & 1.011 & 0.979 & 0.944 & 0.919 & 0.902 & 0.892 & 0.832 & 0.844 & 0.861 & 0.873 \\
75\% prct & 2.746 & 1.959 & 1.629 & 1.504 & 1.451 & 1.402 & 1.327 & 1.261 & 1.231 & 1.227 & 1.216 & 1.208 & 1.205 & 1.211 \\
90\% prct & 4.046 & 3.264 & 2.666 & 2.257 & 2.003 & 1.902 & 1.775 & 1.704 & 1.659 & 1.720 & 1.777 & 1.847 & 1.908 & 1.966 \\
95\% prct & 5.366 & 3.543 & 3.010 & 2.854 & 2.773 & 2.666 & 2.593 & 2.521 & 2.470 & 2.392 & 2.385 & 2.424 & 2.482 & 2.553 \\
Max       &18.957 & 9.563 & 6.547 & 5.165 & 4.378 & 3.890 & 3.566 & 3.333 & 3.154 & 3.163 & 3.263 & 3.403 & 3.524 & 3.655 \\
Mean      & 2.608 & 1.726 & 1.415 & 1.301 & 1.211 & 1.157 & 1.120 & 1.088 & 1.063 & 1.038 & 1.014 & 0.995 & 0.988 & 0.993 \\
Std       & 2.732 & 1.429 & 1.024 & 0.823 & 0.730 & 0.680 & 0.654 & 0.640 & 0.634 & 0.635 & 0.651 & 0.683 & 0.711 & 0.733 \\
\hline
\end{tabular}
\end{threeparttable}}
\end{table}

% ********************************************************************
\begin{table}
\caption{Forecasting Results for Cases: Mean Absolute Percentage Error Ratios}
\label{tab_cases}
\centering
\begin{minipage}{\linewidth}
\begin{footnotesize}
The table shows the ratios of the forecasting mean absolute percentage error (MAPE) of either the \texttt{ECM} or the \texttt{trend} model over the \texttt{AR} benchmark. Numbers below one indicates that the ECM or the trend model outperforms the AR.
The table presents results for forecasting horizons of 1 to 14 days ahead of the Covid-19 accumulated number of cases.
The models were computed on a rolling window scheme with 28 in-sample observations per window.
For each country, the model estimation started when the number of cases of Covid-19 reached 20,000.
The last out-of-sample day for every country is July 1, 2021.
Values in parenthesis are $p$-values for the Giacomini \& White test. \citep{rGhW2006}.
\end{footnotesize}
\end{minipage}
\resizebox{\linewidth}{!}{
\begin{threeparttable}
\begin{tabular}{lcccccccccccccc}
\hline
\multicolumn{15}{c}{{\ul \textbf{Cases: Forecasting Mean Absolute Percentage Errors}}}\\
& \multicolumn{14}{c}{{\ul \textbf{Brazil}}} \\
\textbf{Days ahead} & \textbf{1} & \textbf{2} & \textbf{3} & \textbf{4} & \textbf{5} & \textbf{6} & \textbf{7} & \textbf{8} & \textbf{9} & \textbf{10} & \textbf{11} & \textbf{12} & \textbf{13} & \textbf{14} \\ \hline
Trend & 1.983   & 1.126   & 0.888   & 0.813   & 0.813   & 0.832   & 0.847   & 0.825   & 0.788   & 0.767   & 0.761   & 0.761   & 0.760   & 0.759   \\
      & (0.000) & (0.180) & (0.184) & (0.071) & (0.088) & (0.114) & (0.131) & (0.093) & (0.051) & (0.029) & (0.021) & (0.016) & (0.012) & (0.014) \\
ECM   & 1.157   & 0.899   & 0.739   & 0.656   & 0.638   & 0.621   & 0.616   & 0.610   & 0.590   & 0.586   & 0.592   & 0.599   & 0.607   & 0.615   \\
      & (0.024) & (0.074) & (0.000) & (0.000) & (0.000) & (0.001) & (0.004) & (0.011) & (0.016) & (0.018) & (0.020) & (0.022) & (0.027) & (0.034) \\
\\
& \multicolumn{14}{c}{{\ul \textbf{Chile}}} \\
\textbf{Days ahead} & \textbf{1} & \textbf{2} & \textbf{3} & \textbf{4} & \textbf{5} & \textbf{6} & \textbf{7} & \textbf{8} & \textbf{9} & \textbf{10} & \textbf{11} & \textbf{12} & \textbf{13} & \textbf{14} \\ \hline
Trend & 1.476   &  0.869  & 0.671   & 0.587   & 0.549   & 0.543   & 0.598   & 0.616   & 0.631   & 0.646   & 0.670   & 0.700   & 0.736   & 0.783  \\
      & (0.003) & (0.203) & (0.017) & (0.010) & (0.017) & (0.030) & (0.054) & (0.071) & (0.081) & (0.089) & (0.102) & (0.119) & (0.144) & (0.181)\\
ECM   & 0.994   & 0.758   & 0.607   & 0.536   & 0.498   & 0.489   & 0.541   & 0.579   & 0.603   & 0.628   & 0.659   & 0.688   & 0.720   & 0.764  \\
      & (0.479) & (0.037) & (0.004) & (0.003) & (0.006) & (0.014) & (0.025) & (0.034) & (0.039) & (0.045) & (0.054) & (0.067) & (0.082) & (0.106)\\
\\
& \multicolumn{14}{c}{{\ul \textbf{Mexico}}}\\
                    \textbf{Days ahead} & \textbf{1} & \textbf{2} & \textbf{3} & \textbf{4} & \textbf{5} & \textbf{6} & \textbf{7} & \textbf{8} & \textbf{9} & \textbf{10} & \textbf{11} & \textbf{12} & \textbf{13} & \textbf{14} \\
                    \hline

Trend &  1.800  & 1.107   & 0.850   & 0.745   & 0.710   & 0.704   & 0.719   & 0.714   & 0.692   & 0.674   & 0.666   & 0.661   & 0.662   &   0.668 \\
      & (0.000) & (0.116) & (0.038) & (0.002) & (0.001) & (0.001) & (0.003) & (0.003) & (0.002) & (0.001) & (0.001) & (0.002) & (0.003) & (0.008) \\
ECM   & 1.003   & 0.831   & 0.668   & 0.599   & 0.556   & 0.550   & 0.568   & 0.577   & 0.573   & 0.566   & 0.572   & 0.572   & 0.582   & 0.580   \\
      & (0.478) & (0.000) & (0.000) & (0.000) & (0.000) & (0.000) & (0.000) & (0.000) & (0.000) & (0.000) & (0.000) & (0.000) & (0.000) & (0.000) \\
\\
& \multicolumn{14}{c}{{\ul \textbf{Portugal}}} \\
                    \textbf{Days ahead} & \textbf{1} & \textbf{2} & \textbf{3} & \textbf{4} & \textbf{5} & \textbf{6} & \textbf{7} & \textbf{8} & \textbf{9} & \textbf{10} & \textbf{11} & \textbf{12} & \textbf{13} & \textbf{14} \\
                    \hline
Trend & 3.353   & 2.079   & 1.677   & 1.536   & 1.540   & 1.605   & 1.664   & 1.681   & 1.645   & 1.602   & 1.589   & 1.601   & 1.621   & 1.638 \\
      & (0.000) & (0.000) & (0.005) & (0.023) & (0.031) & (0.026) & (0.023) & (0.025) & (0.031) & (0.040) & (0.045) & (0.045) & (0.044) & ( 0.043) \\
ECM   &  1.147  & 0.965   & 0.891   & 0.881   & 0.906   & 0.962   & 1.024   & 1.072   & 1.090   & 1.098   & 1.117   & 1.153   & 1.195   & 1.241 \\
      & (0.011) & (0.315) & (0.072) & (0.078) & (0.174) & (0.375) & (0.435) & (0.336) & (0.315) & (0.310) & (0.288) & (0.250) & (0.212) & (0.182) \\
 \hline
\end{tabular}
\end{threeparttable}}
\end{table}

\begin{table}
\caption{Forecasting Results for Deaths: Mean Absolute Percentage Error Ratios}
\label{tab_deaths}
\centering
\begin{minipage}{\linewidth}
\begin{footnotesize}
The table shows the ratios of the forecasting mean absolute percentage error (MAPE) of either the \texttt{ECM} or the \texttt{trend} model over the \texttt{AR} benchmark. Numbers below one indicates that the ECM or the trend model outperforms the AR.
The table presents results for forecasting horizons of 1 to 14 days ahead of the Covid-19 accumulated number of deaths.
The models were computed on a rolling window scheme with 28 in-sample observations per window.
For each country, the model estimation started when the number of cases of Covid-19 reached 20,000.
The last out-of-sample day for every country is July 1, 2021.
Values in parenthesis are $p$-values for the Giacomini \& White test. \citep{rGhW2006}.
\end{footnotesize}
\end{minipage}
\resizebox{\linewidth}{!}{
\begin{threeparttable}
\begin{tabular}{lcccccccccccccc}
\hline
\multicolumn{15}{c}{{\ul \textbf{Deaths: Forecasting Mean Absolute Percentage Errors}}}\\
& \multicolumn{14}{c}{{\ul \textbf{Brazil}}}\\  \textbf{Days ahead} & \textbf{1} & \textbf{2} & \textbf{3} & \textbf{4} & \textbf{5} & \textbf{6} & \textbf{7} & \textbf{8} & \textbf{9} & \textbf{10} & \textbf{11} & \textbf{12} & \textbf{13} & \textbf{14} \\ \hline
Trend &  2.369  & 1.259   & 1.021   & 0.969   & 0.973   & 1.007   & 1.054   & 1.043   & 1.006   & 0.972   & 0.956   & 0.946   & 0.933   & 0.920   \\
      & (0.000) & (0.141) & (0.463) & (0.444) & (0.453) & (0.489) & (0.420) & (0.437) & (0.490) & (0.457) & (0.433) & (0.417) & (0.398) & (0.375) \\
ECM   &  1.400  & 0.881   & 0.745   & 0.695   & 0.686   & 0.696   & 0.725   & 0.738   & 0.738   & 0.732   & 0.738   & 0.743   & 0.751   & 0.766   \\
      & (0.000) & (0.138) & (0.027) & (0.010) & (0.002) & (0.000) & (0.000) & (0.001) & (0.016) & (0.024) & (0.032) & (0.036) & (0.039) & (0.059) \\
\\
{\ul \textbf{}}     & \multicolumn{14}{c}{{\ul \textbf{Chile}}}\\
\textbf{Days ahead} & \textbf{1} & \textbf{2} & \textbf{3} & \textbf{4} & \textbf{5} & \textbf{6} & \textbf{7} & \textbf{8} & \textbf{9} & \textbf{10} & \textbf{11} & \textbf{12} & \textbf{13} & \textbf{14} \\ \hline
Trend &  2.471  & 1.511   & 1.207   & 0.966   & 0.745    & 0.522   & 0.296   & 0.131   & 0.047   & 0.013   & 0.003   & 0.000   & 0.000   & 0.000   \\
      & (0.000) & (0.031) & (0.268) & (0.465) &  (0.294) & (0.210) & (0.175) & (0.162) & (0.158) & (0.156) & (0.156) & (0.156) & (0.155) & (0.155) \\
ECM   &  1.256  & 0.818   & 0.668   & 0.542   & 0.418    & 0.295   & 0.169   & 0.076   & 0.027   & 0.007   & 0.002   & 0.000   & 0.000   & 0.000   \\
      & (0.079) & (0.208) & (0.137) & (0.103) & (0.086)  & (0.103) & (0.130) & (0.146) & (0.153) & (0.155) & (0.156) & (0.156) & (0.155) & (0.155) \\
\\
& \multicolumn{14}{c}{{\ul \textbf{Mexico}}}\\ \textbf{Days ahead} & \textbf{1} & \textbf{2} & \textbf{3} & \textbf{4} & \textbf{5} & \textbf{6} & \textbf{7} & \textbf{8} & \textbf{9} & \textbf{10} & \textbf{11} & \textbf{12} & \textbf{13} & \textbf{14} \\ \hline

Trend &  1.799  & 0.987   & 0.796   & 0.719   & 0.677    & 0.656    & 0.649   & 0.634   & 0.610   & 0.591   & 0.576   & 0.565   & 0.557   & 0.552   \\
      & (0.000) & (0.453) & (0.024) & (0.006) & (0.002)  & (0.000)  & (0.000) & (0.000) & (0.000) & (0.001) & (0.001) & (0.001) & (0.001) & (0.001) \\
ECM   &  1.552  & 0.983   & 0.756   & 0.627   & 0.528    & 0.481    & 0.492   & 0.511   & 0.513   & 0.494   & 0.476   & 0.455   & 0.450   & 0.454   \\
      & (0.000) & (0.418) & (0.000) & (0.000) & (0.000)  & (0.000)  & (0.000) & (0.000) & (0.000) & (0.000) & (0.001) & (0.001) & (0.001) & (0.001) \\
\\
& \multicolumn{14}{c}{{\ul \textbf{Portugal}}}\\ \textbf{Days ahead} & \textbf{1} & \textbf{2} & \textbf{3} & \textbf{4} & \textbf{5} & \textbf{6} & \textbf{7} & \textbf{8} & \textbf{9} & \textbf{10} & \textbf{11} & \textbf{12} & \textbf{13} & \textbf{14} \\ \hline

Trend & 5.425   & 3.881   & 3.341   & 2.983   & 2.736   & 2.570   & 2.445   & 2.340   & 2.258   & 2.182   & 2.113   & 2.061   & 2.016   & 1.974   \\
      & (0.000) & (0.000) & (0.000) & (0.000) & (0.001) & (0.002) & (0.004) & (0.006) & (0.009) & (0.012) & (0.015) & (0.018) & (0.021) & (0.024) \\
ECM   &  2.114  & 1.640   & 1.526   & 1.419   & 1.347   & 1.292   & 1.267   & 1.237   & 1.217   & 1.197   & 1.183   & 1.172   & 1.161   & 1.153   \\
      & (0.000) & (0.000) & (0.000) & (0.001) & (0.010) & (0.034) & (0.057) & (0.092) & (0.122) & (0.151) & (0.171) & (0.191) & (0.212) & (0.227) \\
\hline
\end{tabular}
\end{threeparttable}}
\end{table}

\end{landscape}

\begin{landscape}
\begin{table}[H]
\caption{Proportion of times each variable is selected by the LASSO}
\label{T:select}
\begin{minipage}{\linewidth}
\begin{footnotesize}
The table shows the frequency of times each variable is selected by the LASSO in the first-stage regression.
\end{footnotesize}
\end{minipage}
\begin{threeparttable}
\begin{tabular}{lcccccccccccc}
\hline
Target   & $t$    & $t^2$   & France & Iran & Italy & Japan & South Korea & Singapore & Germany & Spain & United Kingdom & US   \\ \hline
Brazil   & 0.52 & 0.07 & 0.50   & 0.64 & 0.70  & 0.37  & 0.53        & 0.57      & -       & -     & -              & -    \\
Chile    & 0.63 & 0.16 & 0.55   & 0.55 & 0.38  & 0.45  & 0.32        & 0.69      & 0.25    & -     & -              & -    \\
Mexico   & 0.41 & 0.09 & 0.08   & 0.66 & 0.17  & 0.31  & 0.32        & 0.25      & 0.39    & 0.29  & 0.33           & 0.67 \\
Portugal & 0.54 & 0.42 & -      & 0.69 & 0.69  & 0.40  & 0.52        & -         & -       & -     & -              & -    \\ \hline
\end{tabular}%
\end{threeparttable}
\end{table}
\end{landscape}

\begin{table}[]
\caption{Effects of data inflation.}
\begin{minipage}{\linewidth}
\begin{footnotesize}
Forecasting MAPEs of the ECM with data inflation divided by the forecasting MAPEs of the ECM without data inflation. Numbers lower than one favors the inflation heuristic.
\end{footnotesize}
\end{minipage}
\label{T:inflation}
\resizebox{\textwidth}{!}{%
\begin{tabular}{lccccccccccc}
\hline
{\ul }        & \multicolumn{11}{c}{{\ul Country}}                                                                                                                                    \\
{\ul }        & \multicolumn{2}{c}{{\ul Brazil}} & {\ul } & \multicolumn{2}{c}{{\ul Chile}} & {\ul } & \multicolumn{2}{c}{{\ul Mexico}} & {\ul } & \multicolumn{2}{c}{{\ul Portugal}} \\
{\ul horizon} & {\ul cases}    & {\ul deaths}    & {\ul } & {\ul cases}    & {\ul deaths}   & {\ul } & {\ul cases}    & {\ul deaths}    & {\ul } & {\ul cases}     & {\ul deaths}     \\ \hline
1             & 0.868          & 1.040           &        & 0.828          & 0.942          &        & 0.915          & 1.009           &        & 0.807           & 0.880            \\
2             & 0.950          & 0.999           &        & 0.899          & 0.912          &        & 0.971          & 1.036           &        & 0.814           & 0.833            \\
3             & 0.971          & 0.964           &        & 0.847          & 0.915          &        & 0.989          & 1.044           &        & 0.834           & 0.820            \\
4             & 0.944          & 0.925           &        & 0.828          & 0.914          &        & 0.989          & 1.066           &        & 0.860           & 0.822            \\
5             & 0.907          & 0.889           &        & 0.827          & 0.915          &        & 0.974          & 1.083           &        & 0.873           & 0.852            \\
6             & 0.870          & 0.854           &        & 0.817          & 0.925          &        & 0.946          & 1.052           &        & 0.878           & 0.865            \\
7             & 0.857          & 0.840           &        & 0.818          & 0.922          &        & 0.948          & 1.051           &        & 0.883           & 0.875            \\
8             & 0.859          & 0.825           &        & 0.834          & 0.922          &        & 0.959          & 1.054           &        & 0.892           & 0.873            \\
9             & 0.849          & 0.827           &        & 0.842          & 0.925          &        & 0.966          & 1.052           &        & 0.900           & 0.861            \\
10            & 0.854          & 0.825           &        & 0.848          & 0.929          &        & 0.974          & 1.060           &        & 0.916           & 0.855            \\
11            & 0.866          & 0.820           &        & 0.852          & 0.927          &        & 0.992          & 1.064           &        & 0.932           & 0.862            \\
12            & 0.864          & 0.811           &        & 0.850          & 0.931          &        & 0.991          & 1.056           &        & 0.947           & 0.872            \\
13            & 0.865          & 0.811           &        & 0.844          & 0.929          &        & 1.001          & 1.056           &        & 0.958           & 0.878            \\
14            & 0.860          & 0.813           &        & 0.837          & 0.941          &        & 0.983          & 1.050           &        & 0.970           & 0.882            \\ \hline
\end{tabular}%
}
\end{table}

% ************************************************************************
% Robustness
% ************************************************************************

\begin{landscape}

\begin{table}
\caption{Forecasting Results for Cases: Distribution of Median Absolute Deviation from the Median Ratios}
\label{tab_cases_mad}
\centering
\begin{minipage}{0.9\linewidth}
\begin{footnotesize}
The table presents results with respect forecasting models for the number of cases of Covid-19.
The table shows descriptive statistics for the ratio of the forecasting median absolute deviation from the median (MAD)
of either ECM or Trend models and the AR benchmark.
The results are shown for forecasting horizons of 1 to 14 days ahead.
The models were computed on a rolling window scheme with 28 in-sample observations per window.
\end{footnotesize}
\end{minipage}
\resizebox{0.9\linewidth}{!}{
\begin{threeparttable}
\begin{tabular}{lcccccccccccccc}
\hline
\multicolumn{15}{c}{{\ul \textbf{Descriptive Statistics: Ratio of Forecasting Median Absolute Deviation from the Median Ratios}}}\\
{\ul \textbf{}}
& \multicolumn{14}{c}{{\ul \textbf{Cases: ECM x AR}}}\\
\textbf{Days ahead} & \textbf{1} & \textbf{2} & \textbf{3} & \textbf{4} & \textbf{5} & \textbf{6} & \textbf{7} & \textbf{8} & \textbf{9} & \textbf{10} & \textbf{11} & \textbf{12} & \textbf{13} & \textbf{14} \\ \hline
Min       &  0.910 & 0.686 & 0.480 & 0.481 & 0.536 & 0.523 & 0.478 & 0.546 & 0.569 & 0.548 & 0.570 & 0.587 & 0.622 & 0.604 \\
5\% prct  &  0.972 & 0.773 & 0.625 & 0.542 & 0.541 & 0.583 & 0.586 & 0.588 & 0.580 & 0.608 & 0.634 & 0.666 & 0.706 & 0.699 \\
10\% prct &  1.027 & 0.810 & 0.673 & 0.570 & 0.589 & 0.612 & 0.652 & 0.681 & 0.694 & 0.674 & 0.694 & 0.721 & 0.736 & 0.771 \\
25\% prct &  1.088 & 0.860 & 0.748 & 0.720 & 0.724 & 0.699 & 0.832 & 0.824 & 0.800 & 0.824 & 0.833 & 0.870 & 0.894 & 0.922 \\
Median    &  1.193 & 1.021 & 0.869 & 0.904 & 0.891 & 0.895 & 0.942 & 0.950 & 0.928 & 0.998 & 0.963 & 0.994 & 1.042 & 1.102 \\
75\% prct &  1.441 & 1.255 & 1.094 & 1.098 & 1.122 & 1.194 & 1.330 & 1.282 & 1.385 & 1.366 & 1.335 & 1.416 & 1.459 & 1.506 \\
90\% prct &  1.906 & 1.593 & 1.398 & 1.370 & 1.340 & 1.574 & 1.524 & 1.565 & 1.791 & 1.732 & 1.696 & 1.794 & 1.885 & 1.941 \\
95\% prct &  2.158 & 1.731 & 1.665 & 1.620 & 1.753 & 1.902 & 1.817 & 1.838 & 1.883 & 1.993 & 2.040 & 1.912 & 1.946 & 2.238 \\
Max       &  2.656 & 3.001 & 2.622 & 2.431 & 2.293 & 2.290 & 2.182 & 2.081 & 2.550 & 2.420 & 2.566 & 2.290 & 2.231 & 2.340 \\
Mean      &  1.320 & 1.125 & 0.999 & 0.964 & 0.960 & 1.019 & 1.058 & 1.073 & 1.106 & 1.107 & 1.127 & 1.139 & 1.172 & 1.232 \\
Std       &  0.373 & 0.406 & 0.378 & 0.371 & 0.364 & 0.407 & 0.380 & 0.368 & 0.428 & 0.419 & 0.432 & 0.402 & 0.399 & 0.452 \\
\\
& \multicolumn{14}{c}{{\ul \textbf{Cases: Trend x AR}}}\\
\textbf{Days ahead} & \textbf{1} & \textbf{2} & \textbf{3} & \textbf{4} & \textbf{5} & \textbf{6} & \textbf{7} & \textbf{8} & \textbf{9} & \textbf{10} & \textbf{11} & \textbf{12} & \textbf{13} & \textbf{14} \\ \hline
Min       &   1.230 & 0.904 & 0.636 & 0.561 & 0.539 & 0.580 & 0.589 & 0.569 & 0.560 & 0.531 & 0.533 & 0.537 & 0.543 & 0.570\\
5\% prct  &   1.624 & 0.985 & 0.740 & 0.670 & 0.737 & 0.735 & 0.721 & 0.766 & 0.713 & 0.735 & 0.753 & 0.725 & 0.720 & 0.702\\
10\% prct &   1.684 & 1.093 & 0.835 & 0.722 & 0.745 & 0.797 & 0.828 & 0.896 & 0.837 & 0.807 & 0.843 & 0.839 & 0.893 & 0.881\\
25\% prct &   2.154 & 1.304 & 1.038 & 0.938 & 0.966 & 1.026 & 1.069 & 1.041 & 1.046 & 1.019 & 1.017 & 1.021 & 1.056 & 1.116\\
Median    &   2.651 & 1.622 & 1.349 & 1.263 & 1.277 & 1.318 & 1.367 & 1.365 & 1.318 & 1.292 & 1.328 & 1.345 & 1.390 & 1.419\\
75\% prct &   3.629 & 2.185 & 1.915 & 1.647 & 1.625 & 1.733 & 1.735 & 1.745 & 1.706 & 1.685 & 1.675 & 1.716 & 1.738 & 1.806\\
90\% prct &   5.297 & 3.251 & 2.370 & 2.203 & 2.152 & 2.138 & 2.070 & 2.033 & 2.085 & 2.074 & 2.011 & 1.978 & 1.959 & 2.052\\
95\% prct &   7.304 & 4.005 & 3.325 & 2.626 & 2.608 & 2.338 & 2.194 & 2.130 & 2.298 & 2.451 & 2.597 & 2.348 & 2.257 & 2.544\\
Max       &  12.237 & 7.271 & 5.975 & 5.265 & 4.796 & 4.460 & 3.938 & 3.832 & 3.762 & 3.773 & 3.744 & 3.699 & 3.637 & 3.563\\
Mean      &   3.236 & 2.007 & 1.600 & 1.431 & 1.405 & 1.436 & 1.440 & 1.436 & 1.443 & 1.434 & 1.436 & 1.437 & 1.450 & 1.499\\
Std       &   2.059 & 1.190 & 0.931 & 0.797 & 0.716 & 0.656 & 0.582 & 0.558 & 0.585 & 0.590 & 0.592 & 0.566 & 0.548 & 0.562\\
\\
\hline
\end{tabular}
\end{threeparttable}}
\end{table}

\begin{table}
\caption{Forecasting Results for Deaths: Distribution Median Absolute Deviation from the Median Ratios}
\label{tab_deaths_mad}
\centering
\begin{minipage}{0.9\linewidth}
\begin{footnotesize}
The table presents results with respect forecasting models for the number of deaths by Covid-19.
The table shows descriptive statistics for the ratio of the forecasting median absolute deviation from the median (MAD)
of either ECM or Trend models and the AR benchmark.
The results are shown for forecasting horizons of 1 to 14 days ahead.
The models were computed on a rolling window scheme with 28 in-sample observations per window.
\end{footnotesize}
\end{minipage}
\resizebox{0.9\linewidth}{!}{
\begin{threeparttable}
\begin{tabular}{lcccccccccccccc}
\hline
\multicolumn{15}{c}{{\ul \textbf{Descriptive Statistics: Ratio of Forecasting Median Absolute Deviation from the Median Ratios}}}\\
{\ul \textbf{}}
& \multicolumn{14}{c}{{\ul \textbf{Deaths: ECM x AR}}}\\
\textbf{Days ahead} & \textbf{1} & \textbf{2} & \textbf{3} & \textbf{4} & \textbf{5} & \textbf{6} & \textbf{7} & \textbf{8} & \textbf{9} & \textbf{10} & \textbf{11} & \textbf{12} & \textbf{13} & \textbf{14} \\ \hline
Min       &  0.944 & 0.666 & 0.676 & 0.514 & 0.473 & 0.398 & 0.471 & 0.486 & 0.480 & 0.420 & 0.440 & 0.450 & 0.456 & 0.433 \\
5\% prct  &  1.040 & 0.816 & 0.689 & 0.643 & 0.581 & 0.596 & 0.580 & 0.568 & 0.568 & 0.561 & 0.532 & 0.520 & 0.527 & 0.562 \\
10\% prct &  1.128 & 0.904 & 0.747 & 0.663 & 0.614 & 0.609 & 0.654 & 0.637 & 0.618 & 0.628 & 0.622 & 0.601 & 0.589 & 0.598 \\
25\% prct &  1.290 & 0.963 & 0.843 & 0.744 & 0.700 & 0.701 & 0.719 & 0.734 & 0.716 & 0.736 & 0.713 & 0.733 & 0.749 & 0.767 \\
Median    &  1.455 & 1.162 & 0.986 & 0.898 & 0.900 & 0.910 & 0.899 & 0.926 & 0.930 & 0.927 & 0.894 & 0.891 & 0.931 & 0.935 \\
75\% prct &  1.827 & 1.260 & 1.209 & 1.077 & 1.049 & 1.070 & 1.116 & 1.152 & 1.185 & 1.206 & 1.197 & 1.156 & 1.208 & 1.198 \\
90\% prct &  2.216 & 1.552 & 1.326 & 1.356 & 1.410 & 1.305 & 1.417 & 1.355 & 1.377 & 1.347 & 1.418 & 1.428 & 1.480 & 1.492 \\
95\% prct &  3.934 & 1.915 & 2.230 & 1.616 & 1.465 & 1.465 & 1.441 & 1.419 & 1.453 & 1.523 & 1.542 & 1.499 & 1.585 & 1.586 \\
Max       &  5.091 & 2.660 & 2.305 & 2.121 & 1.759 & 1.577 & 1.558 & 1.572 & 1.545 & 1.543 & 1.666 & 1.597 & 1.691 & 1.744 \\
Mean      &  1.732 & 1.209 & 1.080 & 0.978 & 0.941 & 0.928 & 0.952 & 0.969 & 0.961 & 0.960 & 0.968 & 0.957 & 0.983 & 0.998 \\
Std       &  0.869 & 0.382 & 0.376 & 0.317 & 0.292 & 0.277 & 0.287 & 0.273 & 0.278 & 0.289 & 0.309 & 0.302 & 0.319 & 0.319 \\
\\
& \multicolumn{14}{c}{{\ul \textbf{Deaths: Trend x AR}}}\\
\textbf{Days ahead} & \textbf{1} & \textbf{2} & \textbf{3} & \textbf{4} & \textbf{5} & \textbf{6} & \textbf{7} & \textbf{8} & \textbf{9} & \textbf{10} & \textbf{11} & \textbf{12} & \textbf{13} & \textbf{14} \\ \hline
Min       &  1.258 & 0.661 & 0.592 & 0.519 & 0.461 & 0.425 & 0.432 & 0.425 & 0.448 & 0.451 & 0.444 & 0.472 & 0.472 & 0.497 \\
5\% prct  &  1.449 & 0.842 & 0.684 & 0.570 & 0.573 & 0.545 & 0.571 & 0.588 & 0.619 & 0.610 & 0.600 & 0.603 & 0.640 & 0.534 \\
10\% prct &  1.517 & 1.097 & 0.832 & 0.641 & 0.723 & 0.731 & 0.712 & 0.766 & 0.760 & 0.796 & 0.798 & 0.841 & 0.723 & 0.737 \\
25\% prct &  1.817 & 1.261 & 1.040 & 0.981 & 0.968 & 0.972 & 0.992 & 1.021 & 0.969 & 0.981 & 0.980 & 0.973 & 0.994 & 1.020 \\
Median    &  2.238 & 1.592 & 1.393 & 1.258 & 1.236 & 1.242 & 1.225 & 1.198 & 1.212 & 1.191 & 1.214 & 1.178 & 1.190 & 1.205 \\
75\% prct &  2.649 & 1.973 & 1.668 & 1.486 & 1.470 & 1.421 & 1.468 & 1.461 & 1.487 & 1.517 & 1.552 & 1.485 & 1.478 & 1.545 \\
90\% prct &  3.514 & 2.290 & 2.103 & 2.134 & 1.791 & 1.825 & 1.914 & 1.886 & 1.857 & 1.866 & 1.954 & 2.043 & 2.121 & 2.127 \\
95\% prct &  9.170 & 2.795 & 2.531 & 2.372 & 2.401 & 2.368 & 2.409 & 2.364 & 2.412 & 2.378 & 2.348 & 2.402 & 2.439 & 2.517 \\
Max       & 10.018 & 6.218 & 4.816 & 4.210 & 3.502 & 3.003 & 3.051 & 3.137 & 3.039 & 2.906 & 2.992 & 3.063 & 3.109 & 3.326 \\
Mean      &  2.733 & 1.706 & 1.464 & 1.340 & 1.295 & 1.275 & 1.290 & 1.292 & 1.276 & 1.275 & 1.288 & 1.299 & 1.311 & 1.341 \\
Std       &  1.935 & 0.852 & 0.711 & 0.634 & 0.565 & 0.517 & 0.533 & 0.514 & 0.514 & 0.495 & 0.509 & 0.519 & 0.540 & 0.576 \\
\hline
\end{tabular}
\end{threeparttable}}
\end{table}

\begin{table}
\caption{Forecasting Results for Cases: Distribution of Mean Absolute Error Ratios}
\label{tab_cases_mae}
\centering
\begin{minipage}{0.9\linewidth}
\begin{footnotesize}
The table presents results with respect forecasting models for the number of cases of Covid-19.
The table shows descriptive statistics for the ratio of the forecasting mean absolute error (MAE)
of either ECM or Trend models and the AR benchmark.
The results are shown for forecasting horizons of 1 to 14 days ahead.
The models were computed on a rolling window scheme with 28 in-sample observations per window.
\end{footnotesize}
\end{minipage}
\resizebox{0.9\linewidth}{!}{
\begin{threeparttable}
\begin{tabular}{lcccccccccccccc}
\hline
\multicolumn{15}{c}{{\ul \textbf{Descriptive Statistics: Ratio of Forecasting Mean Absolute Error Ratios}}}\\
{\ul \textbf{}}
& \multicolumn{14}{c}{{\ul \textbf{Cases: ECM x AR}}}\\
\textbf{Days ahead} & \textbf{1} & \textbf{2} & \textbf{3} & \textbf{4} & \textbf{5} & \textbf{6} & \textbf{7} & \textbf{8} & \textbf{9} & \textbf{10} & \textbf{11} & \textbf{12} & \textbf{13} & \textbf{14} \\ \hline
Min       &   0.150 & 0.014 & 0.000 & 0.000 & 0.000 & 0.000 & 0.000 & 0.000 & 0.000&  0.000 & 0.000  &0.000   &0.000   &0.000 \\
5\% prct  &   0.528 & 0.339 & 0.203 & 0.152 & 0.133 & 0.124 & 0.125 & 0.135 & 0.143&  0.147 & 0.141  &0.117   &0.114   &0.064 \\
10\% prct &   0.709 & 0.569 & 0.407 & 0.338 & 0.263 & 0.252 & 0.243 & 0.239 & 0.267&  0.320 & 0.389  &0.270   &0.221   &0.233 \\
25\% prct &   0.943 & 0.750 & 0.638 & 0.595 & 0.614 & 0.654 & 0.645 & 0.705 & 0.670&  0.744 & 0.779  &0.880   &0.905   &0.940 \\
Median    &   1.517 & 1.140 & 1.004 & 0.931 & 0.949 & 1.036 & 1.073 & 1.161 & 1.166&  1.164 & 1.183  &1.287   &1.296   &1.615 \\
75\% prct &   2.097 & 1.530 & 1.388 & 1.430 & 1.570 & 1.580 & 1.769 & 1.817 & 1.840&  1.891 & 1.958  &2.017   &2.141   &2.342 \\
90\% prct &   3.389 & 3.549 & 2.761 & 3.676 & 4.442 & 3.996 & 3.585 & 3.653 & 3.009&  2.837 & 2.719  &2.935   &3.635   &4.516 \\
95\% prct &   4.256 & 4.746 & 4.301 & 8.601 & 8.001 & 7.736 & 6.135 & 5.110 & 5.160&  5.751 & 6.406  &7.485   &9.883   &14.446 \\
Max       &   7.488 &10.528 &27.357 &19.753 &16.453 & 9.445 &15.989 &20.139 &44.067& 84.623 &186.017 &365.585 &820.855 &1424.237 \\
Mean      &   1.783 & 1.598 & 1.816 & 1.970 & 1.857 & 1.684 & 1.818 & 1.906 & 2.438&  3.384 & 5.713  &9.842   &20.144  &33.878 \\
Std       &   1.312 & 1.784 & 4.031 & 3.926 & 3.082 & 2.117 & 2.600 & 3.085 & 6.511& 12.499 &27.559  &54.285  &122.105 &212.001 \\
\\
& \multicolumn{14}{c}{{\ul \textbf{Cases: Trend x AR}}}\\
\textbf{Days ahead} & \textbf{1} & \textbf{2} & \textbf{3} & \textbf{4} & \textbf{5} & \textbf{6} & \textbf{7} & \textbf{8} & \textbf{9} & \textbf{10} & \textbf{11} & \textbf{12} & \textbf{13} & \textbf{14} \\ \hline
Min       &  0.551 & 0.027 & 0.000 & 0.000 & 0.000 & 0.000 & 0.000 & 0.000 & 0.000 & 0.000 & 0.000 & 0.000 & 0.000 & 0.000\\
5\% prct  &  0.805 & 0.326 & 0.196 & 0.155 & 0.132 & 0.123 & 0.118 & 0.117 & 0.118 & 0.120 & 0.123 & 0.127 & 0.133 & 0.095\\
10\% prct &  1.184 & 0.620 & 0.444 & 0.338 & 0.294 & 0.299 & 0.334 & 0.333 & 0.307 & 0.285 & 0.276 & 0.287 & 0.282 & 0.252\\
25\% prct &  2.632 & 1.247 & 0.848 & 0.690 & 0.624 & 0.626 & 0.712 & 0.743 & 0.725 & 0.716 & 0.703 & 0.728 & 0.767 & 0.790\\
Median    &  5.413 & 2.077 & 1.554 & 1.316 & 1.228 & 1.389 & 1.434 & 1.485 & 1.421 & 1.310 & 1.297 & 1.339 & 1.367 & 1.469\\
75\% prct & 12.089 & 5.561 & 3.705 & 3.072 & 2.802 & 2.509 & 2.567 & 2.632 & 2.556 & 2.365 & 2.191 & 2.137 & 2.213 & 2.215\\
90\% prct & 31.339 &14.599 & 9.440 & 7.467 & 6.548 & 5.534 & 5.188 & 4.514 & 3.915 & 3.523 & 3.203 & 2.964 & 3.016 & 3.100\\
95\% prct & 46.282 &19.318 &11.897 & 8.977 & 7.203 & 6.426 & 5.642 & 5.245 & 4.972 & 4.726 & 4.458 & 4.296 & 4.211 & 4.168\\
Max       & 65.541 &23.317 &16.256 &12.490 &10.253 & 9.135 & 8.424 & 7.961 & 7.661 & 7.461 & 7.415 & 7.420 & 7.461 & 7.557\\
Mean      & 10.980 & 4.794 & 3.216 & 2.558 & 2.239 & 2.120 & 2.078 & 1.973 & 1.839 & 1.738 & 1.673 & 1.643 & 1.644 & 1.662\\
Std       & 14.129 & 5.987 & 3.873 & 2.908 & 2.376 & 2.076 & 1.878 & 1.696 & 1.563 & 1.469 & 1.414 & 1.388 & 1.378 & 1.395\\
\\
\hline
\end{tabular}
\end{threeparttable}}
\end{table}

\begin{table}
\caption{Forecasting Results for Deaths: Distribution Mean Absolute Error Ratios}
\label{tab_deaths_mae}
\centering
\begin{minipage}{0.9\linewidth}
\begin{footnotesize}
The table presents results with respect forecasting models for the number of deaths by Covid-19.
The table shows descriptive statistics for the ratio of the forecasting mean absolute ratio (MAE)
of either ECM or Trend models and the AR benchmark.
The results are shown for forecasting horizons of 1 to 14 days ahead.
The models were computed on a rolling window scheme with 28 in-sample observations per window.
\end{footnotesize}
\end{minipage}
\resizebox{0.9\linewidth}{!}{
\begin{threeparttable}
\begin{tabular}{lcccccccccccccc}
\hline
\multicolumn{15}{c}{{\ul \textbf{Descriptive Statistics: Ratio of Forecasting Mean Absolute Error Ratios}}}\\
{\ul \textbf{}}
& \multicolumn{14}{c}{{\ul \textbf{Deaths: ECM x AR}}}\\
\textbf{Days ahead} & \textbf{1} & \textbf{2} & \textbf{3} & \textbf{4} & \textbf{5} & \textbf{6} & \textbf{7} & \textbf{8} & \textbf{9} & \textbf{10} & \textbf{11} & \textbf{12} & \textbf{13} & \textbf{14} \\ \hline
Min       &   0.034 & 0.000 & 0.000 & 0.074 & 0.086 & 0.100 & 0.114 & 0.107 & 0.039 & 0.009 & 0.001 & 0.000 & 0.000 & 0.000\\
5\% prct  &   0.093 & 0.085 & 0.118 & 0.345 & 0.320 & 0.299 & 0.212 & 0.127 & 0.119 & 0.121 & 0.091 & 0.024 & 0.003 & 0.000\\
10\% prct &   0.851 & 0.661 & 0.503 & 0.512 & 0.410 & 0.324 & 0.312 & 0.377 & 0.433 & 0.298 & 0.273 & 0.154 & 0.060 & 0.015\\
25\% prct &   1.151 & 0.878 & 0.773 & 0.706 & 0.643 & 0.617 & 0.601 & 0.607 & 0.588 & 0.573 & 0.592 & 0.566 & 0.554 & 0.551\\
Median    &   1.399 & 1.089 & 0.907 & 0.828 & 0.802 & 0.770 & 0.773 & 0.761 & 0.761 & 0.749 & 0.710 & 0.720 & 0.741 & 0.749\\
75\% prct &   1.755 & 1.354 & 1.223 & 1.178 & 1.118 & 1.110 & 1.147 & 1.110 & 1.059 & 1.081 & 1.098 & 1.106 & 1.114 & 1.141\\
90\% prct &   1.978 & 1.789 & 1.848 & 1.709 & 1.561 & 1.573 & 1.598 & 1.621 & 1.694 & 1.672 & 1.634 & 1.633 & 1.615 & 1.601\\
95\% prct &   2.744 & 2.213 & 2.044 & 2.014 & 1.964 & 1.897 & 1.807 & 1.745 & 1.741 & 1.771 & 1.869 & 2.031 & 2.185 & 2.628\\
Max       &  10.692 &10.383 & 7.856 & 7.410 & 6.599 & 6.406 & 6.785 & 6.816 & 6.782 & 6.755 & 6.754 & 7.021 & 7.746 & 8.455\\
Mean      &   1.595 & 1.309 & 1.135 & 1.077 & 1.009 & 0.979 & 0.977 & 0.974 & 0.967 & 0.953 & 0.948 & 0.954 & 0.978 & 1.023\\
Std       &   1.499 & 1.463 & 1.124 & 1.056 & 0.948 & 0.922 & 0.976 & 0.985 & 0.990 & 0.993 & 1.008 & 1.062 & 1.178 & 1.304\\
\\
& \multicolumn{14}{c}{{\ul \textbf{Deaths: Trend x AR}}}\\
\textbf{Days ahead} & \textbf{1} & \textbf{2} & \textbf{3} & \textbf{4} & \textbf{5} & \textbf{6} & \textbf{7} & \textbf{8} & \textbf{9} & \textbf{10} & \textbf{11} & \textbf{12} & \textbf{13} & \textbf{14} \\ \hline
Min       &   0.056 & 0.000 & 0.000 & 0.102 & 0.118 & 0.134 & 0.150 & 0.126 & 0.046 & 0.011 & 0.002 & 0.000 & 0.000 & 0.000\\
5\% prct  &   0.278 & 0.148 & 0.186 & 0.445 & 0.444 & 0.378 & 0.300 & 0.168 & 0.154 & 0.156 & 0.165 & 0.060 & 0.009 & 0.001\\
10\% prct &   1.015 & 0.816 & 0.660 & 0.609 & 0.593 & 0.503 & 0.540 & 0.541 & 0.528 & 0.509 & 0.347 & 0.201 & 0.079 & 0.020\\
25\% prct &   1.574 & 1.001 & 0.835 & 0.748 & 0.708 & 0.684 & 0.698 & 0.711 & 0.680 & 0.659 & 0.603 & 0.590 & 0.589 & 0.591\\
Median    &   2.027 & 1.451 & 1.217 & 1.091 & 1.012 & 0.981 & 0.940 & 0.917 & 0.901 & 0.843 & 0.816 & 0.819 & 0.823 & 0.826\\
75\% prct &   2.563 & 1.835 & 1.647 & 1.582 & 1.517 & 1.484 & 1.471 & 1.478 & 1.344 & 1.315 & 1.291 & 1.273 & 1.262 & 1.288\\
90\% prct &   4.111 & 2.622 & 2.327 & 2.203 & 2.054 & 1.962 & 1.965 & 1.886 & 1.878 & 1.844 & 1.813 & 1.795 & 1.771 & 1.772\\
95\% prct &   4.749 & 3.298 & 2.797 & 3.034 & 2.717 & 2.444 & 2.164 & 2.141 & 2.168 & 2.187 & 2.157 & 2.111 & 2.122 & 2.142\\
Max       &  10.753 & 6.477 & 4.858 & 4.072 & 3.592 & 3.257 & 2.997 & 2.800 & 2.628 & 2.493 & 2.389 & 2.330 & 2.426 & 2.552\\
Mean      &   2.349 & 1.611 & 1.347 & 1.281 & 1.194 & 1.144 & 1.106 & 1.068 & 1.031 & 0.997 & 0.968 & 0.946 & 0.937 & 0.937\\
Std       &   1.689 & 1.085 & 0.863 & 0.773 & 0.694 & 0.640 & 0.600 & 0.574 & 0.559 & 0.556 & 0.569 & 0.590 & 0.610 & 0.625\\
\hline
\end{tabular}
\end{threeparttable}}
\end{table}

\begin{table}
\caption{Forecasting Results for Cases: Distribution of Mean Squared Error Ratios}
\label{tab_cases_mse}
\centering
\begin{minipage}{0.9\linewidth}
\begin{footnotesize}
The table presents results with respect forecasting models for the number of cases of Covid-19.
The table shows descriptive statistics for the ratio of the forecasting mean squared error (MSE)
of either ECM or Trend models and the AR benchmark.
The results are shown for forecasting horizons of 1 to 14 days ahead.
The models were computed on a rolling window scheme with 28 in-sample observations per window.
\end{footnotesize}
\end{minipage}
\resizebox{0.9\linewidth}{!}{
\begin{threeparttable}
\begin{tabular}{lcccccccccccccc}
\hline
\multicolumn{15}{c}{{\ul \textbf{Descriptive Statistics: Ratio of Forecasting Mean Squared Error Ratios}}}\\
{\ul \textbf{}}
& \multicolumn{14}{c}{{\ul \textbf{Cases: ECM x AR}}}\\
\textbf{Days ahead} & \textbf{1} & \textbf{2} & \textbf{3} & \textbf{4} & \textbf{5} & \textbf{6} & \textbf{7} & \textbf{8} & \textbf{9} & \textbf{10} & \textbf{11} & \textbf{12} & \textbf{13} & \textbf{14} \\ \hline
Min       &  0.150 & 0.014&  0.000&  0.000&  0.000&  0.000&  0.000&  0.000&  0.000&  0.000&   0.000&   0.000&   0.000&    0.000\\
5\% prct  &  0.528 & 0.339&  0.203&  0.152&  0.133&  0.124&  0.125&  0.135&  0.143&  0.147&   0.141&   0.117&   0.114&    0.064\\
10\% prct &  0.709 & 0.569&  0.407&  0.338&  0.263&  0.252&  0.243&  0.239&  0.267&  0.320&   0.389&   0.270&   0.221&    0.233\\
25\% prct &  0.943 & 0.750&  0.638&  0.595&  0.614&  0.654&  0.645&  0.705&  0.670&  0.744&   0.779&   0.880&   0.905&    0.940\\
Median    &  1.517 & 1.140&  1.004&  0.931&  0.949&  1.036&  1.073&  1.161&  1.166&  1.164&   1.183&   1.287&   1.296&    1.615\\
75\% prct &  2.097 & 1.530&  1.388&  1.430&  1.570&  1.580&  1.769&  1.817&  1.840&  1.891&   1.958&   2.017&   2.141&    2.342\\
90\% prct &  3.389 & 3.549&  2.761&  3.676&  4.442&  3.996&  3.585&  3.653&  3.009&  2.837&   2.719&   2.935&   3.635&    4.516\\
95\% prct &  4.256 & 4.746&  4.301&  8.601&  8.001&  7.736&  6.135&  5.110&  5.160&  5.751&   6.406&   7.485&   9.883&   14.446\\
Max       &  7.488 &10.528& 27.357& 19.753& 16.453&  9.445& 15.989& 20.139& 44.067& 84.623& 186.017& 365.585& 820.855& 1424.237\\
Mean      &  1.783 & 1.598&  1.816&  1.970&  1.857&  1.684&  1.818&  1.906&  2.438&  3.384&   5.713&   9.842&  20.144&   33.878\\
Std       &  1.312 & 1.784&  4.031&  3.926&  3.082&  2.117&  2.600&  3.085&  6.511& 12.499&  27.559&  54.285& 122.105&  212.001\\
\\
& \multicolumn{14}{c}{{\ul \textbf{Cases: Trend x AR}}}\\
\textbf{Days ahead} & \textbf{1} & \textbf{2} & \textbf{3} & \textbf{4} & \textbf{5} & \textbf{6} & \textbf{7} & \textbf{8} & \textbf{9} & \textbf{10} & \textbf{11} & \textbf{12} & \textbf{13} & \textbf{14} \\ \hline
Min       & 0.551 & 0.027 & 0.000 & 0.000 & 0.000 & 0.000 & 0.000 & 0.000 & 0.000 & 0.000 & 0.000 & 0.000 & 0.000 & 0.000\\
5\% prct  & 0.805 & 0.326 & 0.196 & 0.155 & 0.132 & 0.123 & 0.118 & 0.117 & 0.118 & 0.120 & 0.123 & 0.127 & 0.133 & 0.095\\
10\% prct & 1.184 & 0.620 & 0.444 & 0.338 & 0.294 & 0.299 & 0.334 & 0.333 & 0.307 & 0.285 & 0.276 & 0.287 & 0.282 & 0.252\\
25\% prct & 2.632 & 1.247 & 0.848 & 0.690 & 0.624 & 0.626 & 0.712 & 0.743 & 0.725 & 0.716 & 0.703 & 0.728 & 0.767 & 0.790\\
Median    & 5.413 & 2.077 & 1.554 & 1.316 & 1.228 & 1.389 & 1.434 & 1.485 & 1.421 & 1.310 & 1.297 & 1.339 & 1.367 & 1.469\\
75\% prct &12.089 & 5.561 & 3.705 & 3.072 & 2.802 & 2.509 & 2.567 & 2.632 & 2.556 & 2.365 & 2.191 & 2.137 & 2.213 & 2.215\\
90\% prct &31.339 &14.599 & 9.440 & 7.467 & 6.548 & 5.534 & 5.188 & 4.514 & 3.915 & 3.523 & 3.203 & 2.964 & 3.016 & 3.100\\
95\% prct &46.282 &19.318 &11.897 & 8.977 & 7.203 & 6.426 & 5.642 & 5.245 & 4.972 & 4.726 & 4.458 & 4.296 & 4.211 & 4.168\\
Max       &65.541 &23.317 &16.256 &12.490 &10.253 & 9.135 & 8.424 & 7.961 & 7.661 & 7.461 & 7.415 & 7.420 & 7.461 & 7.557\\
Mean      &10.980 & 4.794 & 3.216 & 2.558 & 2.239 & 2.120 & 2.078 & 1.973 & 1.839 & 1.738 & 1.673 & 1.643 & 1.644 & 1.662\\
Std       &14.129 & 5.987 & 3.873 & 2.908 & 2.376 & 2.076 & 1.878 & 1.696 & 1.563 & 1.469 & 1.414 & 1.388 & 1.378 & 1.395\\
\\
\hline
\end{tabular}
\end{threeparttable}}
\end{table}

\begin{table}
\caption{Forecasting Results for Deaths: Distribution Mean Squared Error Ratios}
\label{tab_deaths_mse}
\centering
\begin{minipage}{0.9\linewidth}
\begin{footnotesize}
The table presents results with respect forecasting models for the number of deaths by Covid-19.
The table shows descriptive statistics for the ratio of the forecasting mean squared error ratio (MSE)
of either ECM or Trend models and the AR benchmark.
The results are shown for forecasting horizons of 1 to 14 days ahead.
The models were computed on a rolling window scheme with 28 in-sample observations per window.
\end{footnotesize}
\end{minipage}
\resizebox{0.9\linewidth}{!}{
\begin{threeparttable}
\begin{tabular}{lcccccccccccccc}
\hline
\multicolumn{15}{c}{{\ul \textbf{Descriptive Statistics: Ratio of Forecasting Mean Squared Error Ratios}}}\\
{\ul \textbf{}}
& \multicolumn{14}{c}{{\ul \textbf{Deaths: ECM x AR}}}\\
\textbf{Days ahead} & \textbf{1} & \textbf{2} & \textbf{3} & \textbf{4} & \textbf{5} & \textbf{6} & \textbf{7} & \textbf{8} & \textbf{9} & \textbf{10} & \textbf{11} & \textbf{12} & \textbf{13} & \textbf{14} \\ \hline
Min       &    0.000&   0.000&   0.000&   0.001&   0.002&   0.002&   0.001&   0.000&   0.000&   0.000&   0.000&   0.000&    0.000&    0.000\\
5\% prct  &    0.000&   0.001&   0.001&   0.035&   0.016&   0.009&   0.003&   0.003&   0.004&   0.002&   0.000&   0.000&    0.000&    0.000\\
10\% prct &    0.615&   0.249&   0.107&   0.114&   0.118&   0.109&   0.080&   0.059&   0.019&   0.008&   0.005&   0.001&    0.000&    0.000\\
25\% prct &    0.956&   0.666&   0.480&   0.447&   0.370&   0.350&   0.327&   0.342&   0.336&   0.261&   0.292&   0.345&    0.380&    0.373\\
Median    &    1.591&   1.107&   0.842&   0.719&   0.664&   0.675&   0.681&   0.637&   0.634&   0.720&   0.764&   0.807&    0.859&    0.921\\
75\% prct &    2.793&   2.532&   2.168&   1.868&   1.613&   1.617&   1.593&   1.672&   1.733&   1.770&   1.890&   2.113&    2.128&    2.102\\
90\% prct &    4.321&   3.937&   4.092&   3.907&   3.782&   4.145&   4.228&   3.807&   4.660&   4.248&   4.004&   4.020&    3.865&    5.081\\
95\% prct &    7.341&   7.368&   8.944&   8.674&   6.653&   5.329&   5.556&   5.589&   7.230&   6.647&  10.367&  14.764&   18.939&   30.781\\
Max       & 1201.247& 603.935& 273.855& 237.143& 189.338& 194.820& 323.517& 387.550& 381.101& 461.763& 575.176& 753.366& 1072.266& 1496.674\\
Mean      &   28.737&  15.049&   7.724&   6.759&   5.534&   5.564&   8.387&   9.831&   9.782&  11.577&  14.260&  18.450&   25.827&   35.895\\
Std       &  178.770&  89.796&  40.639&  35.182&  28.076&  28.893&  48.066&  57.606&  56.639&  68.660&  85.550& 112.088&  159.595&  222.807\\
\\
& \multicolumn{14}{c}{{\ul \textbf{Deaths: Trend x AR}}}\\
\textbf{Days ahead} & \textbf{1} & \textbf{2} & \textbf{3} & \textbf{4} & \textbf{5} & \textbf{6} & \textbf{7} & \textbf{8} & \textbf{9} & \textbf{10} & \textbf{11} & \textbf{12} & \textbf{13} & \textbf{14} \\ \hline
Min       &    0.000&  0.000&  0.000&  0.001&  0.002&  0.003&  0.002&  0.000&  0.000&  0.000&  0.000&  0.000&  0.000&  0.000\\
5\% prct  &    0.002&  0.002&  0.003&  0.057&  0.023&  0.013&  0.004&  0.004&  0.005&  0.005&  0.001&  0.000&  0.000&  0.000\\
10\% prct &    0.660&  0.421&  0.299&  0.226&  0.193&  0.137&  0.080&  0.041&  0.039&  0.016&  0.007&  0.000&  0.000&  0.000\\
25\% prct &    1.752&  0.801&  0.492&  0.464&  0.420&  0.411&  0.433&  0.371&  0.252&  0.254&  0.257&  0.261&  0.267&  0.279\\
Median    &    2.867&  1.555&  1.163&  1.079&  0.964&  0.896&  0.719&  0.662&  0.652&  0.630&  0.622&  0.620&  0.645&  0.676\\
75\% prct &    6.354&  3.223&  2.507&  2.315&  2.192&  1.724&  1.725&  1.673&  1.590&  1.533&  1.550&  1.580&  1.625&  1.684\\
90\% prct &   11.250&  7.312&  5.380&  4.685&  4.720&  4.699&  4.262&  3.917&  4.061&  4.258&  4.178&  3.795&  3.472&  3.496\\
95\% prct &   35.120& 16.400& 10.692&  8.296&  7.069&  6.699&  6.703&  6.568&  6.129&  5.833&  5.665&  5.995&  6.480&  6.985\\
Max       &  114.433& 37.694& 20.869& 14.365& 11.058&  8.947&  7.468&  7.557&  8.922& 10.524& 12.534& 15.027& 18.136& 22.141\\
Mean      &    8.033&  3.675&  2.508&  2.131&  1.821&  1.658&  1.568&  1.502&  1.472&  1.477&  1.514&  1.580&  1.673&  1.793\\
Std       &   19.206&  6.947&  4.192&  3.092&  2.463&  2.131&  1.994&  1.960&  2.026&  2.169&  2.398&  2.722&  3.150&  3.716\\
\hline
\end{tabular}
\end{threeparttable}}
\end{table}

\begin{table}
\caption{Forecasting Combination Results for Cases and Deaths: Distribution of Mean Absolute Percentage Error Ratios}
\label{tab_comb}
\centering
\begin{minipage}{0.9\linewidth}
\begin{footnotesize}
The table presents results with respect forecasting models for the number of cases and deaths.
The table shows descriptive statistics for the ratio of the forecasting mean absolute percentage error (MAPE)
of either the combination ECM and AR or ECM and Trend models and the AR benchmark.
The results are shown for forecasting horizons of 1 to 14 days ahead.
The models were computed on a rolling window scheme with 28 in-sample observations per window.
\end{footnotesize}
\end{minipage}
\resizebox{0.9\linewidth}{!}{
\begin{threeparttable}
\begin{tabular}{lcccccccccccccc}
\hline
\multicolumn{15}{c}{{\ul \textbf{Descriptive Statistics: Ratio of Forecasting Mean Absolute Percentage Errors}}}\\
{\ul \textbf{}}
& \multicolumn{14}{c}{{\ul \textbf{Cases: ECM and AR x AR}}}\\
\textbf{Days ahead} & \textbf{1} & \textbf{2} & \textbf{3} & \textbf{4} & \textbf{5} & \textbf{6} & \textbf{7} & \textbf{8} & \textbf{9} & \textbf{10} & \textbf{11} & \textbf{12} & \textbf{13} & \textbf{14} \\ \hline
Min       &   0.821 & 0.679 & 0.541 & 0.502 & 0.500 & 0.495 & 0.455 & 0.461 & 0.469 & 0.478 & 0.475 & 0.466 & 0.458 & 0.446\\
5\% prct  &   0.836 & 0.705 & 0.585 & 0.546 & 0.518 & 0.500 & 0.500 & 0.500 & 0.499 & 0.497 & 0.497 & 0.496 & 0.495 & 0.495\\
10\% prct &   0.867 & 0.718 & 0.626 & 0.570 & 0.559 & 0.534 & 0.527 & 0.528 & 0.517 & 0.532 & 0.542 & 0.562 & 0.559 & 0.543\\
25\% prct &   0.892 & 0.790 & 0.717 & 0.677 & 0.651 & 0.631 & 0.622 & 0.616 & 0.614 & 0.609 & 0.615 & 0.614 & 0.618 & 0.622\\
Median    &   0.937 & 0.835 & 0.791 & 0.768 & 0.767 & 0.773 & 0.762 & 0.759 & 0.764 & 0.771 & 0.770 & 0.762 & 0.769 & 0.775\\
75\% prct &   1.014 & 0.890 & 0.848 & 0.843 & 0.843 & 0.857 & 0.873 & 0.885 & 0.898 & 0.910 & 0.929 & 0.944 & 0.958 & 0.976\\
90\% prct &   1.137 & 1.099 & 1.068 & 1.083 & 1.117 & 1.099 & 1.091 & 1.142 & 1.186 & 1.199 & 1.210 & 1.240 & 1.279 & 1.322\\
95\% prct &   1.392 & 1.347 & 1.272 & 1.215 & 1.194 & 1.208 & 1.298 & 1.334 & 1.340 & 1.343 & 1.343 & 1.361 & 1.399 & 1.433\\
Max       &   1.671 & 1.611 & 1.558 & 1.478 & 1.447 & 1.411 & 1.388 & 1.362 & 1.603 & 1.973 & 2.596 & 3.472 & 5.001 & 6.279\\
Mean      &   0.991 & 0.888 & 0.832 & 0.805 & 0.795 & 0.790 & 0.794 & 0.797 & 0.806 & 0.818 & 0.835 & 0.861 & 0.902 & 0.941\\
Std       &   0.173 & 0.191 & 0.198 & 0.203 & 0.209 & 0.212 & 0.225 & 0.230 & 0.249 & 0.283 & 0.351 & 0.462 & 0.671 & 0.852\\
\\
& \multicolumn{14}{c}{{\ul \textbf{Deaths: ECM and AR x AR}}}\\
\textbf{Days ahead} & \textbf{1} & \textbf{2} & \textbf{3} & \textbf{4} & \textbf{5} & \textbf{6} & \textbf{7} & \textbf{8} & \textbf{9} & \textbf{10} & \textbf{11} & \textbf{12} & \textbf{13} & \textbf{14} \\ \hline
Min       &  0.517 & 0.500 & 0.487 & 0.459 & 0.443 & 0.448 & 0.461 & 0.467 & 0.438 & 0.445 & 0.445 & 0.472 & 0.491 & 0.489  \\
5\% prct  &  0.546 & 0.541 & 0.519 & 0.525 & 0.527 & 0.516 & 0.509 & 0.505 & 0.508 & 0.502 & 0.500 & 0.499 & 0.498 & 0.497  \\
10\% prct &  0.721 & 0.625 & 0.561 & 0.580 & 0.570 & 0.548 & 0.533 & 0.522 & 0.519 & 0.519 & 0.507 & 0.502 & 0.501 & 0.500  \\
25\% prct &  0.858 & 0.736 & 0.680 & 0.644 & 0.608 & 0.594 & 0.579 & 0.569 & 0.570 & 0.564 & 0.555 & 0.544 & 0.527 & 0.525  \\
Median    &  0.936 & 0.809 & 0.735 & 0.697 & 0.668 & 0.661 & 0.647 & 0.641 & 0.643 & 0.639 & 0.629 & 0.621 & 0.611 & 0.609  \\
75\% prct &  1.086 & 0.976 & 0.891 & 0.838 & 0.787 & 0.754 & 0.732 & 0.730 & 0.728 & 0.727 & 0.730 & 0.748 & 0.766 & 0.780  \\
90\% prct &  1.318 & 1.201 & 1.093 & 1.105 & 1.111 & 1.000 & 0.990 & 0.985 & 1.027 & 1.003 & 0.983 & 0.979 & 0.985 & 0.992  \\
95\% prct &  1.555 & 1.308 & 1.258 & 1.208 & 1.251 & 1.344 & 1.476 & 1.502 & 1.521 & 1.543 & 1.583 & 1.652 & 1.736 & 1.828  \\
Max       &  2.065 & 1.856 & 1.532 & 1.481 & 1.390 & 1.427 & 1.578 & 1.963 & 2.431 & 2.053 & 2.503 & 2.823 & 3.276 & 3.891  \\
Mean      &  0.995 & 0.864 & 0.801 & 0.771 & 0.746 & 0.731 & 0.726 & 0.731 & 0.742 & 0.732 & 0.740 & 0.750 & 0.763 & 0.794  \\
Std       &  0.293 & 0.251 & 0.221 & 0.213 & 0.215 & 0.226 & 0.255 & 0.293 & 0.346 & 0.312 & 0.366 & 0.415 & 0.484 & 0.584  \\
\\
\hline
\end{tabular}
\end{threeparttable}}
\end{table}

\begin{table}
\caption{Forecasting Combination Results for Cases and Deaths: Distribution of Median Absolute Deviation from the Median Ratios}
\label{tab_comb_mad}
\centering
\begin{minipage}{0.9\linewidth}
\begin{footnotesize}
The table presents results with respect forecasting models for the number of cases and deaths.
The table shows descriptive statistics for the ratio of the forecasting median absolute deviation from the median (MAD)
of either the combination ECM and AR or ECM and Trend models and the AR benchmark.
The results are shown for forecasting horizons of 1 to 14 days ahead.
The models were computed on a rolling window scheme with 28 in-sample observations per window.
\end{footnotesize}
\end{minipage}
\resizebox{0.9\linewidth}{!}{
\begin{threeparttable}
\begin{tabular}{lcccccccccccccc}
\hline
\multicolumn{15}{c}{{\ul \textbf{Descriptive Statistics: Ratio of Forecasting Median Absolute Deviation from the Median}}}\\
{\ul \textbf{}}
& \multicolumn{14}{c}{{\ul \textbf{Cases: ECM and AR x AR}}}\\
\textbf{Days ahead} & \textbf{1} & \textbf{2} & \textbf{3} & \textbf{4} & \textbf{5} & \textbf{6} & \textbf{7} & \textbf{8} & \textbf{9} & \textbf{10} & \textbf{11} & \textbf{12} & \textbf{13} & \textbf{14} \\ \hline
Min       &  0.374 & 0.280 & 0.256 & 0.250 & 0.250 & 0.250 & 0.250 & 0.250 & 0.250 & 0.249 & 0.247 & 0.243 & 0.241 & 0.240\\
5\% prct  &  0.585 & 0.397 & 0.331 & 0.289 & 0.278 & 0.263 & 0.260 & 0.255 & 0.254 & 0.250 & 0.250 & 0.250 & 0.280 & 0.250\\
10\% prct &  0.646 & 0.508 & 0.453 & 0.397 & 0.368 & 0.315 & 0.338 & 0.293 & 0.323 & 0.311 & 0.332 & 0.310 & 0.348 & 0.301\\
25\% prct &  0.702 & 0.556 & 0.500 & 0.472 & 0.446 & 0.441 & 0.450 & 0.466 & 0.460 & 0.461 & 0.466 & 0.465 & 0.488 & 0.498\\
Median    &  0.854 & 0.685 & 0.630 & 0.599 & 0.563 & 0.585 & 0.664 & 0.683 & 0.686 & 0.696 & 0.703 & 0.701 & 0.746 & 0.802\\
75\% prct &  1.040 & 0.869 & 0.812 & 0.816 & 0.871 & 0.930 & 0.928 & 0.938 & 0.928 & 0.907 & 0.931 & 0.977 & 1.042 & 1.090\\
90\% prct &  1.441 & 1.357 & 1.221 & 1.396 & 1.587 & 1.603 & 1.703 & 1.603 & 1.514 & 1.485 & 1.447 & 1.414 & 1.460 & 1.562\\
95\% prct &  1.693 & 1.933 & 1.762 & 2.908 & 2.450 & 2.284 & 1.953 & 1.885 & 1.926 & 2.077 & 2.248 & 2.637 & 3.401 & 4.638\\
Max       &  2.730 & 3.374 & 7.722 & 5.560 & 4.491 & 2.927 & 4.263 & 5.261 &11.158 &21.248 &46.520 &91.392 &205.177& 355.921\\
Mean      &  0.958 & 0.841 & 0.876 & 0.894 & 0.855 & 0.805 & 0.840 & 0.862 & 0.997 & 1.229 & 1.810 & 2.838 & 5.536 & 9.048\\
Std       &  0.407 & 0.546 & 1.104 & 1.054 & 0.835 & 0.604 & 0.695 & 0.797 & 1.612 & 3.091 & 6.838 &13.516 &30.806 &53.518\\
\\
& \multicolumn{14}{c}{{\ul \textbf{Deaths: ECM and AR x AR}}}\\
\textbf{Days ahead} & \textbf{1} & \textbf{2} & \textbf{3} & \textbf{4} & \textbf{5} & \textbf{6} & \textbf{7} & \textbf{8} & \textbf{9} & \textbf{10} & \textbf{11} & \textbf{12} & \textbf{13} & \textbf{14} \\ \hline
Min       &   0.766 & 0.663 & 0.597 & 0.593 & 0.543 & 0.508 & 0.516 & 0.502 & 0.542 & 0.535 & 0.528 & 0.522 & 0.556 & 0.536\\
5\% prct  &   0.824 & 0.713 & 0.698 & 0.624 & 0.597 & 0.572 & 0.573 & 0.600 & 0.582 & 0.573 & 0.555 & 0.566 & 0.563 & 0.581\\
10\% prct &   0.856 & 0.773 & 0.713 & 0.649 & 0.621 & 0.628 & 0.612 & 0.606 & 0.605 & 0.598 & 0.589 & 0.595 & 0.583 & 0.601\\
25\% prct &   0.965 & 0.861 & 0.775 & 0.743 & 0.717 & 0.685 & 0.682 & 0.675 & 0.694 & 0.693 & 0.690 & 0.699 & 0.717 & 0.671\\
Median    &   1.016 & 0.921 & 0.887 & 0.840 & 0.820 & 0.811 & 0.784 & 0.794 & 0.838 & 0.818 & 0.801 & 0.782 & 0.782 & 0.785\\
75\% prct &   1.283 & 1.023 & 0.945 & 0.940 & 0.924 & 0.933 & 0.935 & 0.970 & 0.970 & 0.993 & 1.005 & 0.960 & 0.960 & 0.978\\
90\% prct &   1.649 & 1.098 & 1.049 & 1.035 & 0.984 & 1.040 & 1.136 & 1.067 & 1.077 & 1.098 & 1.090 & 1.097 & 1.071 & 1.117\\
95\% prct &   2.312 & 1.412 & 1.464 & 1.235 & 1.123 & 1.185 & 1.248 & 1.104 & 1.131 & 1.185 & 1.156 & 1.137 & 1.167 & 1.174\\
Max       &   4.006 & 1.758 & 1.641 & 1.527 & 1.351 & 1.233 & 1.448 & 1.311 & 1.316 & 1.387 & 1.275 & 1.349 & 1.481 & 1.506\\
Mean      &   1.222 & 0.963 & 0.905 & 0.860 & 0.826 & 0.823 & 0.830 & 0.829 & 0.841 & 0.843 & 0.841 & 0.826 & 0.833 & 0.831\\
Std       &   0.584 & 0.209 & 0.204 & 0.182 & 0.167 & 0.176 & 0.205 & 0.181 & 0.184 & 0.198 & 0.193 & 0.189 & 0.201 & 0.208\\
\\
\hline
\end{tabular}
\end{threeparttable}}
\end{table}

\begin{table}
\caption{Forecasting Combination Results for Cases and Deaths: Distribution of Mean Absolute Error Ratios}
\label{tab_comb_mae}
\centering
\begin{minipage}{0.9\linewidth}
\begin{footnotesize}
The table presents results with respect forecasting models for the number of cases and deaths.
The table shows descriptive statistics for the ratio of the forecasting mean absolute error (MAE)
of either the combination ECM and AR or ECM and Trend models and the AR benchmark.
The results are shown for forecasting horizons of 1 to 14 days ahead.
The models were computed on a rolling window scheme with 28 in-sample observations per window.
\end{footnotesize}
\end{minipage}
\resizebox{0.9\linewidth}{!}{
\begin{threeparttable}
\begin{tabular}{lcccccccccccccc}
\hline
\multicolumn{15}{c}{{\ul \textbf{Descriptive Statistics: Ratio of Forecasting Mean Absolute Error}}}\\
{\ul \textbf{}}
& \multicolumn{14}{c}{{\ul \textbf{Cases: ECM and AR x AR}}}\\
\textbf{Days ahead} & \textbf{1} & \textbf{2} & \textbf{3} & \textbf{4} & \textbf{5} & \textbf{6} & \textbf{7} & \textbf{8} & \textbf{9} & \textbf{10} & \textbf{11} & \textbf{12} & \textbf{13} & \textbf{14} \\ \hline
Min       &  0.809 & 0.645 & 0.539 & 0.502 & 0.500 & 0.491 & 0.483 & 0.475 & 0.471 & 0.464 & 0.462 & 0.462 & 0.458 & 0.457\\
5\% prct  &  0.832 & 0.704 & 0.595 & 0.526 & 0.506 & 0.500 & 0.500 & 0.500 & 0.500 & 0.500 & 0.500 & 0.500 & 0.500 & 0.500\\
10\% prct &  0.852 & 0.716 & 0.671 & 0.632 & 0.589 & 0.566 & 0.550 & 0.561 & 0.570 & 0.574 & 0.574 & 0.569 & 0.565 & 0.563\\
25\% prct &  0.880 & 0.760 & 0.716 & 0.689 & 0.668 & 0.664 & 0.659 & 0.662 & 0.663 & 0.664 & 0.665 & 0.665 & 0.653 & 0.644\\
Median    &  0.943 & 0.839 & 0.777 & 0.753 & 0.763 & 0.755 & 0.753 & 0.766 & 0.775 & 0.771 & 0.771 & 0.778 & 0.809 & 0.825\\
75\% prct &  1.030 & 0.932 & 0.895 & 0.883 & 0.896 & 0.929 & 0.958 & 0.966 & 0.967 & 0.968 & 0.978 & 0.979 & 1.000 & 1.040\\
90\% prct &  1.228 & 1.152 & 1.068 & 1.033 & 1.052 & 1.046 & 1.068 & 1.077 & 1.093 & 1.098 & 1.110 & 1.133 & 1.182 & 1.262\\
95\% prct &  1.294 & 1.305 & 1.292 & 1.274 & 1.248 & 1.237 & 1.227 & 1.211 & 1.191 & 1.187 & 1.211 & 1.252 & 1.328 & 1.357\\
Max       &  1.771 & 1.731 & 1.664 & 1.571 & 1.508 & 1.453 & 1.415 & 1.385 & 1.371 & 1.363 & 1.358 & 1.356 & 1.462 & 1.712\\
Mean      &  0.989 & 0.888 & 0.837 & 0.812 & 0.802 & 0.802 & 0.806 & 0.810 & 0.813 & 0.816 & 0.821 & 0.830 & 0.844 & 0.862\\
Std       &  0.174 & 0.199 & 0.213 & 0.217 & 0.214 & 0.211 & 0.210 & 0.207 & 0.207 & 0.211 & 0.216 & 0.228 & 0.248 & 0.273\\
\\
& \multicolumn{14}{c}{{\ul \textbf{Deaths: ECM and AR x AR}}}\\
\textbf{Days ahead} & \textbf{1} & \textbf{2} & \textbf{3} & \textbf{4} & \textbf{5} & \textbf{6} & \textbf{7} & \textbf{8} & \textbf{9} & \textbf{10} & \textbf{11} & \textbf{12} & \textbf{13} & \textbf{14} \\ \hline
Min       &   0.512 & 0.500 & 0.500 & 0.521 & 0.524&  0.524 & 0.513 & 0.507 & 0.503 & 0.501 & 0.500 & 0.500 & 0.500 & 0.500\\
5\% prct  &   0.531 & 0.527 & 0.534 & 0.552 & 0.547&  0.528 & 0.523 & 0.520 & 0.514 & 0.508 & 0.504 & 0.502 & 0.501 & 0.500\\
10\% prct &   0.767 & 0.643 & 0.583 & 0.594 & 0.567&  0.558 & 0.532 & 0.529 & 0.530 & 0.531 & 0.531 & 0.522 & 0.504 & 0.501\\
25\% prct &   0.888 & 0.761 & 0.696 & 0.662 & 0.636&  0.612 & 0.614 & 0.599 & 0.608 & 0.600 & 0.590 & 0.580 & 0.574 & 0.565\\
Median    &   0.984 & 0.853 & 0.783 & 0.744 & 0.712&  0.686 & 0.681 & 0.684 & 0.669 & 0.662 & 0.660 & 0.659 & 0.644 & 0.640\\
75\% prct &   1.125 & 0.968 & 0.904 & 0.861 & 0.867&  0.843 & 0.814 & 0.803 & 0.801 & 0.780 & 0.775 & 0.772 & 0.773 & 0.778\\
90\% prct &   1.345 & 1.132 & 1.190 & 1.169 & 1.091&  1.030 & 1.049 & 1.116 & 1.135 & 1.152 & 1.140 & 1.144 & 1.137 & 1.140\\
95\% prct &   1.607 & 1.394 & 1.357 & 1.282 & 1.257&  1.242 & 1.207 & 1.192 & 1.228 & 1.212 & 1.241 & 1.317 & 1.395 & 1.612\\
Max       &   5.487 & 5.400 & 4.195 & 3.980 & 3.580&  3.485 & 3.677 & 3.690 & 3.670 & 3.655 & 3.662 & 3.796 & 4.157 & 4.514\\
Mean      &   1.103 & 0.973 & 0.892 & 0.859 & 0.823&  0.805 & 0.800 & 0.799 & 0.800 & 0.796 & 0.794 & 0.797 & 0.807 & 0.827\\
Std       &   0.715 & 0.705 & 0.542 & 0.513 & 0.460&  0.449 & 0.476 & 0.480 & 0.481 & 0.481 & 0.488 & 0.514 & 0.572 & 0.637\\
\\
\hline
\end{tabular}
\end{threeparttable}}
\end{table}

\begin{table}
\caption{Forecasting Combination Results for Cases and Deaths: Distribution of Mean Squared Error Ratios}
\label{tab_comb_mse}
\centering
\begin{minipage}{0.9\linewidth}
\begin{footnotesize}
The table presents results with respect forecasting models for the number of cases and deaths.
The table shows descriptive statistics for the ratio of the forecasting mean squared error (MSE)
of either the combination ECM and AR or ECM and Trend models and the AR benchmark.
The results are shown for forecasting horizons of 1 to 14 days ahead.
The models were computed on a rolling window scheme with 28 in-sample observations per window.
\end{footnotesize}
\end{minipage}
\resizebox{0.9\linewidth}{!}{
\begin{threeparttable}
\begin{tabular}{lcccccccccccccc}
\hline
\multicolumn{15}{c}{{\ul \textbf{Descriptive Statistics: Ratio of Forecasting Mean Squared Error}}}\\
{\ul \textbf{}}
& \multicolumn{14}{c}{{\ul \textbf{Cases: ECM and AR x AR}}}\\
\textbf{Days ahead} & \textbf{1} & \textbf{2} & \textbf{3} & \textbf{4} & \textbf{5} & \textbf{6} & \textbf{7} & \textbf{8} & \textbf{9} & \textbf{10} & \textbf{11} & \textbf{12} & \textbf{13} & \textbf{14} \\ \hline
Min       &  0.374 & 0.280 & 0.256 & 0.250 & 0.250 & 0.250 & 0.250 & 0.250 & 0.250 & 0.249 & 0.247 & 0.243 & 0.241 & 0.240\\
5\% prct  &  0.585 & 0.397 & 0.331 & 0.289 & 0.278 & 0.263 & 0.260 & 0.255 & 0.254 & 0.250 & 0.250 & 0.250 & 0.280 & 0.250\\
10\% prct &  0.646 & 0.508 & 0.453 & 0.397 & 0.368 & 0.315 & 0.338 & 0.293 & 0.323 & 0.311 & 0.332 & 0.310 & 0.348 & 0.301\\
25\% prct &  0.702 & 0.556 & 0.500 & 0.472 & 0.446 & 0.441 & 0.450 & 0.466 & 0.460 & 0.461 & 0.466 & 0.465 & 0.488 & 0.498\\
Median    &  0.854 & 0.685 & 0.630 & 0.599 & 0.563 & 0.585 & 0.664 & 0.683 & 0.686 & 0.696 & 0.703 & 0.701 & 0.746 & 0.802\\
75\% prct &  1.040 & 0.869 & 0.812 & 0.816 & 0.871 & 0.930 & 0.928 & 0.938 & 0.928 & 0.907 & 0.931 & 0.977 & 1.042 & 1.090\\
90\% prct &  1.441 & 1.357 & 1.221 & 1.396 & 1.587 & 1.603 & 1.703 & 1.603 & 1.514 & 1.485 & 1.447 & 1.414 & 1.460 & 1.562\\
95\% prct &  1.693 & 1.933 & 1.762 & 2.908 & 2.450 & 2.284 & 1.953 & 1.885 & 1.926 & 2.077 & 2.248 & 2.637 & 3.401 & 4.638\\
Max       &  2.730 & 3.374 & 7.722 & 5.560 & 4.491 & 2.927 & 4.263 & 5.261 &11.158 &21.248 &46.520 &91.392 &205.177& 355.921\\
Mean      &  0.958 & 0.841 & 0.876 & 0.894 & 0.855 & 0.805 & 0.840 & 0.862 & 0.997 & 1.229 & 1.810 & 2.838 & 5.536 & 9.048\\
Std       &  0.407 & 0.546 & 1.104 & 1.054 & 0.835 & 0.604 & 0.695 & 0.797 & 1.612 & 3.091 & 6.838 &13.516 &30.806 &53.518\\
\\
& \multicolumn{14}{c}{{\ul \textbf{Deaths: ECM and AR x AR}}}\\
\textbf{Days ahead} & \textbf{1} & \textbf{2} & \textbf{3} & \textbf{4} & \textbf{5} & \textbf{6} & \textbf{7} & \textbf{8} & \textbf{9} & \textbf{10} & \textbf{11} & \textbf{12} & \textbf{13} & \textbf{14} \\ \hline
Min       &    0.247&   0.250&  0.250&  0.247&  0.251&  0.250&  0.250&  0.250&  0.249&   0.245&   0.244&   0.244&   0.246&   0.247\\
5\% prct  &    0.250&   0.250&  0.270&  0.251&  0.255&  0.253&  0.252&  0.252&  0.250&   0.250&   0.250&   0.250&   0.250&   0.250\\
10\% prct &    0.473&   0.344&  0.321&  0.276&  0.260&  0.258&  0.257&  0.257&  0.253&   0.252&   0.250&   0.250&   0.250&   0.250\\
25\% prct &    0.679&   0.505&  0.437&  0.424&  0.383&  0.359&  0.302&  0.326&  0.325&   0.300&   0.302&   0.329&   0.345&   0.342\\
Median    &    0.871&   0.685&  0.599&  0.546&  0.509&  0.485&  0.470&  0.461&  0.435&   0.444&   0.441&   0.452&   0.426&   0.431\\
75\% prct &    1.244&   1.087&  0.937&  0.856&  0.788&  0.775&  0.783&  0.745&  0.749&   0.759&   0.785&   0.824&   0.859&   0.914\\
90\% prct &    1.523&   1.351&  1.656&  1.516&  1.539&  1.397&  1.487&  1.582&  1.625&   1.528&   1.466&   1.479&   1.456&   1.657\\
95\% prct &    2.343&   2.299&  2.680&  2.537&  2.173&  1.992&  1.945&  1.893&  2.329&   2.192&   3.198&   4.201&   5.231&   8.144\\
Max       &  303.101& 152.264& 69.165& 59.955& 47.869& 49.245& 81.825& 98.138& 96.613& 116.896& 145.307& 190.037& 270.110& 376.467\\
Mean      &    7.681&   4.169&  2.354&  2.050&  1.726&  1.723&  2.434&  2.802&  2.791&   3.240&   3.908&   4.951&   6.799&   9.317\\
Std       &  4 5.043&  22.583& 10.325&  8.846&  7.053&  7.260& 12.112& 14.543& 14.314&  17.337&  21.569&  28.231&  40.160&  56.002\\
\\
\hline
\end{tabular}
\end{threeparttable}}
\end{table}

\begin{table}
\caption{Forecasting Results: Distribution of Exception Rates (90\% Prediction Interval)}
\label{tab_ic90}
\centering
\begin{minipage}{0.9\linewidth}
\begin{footnotesize}
The table presents results with respect the prediction intervals produced by the \texttt{ECM} for both cases and deaths.
The table shows descriptive statistics for the frequency that the absolute out-of-sample errors of the \texttt{ECM} model exceeds the 90\% prediction interval. 
The results are shown for forecasting horizons of 1 to 14 days ahead.
The models were computed on a rolling window scheme with 28 in-sample observations per window.
\end{footnotesize}
\end{minipage}
\resizebox{0.9\linewidth}{!}{
\begin{threeparttable}
\begin{tabular}{lcccccccccccccc}
\hline
\multicolumn{15}{c}{{\ul \textbf{Descriptive Statistics: Exception Rates (90\% Confidence Interval)}}}\\
{\ul \textbf{}}
& \multicolumn{14}{c}{{\ul \textbf{Cases: ECM}}}\\
\textbf{Days ahead} & \textbf{1} & \textbf{2} & \textbf{3} & \textbf{4} & \textbf{5} & \textbf{6} & \textbf{7} & \textbf{8} & \textbf{9} & \textbf{10} & \textbf{11} & \textbf{12} & \textbf{13} & \textbf{14} \\ \hline
Min       &   0.026 & 0.022 & 0.014 & 0.003 & 0.003 & 0.003 & 0.002 & 0.002 & 0.002 & 0.002 & 0.002 & 0.002 & 0.002 & 0.002\\
5\% prct  &   0.038 & 0.042 & 0.041 & 0.027 & 0.024 & 0.013 & 0.012 & 0.021 & 0.021 & 0.021 & 0.018 & 0.018 & 0.014 & 0.005\\
10\% prct &   0.051 & 0.054 & 0.046 & 0.037 & 0.035 & 0.032 & 0.041 & 0.035 & 0.037 & 0.037 & 0.045 & 0.040 & 0.022 & 0.021\\
25\% prct &   0.072 & 0.068 & 0.068 & 0.066 & 0.063 & 0.070 & 0.071 & 0.068 & 0.066 & 0.067 & 0.067 & 0.062 & 0.060 & 0.061\\
Median    &   0.086 & 0.086 & 0.087 & 0.091 & 0.087 & 0.088 & 0.088 & 0.089 & 0.085 & 0.087 & 0.082 & 0.082 & 0.081 & 0.079\\
75\% prct &   0.098 & 0.102 & 0.101 & 0.101 & 0.103 & 0.104 & 0.100 & 0.097 & 0.100 & 0.102 & 0.102 & 0.095 & 0.097 & 0.099\\
90\% prct &   0.108 & 0.110 & 0.111 & 0.113 & 0.113 & 0.120 & 0.120 & 0.120 & 0.124 & 0.120 & 0.125 & 0.128 & 0.127 & 0.125\\
95\% prct &   0.114 & 0.113 & 0.115 & 0.116 & 0.121 & 0.125 & 0.131 & 0.133 & 0.137 & 0.141 & 0.137 & 0.132 & 0.134 & 0.131\\
Max       &   0.115 & 0.118 & 0.128 & 0.128 & 0.126 & 0.135 & 0.140 & 0.143 & 0.148 & 0.153 & 0.155 & 0.148 & 0.140 & 0.136\\
Mean      &   0.082 & 0.084 & 0.084 & 0.082 & 0.082 & 0.084 & 0.084 & 0.083 & 0.083 & 0.083 & 0.082 & 0.079 & 0.078 & 0.076\\
Std       &   0.021 & 0.022 & 0.025 & 0.028 & 0.029 & 0.031 & 0.031 & 0.031 & 0.032 & 0.032 & 0.032 & 0.033 & 0.034 & 0.035\\
\\
& \multicolumn{14}{c}{{\ul \textbf{Deaths: ECM}}}\\
\textbf{Days ahead} & \textbf{1} & \textbf{2} & \textbf{3} & \textbf{4} & \textbf{5} & \textbf{6} & \textbf{7} & \textbf{8} & \textbf{9} & \textbf{10} & \textbf{11} & \textbf{12} & \textbf{13} & \textbf{14} \\ \hline
Min       &   0.016 & 0.016 & 0.014 & 0.018 & 0.018 & 0.023 & 0.022 & 0.016 & 0.016 & 0.018 & 0.018 & 0.015 & 0.011 & 0.011\\
5\% prct  &   0.021 & 0.027 & 0.027 & 0.027 & 0.030 & 0.030 & 0.029 & 0.024 & 0.025 & 0.023 & 0.019 & 0.018 & 0.017 & 0.015\\
10\% prct &   0.026 & 0.035 & 0.030 & 0.031 & 0.032 & 0.035 & 0.032 & 0.032 & 0.031 & 0.028 & 0.026 & 0.026 & 0.022 & 0.022\\
25\% prct &   0.054 & 0.044 & 0.044 & 0.045 & 0.051 & 0.050 & 0.058 & 0.051 & 0.055 & 0.052 & 0.051 & 0.047 & 0.053 & 0.052\\
Median    &   0.084 & 0.077 & 0.076 & 0.073 & 0.076 & 0.074 & 0.072 & 0.074 & 0.072 & 0.075 & 0.077 & 0.072 & 0.077 & 0.074\\
75\% prct &   0.095 & 0.090 & 0.091 & 0.093 & 0.097 & 0.096 & 0.092 & 0.088 & 0.090 & 0.094 & 0.093 & 0.089 & 0.091 & 0.091\\
90\% prct &   0.106 & 0.107 & 0.103 & 0.111 & 0.105 & 0.105 & 0.105 & 0.105 & 0.102 & 0.104 & 0.102 & 0.100 & 0.097 & 0.099\\
95\% prct &   0.111 & 0.114 & 0.114 & 0.113 & 0.114 & 0.110 & 0.108 & 0.106 & 0.107 & 0.114 & 0.111 & 0.108 & 0.103 & 0.104\\
Max       &   0.134 & 0.124 & 0.150 & 0.135 & 0.122 & 0.124 & 0.122 & 0.129 & 0.116 & 0.129 & 0.135 & 0.150 & 0.154 & 0.150\\
Mean      &   0.076 & 0.071 & 0.071 & 0.070 & 0.072 & 0.072 & 0.072 & 0.071 & 0.071 & 0.072 & 0.072 & 0.070 & 0.069 & 0.069\\
Std       &   0.029 & 0.027 & 0.030 & 0.030 & 0.029 & 0.026 & 0.025 & 0.027 & 0.026 & 0.029 & 0.029 & 0.030 & 0.031 & 0.030\\
\hline
\end{tabular}
\end{threeparttable}}
\end{table}

\begin{table}
\caption{Forecasting Results: Distribution of Exception Rates (95\% Prediction Interval)}
\label{tab_ic95}
\centering
\begin{minipage}{0.9\linewidth}
\begin{footnotesize}
The table presents results with respect the prediction intervals produced by the \texttt{ECM} for both cases and deaths.
The table shows descriptive statistics for the frequency that the absolute out-of-sample errors of the \texttt{ECM} model exceeds the 95\% prediction interval.
The results are shown for forecasting horizons of 1 to 14 days ahead.
The models were computed on a rolling window scheme with 28 in-sample observations per window.
\end{footnotesize}
\end{minipage}
\resizebox{0.9\linewidth}{!}{
\begin{threeparttable}
\begin{tabular}{lcccccccccccccc}
\hline
\multicolumn{15}{c}{{\ul \textbf{Descriptive Statistics: Exception Rates (95\% Confidence Interval)}}}\\
{\ul \textbf{}}
& \multicolumn{14}{c}{{\ul \textbf{Cases: ECM}}}\\
\textbf{Days ahead} & \textbf{1} & \textbf{2} & \textbf{3} & \textbf{4} & \textbf{5} & \textbf{6} & \textbf{7} & \textbf{8} & \textbf{9} & \textbf{10} & \textbf{11} & \textbf{12} & \textbf{13} & \textbf{14} \\ \hline
Min       &  0.026 & 0.013 & 0.011 & 0.003 & 0.003 & 0.002 & 0.002 & 0.002 & 0.002 & 0.002 & 0.002 & 0.002 & 0.002 & 0.002\\
5\% prct  &  0.029 & 0.034 & 0.028 & 0.020 & 0.016 & 0.011 & 0.012 & 0.008 & 0.015 & 0.018 & 0.018 & 0.015 & 0.007 & 0.005\\
10\% prct &  0.033 & 0.042 & 0.038 & 0.031 & 0.024 & 0.028 & 0.029 & 0.027 & 0.029 & 0.024 & 0.033 & 0.024 & 0.021 & 0.019\\
25\% prct &  0.049 & 0.050 & 0.053 & 0.049 & 0.050 & 0.056 & 0.054 & 0.052 & 0.051 & 0.050 & 0.049 & 0.051 & 0.045 & 0.045\\
Median    &  0.059 & 0.061 & 0.063 & 0.064 & 0.064 & 0.066 & 0.067 & 0.066 & 0.064 & 0.062 & 0.064 & 0.062 & 0.061 & 0.061\\
75\% prct &  0.072 & 0.071 & 0.073 & 0.073 & 0.076 & 0.074 & 0.074 & 0.071 & 0.071 & 0.075 & 0.077 & 0.075 & 0.076 & 0.075\\
90\% prct &  0.076 & 0.080 & 0.080 & 0.081 & 0.083 & 0.090 & 0.095 & 0.090 & 0.091 & 0.092 & 0.090 & 0.092 & 0.090 & 0.086\\
95\% prct &  0.078 & 0.085 & 0.085 & 0.092 & 0.093 & 0.097 & 0.102 & 0.098 & 0.098 & 0.096 & 0.099 & 0.099 & 0.098 & 0.097\\
Max       &  0.080 & 0.092 & 0.093 & 0.098 & 0.100 & 0.100 & 0.103 & 0.105 & 0.105 & 0.098 & 0.102 & 0.108 & 0.105 & 0.120\\
Mean      &  0.059 & 0.060 & 0.061 & 0.060 & 0.061 & 0.062 & 0.063 & 0.061 & 0.061 & 0.061 & 0.061 & 0.061 & 0.059 & 0.058\\
Std       &  0.015 & 0.016 & 0.017 & 0.020 & 0.023 & 0.023 & 0.024 & 0.023 & 0.023 & 0.023 & 0.023 & 0.024 & 0.025 & 0.026\\
\\
& \multicolumn{14}{c}{{\ul \textbf{Deaths: ECM}}}\\
\textbf{Days ahead} & \textbf{1} & \textbf{2} & \textbf{3} & \textbf{4} & \textbf{5} & \textbf{6} & \textbf{7} & \textbf{8} & \textbf{9} & \textbf{10} & \textbf{11} & \textbf{12} & \textbf{13} & \textbf{14} \\ \hline
Min       &  0.016 & 0.013 & 0.011 & 0.014 & 0.016 & 0.018 & 0.015 & 0.016 & 0.014 & 0.015 & 0.014 & 0.011 & 0.011 & 0.011\\
5\% prct  &  0.017 & 0.020 & 0.021 & 0.024 & 0.026 & 0.024 & 0.020 & 0.022 & 0.019 & 0.018 & 0.015 & 0.017 & 0.015 & 0.013\\
10\% prct &  0.023 & 0.023 & 0.022 & 0.026 & 0.029 & 0.027 & 0.030 & 0.027 & 0.021 & 0.020 & 0.018 & 0.019 & 0.019 & 0.019\\
25\% prct &  0.042 & 0.037 & 0.035 & 0.033 & 0.035 & 0.033 & 0.035 & 0.035 & 0.036 & 0.037 & 0.037 & 0.033 & 0.041 & 0.038\\
Median    &  0.054 & 0.050 & 0.052 & 0.054 & 0.052 & 0.052 & 0.053 & 0.055 & 0.055 & 0.056 & 0.054 & 0.054 & 0.052 & 0.054\\
75\% prct &  0.065 & 0.065 & 0.063 & 0.064 & 0.063 & 0.062 & 0.064 & 0.064 & 0.064 & 0.063 & 0.064 & 0.067 & 0.066 & 0.068\\
90\% prct &  0.072 & 0.071 & 0.072 & 0.072 & 0.074 & 0.075 & 0.072 & 0.075 & 0.073 & 0.074 & 0.072 & 0.075 & 0.072 & 0.074\\
95\% prct &  0.079 & 0.079 & 0.078 & 0.076 & 0.077 & 0.080 & 0.076 & 0.078 & 0.079 & 0.079 & 0.078 & 0.077 & 0.074 & 0.075\\
Max       &  0.091 & 0.105 & 0.087 & 0.082 & 0.079 & 0.085 & 0.082 & 0.088 & 0.082 & 0.086 & 0.090 & 0.097 & 0.101 & 0.097\\
Mean      &  0.052 & 0.051 & 0.049 & 0.050 & 0.050 & 0.051 & 0.051 & 0.052 & 0.051 & 0.052 & 0.051 & 0.051 & 0.050 & 0.050\\
Std       &  0.018 & 0.020 & 0.019 & 0.018 & 0.017 & 0.017 & 0.017 & 0.018 & 0.018 & 0.019 & 0.019 & 0.021 & 0.020 & 0.021\\
\hline
\end{tabular}
\end{threeparttable}}
\end{table}

\begin{table}
\caption{Forecasting Results: Distribution of Exception Rates (99\% Prediction Interval)}
\label{tab_ic99}
\centering
\begin{minipage}{0.9\linewidth}
\begin{footnotesize}
The table presents results with respect the prediction intervals produced by the \texttt{ECM} for both cases and deaths.
The table shows descriptive statistics for the frequency that the absolute out-of-sample errors of the \texttt{ECM} model exceeds the 99\% prediction interval.
The results are shown for forecasting horizons of 1 to 14 days ahead.
The models were computed on a rolling window scheme with 28 in-sample observations per window.
\end{footnotesize}
\end{minipage}
\resizebox{0.9\linewidth}{!}{
\begin{threeparttable}
\begin{tabular}{lcccccccccccccc}
\hline
\multicolumn{15}{c}{{\ul \textbf{Descriptive Statistics: Exception Rates (99\% Confidence Interval)}}}\\
{\ul \textbf{}}
& \multicolumn{14}{c}{{\ul \textbf{Cases: ECM}}}\\
\textbf{Days ahead} & \textbf{1} & \textbf{2} & \textbf{3} & \textbf{4} & \textbf{5} & \textbf{6} & \textbf{7} & \textbf{8} & \textbf{9} & \textbf{10} & \textbf{11} & \textbf{12} & \textbf{13} & \textbf{14} \\ \hline
Min       &   0.012 & 0.002 & 0.004 & 0.003 & 0.002 & 0.002 & 0.002 & 0.002 & 0.002 & 0.002 & 0.002 & 0.002 & 0.002 & 0.002\\
5\% prct  &   0.015 & 0.014 & 0.013 & 0.011 & 0.009 & 0.011 & 0.009 & 0.006 & 0.008 & 0.008 & 0.009 & 0.005 & 0.005 & 0.005\\
10\% prct &   0.019 & 0.018 & 0.018 & 0.019 & 0.016 & 0.016 & 0.019 & 0.019 & 0.019 & 0.018 & 0.015 & 0.013 & 0.015 & 0.015\\
25\% prct &   0.025 & 0.024 & 0.028 & 0.028 & 0.026 & 0.027 & 0.029 & 0.027 & 0.025 & 0.027 & 0.024 & 0.026 & 0.025 & 0.026\\
Median    &   0.033 & 0.031 & 0.034 & 0.037 & 0.035 & 0.036 & 0.036 & 0.037 & 0.036 & 0.035 & 0.036 & 0.034 & 0.035 & 0.033\\
75\% prct &   0.038 & 0.038 & 0.041 & 0.042 & 0.042 & 0.041 & 0.043 & 0.042 & 0.041 & 0.042 & 0.044 & 0.042 & 0.041 & 0.041\\
90\% prct &   0.042 & 0.045 & 0.048 & 0.046 & 0.048 & 0.047 & 0.048 & 0.050 & 0.048 & 0.053 & 0.053 & 0.054 & 0.053 & 0.054\\
95\% prct &   0.047 & 0.052 & 0.052 & 0.050 & 0.057 & 0.056 & 0.055 & 0.056 & 0.056 & 0.059 & 0.062 & 0.062 & 0.064 & 0.058\\
Max       &   0.053 & 0.055 & 0.056 & 0.052 & 0.060 & 0.067 & 0.067 & 0.082 & 0.082 & 0.079 & 0.082 & 0.075 & 0.071 & 0.064\\
Mean      &   0.032 & 0.031 & 0.034 & 0.034 & 0.034 & 0.034 & 0.035 & 0.035 & 0.034 & 0.035 & 0.035 & 0.034 & 0.034 & 0.033\\
Std       &   0.009 & 0.011 & 0.011 & 0.011 & 0.013 & 0.013 & 0.013 & 0.014 & 0.014 & 0.015 & 0.016 & 0.016 & 0.015 & 0.014\\
\\
& \multicolumn{14}{c}{{\ul \textbf{Deaths: ECM}}}\\
\textbf{Days ahead} & \textbf{1} & \textbf{2} & \textbf{3} & \textbf{4} & \textbf{5} & \textbf{6} & \textbf{7} & \textbf{8} & \textbf{9} & \textbf{10} & \textbf{11} & \textbf{12} & \textbf{13} & \textbf{14} \\ \hline
Min       &   0.010 & 0.005 & 0.007 & 0.011 & 0.011 & 0.011 & 0.011 & 0.011 & 0.011 & 0.011 & 0.007 & 0.007 & 0.007 & 0.009 \\
5\% prct  &   0.011 & 0.010 & 0.011 & 0.012 & 0.014 & 0.014 & 0.012 & 0.013 & 0.013 & 0.012 & 0.011 & 0.011 & 0.013 & 0.012 \\
10\% prct &   0.013 & 0.013 & 0.014 & 0.016 & 0.015 & 0.015 & 0.015 & 0.015 & 0.014 & 0.014 & 0.014 & 0.013 & 0.015 & 0.014 \\
25\% prct &   0.020 & 0.021 & 0.020 & 0.023 & 0.023 & 0.023 & 0.019 & 0.018 & 0.019 & 0.019 & 0.020 & 0.019 & 0.021 & 0.019 \\
Median    &   0.027 & 0.027 & 0.027 & 0.026 & 0.028 & 0.028 & 0.028 & 0.030 & 0.030 & 0.030 & 0.030 & 0.029 & 0.030 & 0.029 \\
75\% prct &   0.034 & 0.033 & 0.034 & 0.033 & 0.034 & 0.033 & 0.036 & 0.038 & 0.036 & 0.034 & 0.035 & 0.037 & 0.038 & 0.040 \\
90\% prct &   0.038 & 0.035 & 0.040 & 0.042 & 0.044 & 0.040 & 0.041 & 0.042 & 0.041 & 0.041 & 0.045 & 0.046 & 0.046 & 0.047 \\
95\% prct &   0.043 & 0.043 & 0.046 & 0.044 & 0.048 & 0.051 & 0.053 & 0.053 & 0.054 & 0.054 & 0.054 & 0.058 & 0.056 & 0.056 \\
Max       &   0.047 & 0.046 & 0.050 & 0.050 & 0.051 & 0.059 & 0.060 & 0.060 & 0.064 & 0.064 & 0.062 & 0.059 & 0.059 & 0.064 \\
Mean      &   0.027 & 0.026 & 0.027 & 0.028 & 0.029 & 0.029 & 0.029 & 0.029 & 0.030 & 0.029 & 0.030 & 0.029 & 0.030 & 0.030 \\
Std       &   0.010 & 0.009 & 0.010 & 0.010 & 0.010 & 0.010 & 0.012 & 0.012 & 0.012 & 0.012 & 0.012 & 0.013 & 0.013 & 0.013 \\
\hline
\end{tabular}
\end{threeparttable}}
\end{table}

\end{landscape}

%----------------------------------------------------------------------------
% Figures
%----------------------------------------------------------------------------

\end{document}